\documentclass[12pt,a4paper]{article}
\usepackage[margin=2cm]{geometry}
\usepackage{comment}
\usepackage{amsmath,amssymb,extarrows,graphicx,subfigure,setspace}
\usepackage{cite}
\usepackage{slashed}
\usepackage{color}
\usepackage{braket}
\usepackage{float}
\makeatother
\usepackage{pdfpages}
\usepackage{blindtext}
\usepackage{amsfonts}
\usepackage{tensor}
\usepackage{graphicx}
\usepackage{pict2e}
\usepackage{amssymb, amsmath,environ}
\usepackage[
colorlinks=true,
linkcolor=blue,
urlcolor=blue,
filecolor=blue,
citecolor=red,
pdfstartview=FitV,
pdftitle={},
pdfauthor={},
pdfsubject={},
pdfkeywords={},
pdfpagemode=None,
bookmarksopen=true
]{hyperref}

\usepackage{epsfig}

\usepackage{hyperref}

\newcommand{\be}{\begin{equation}}
	\newcommand{\bea}{\begin{eqnarray}}
		\newcommand{\eea}{\end{eqnarray}}
	\newcommand{\ba}{\begin{array}}
		\newcommand{\ea}{\end{array}}
	\newcommand{\ee}{\end{equation}}
\newcommand{\bes}{\begin{equation*}}
	\newcommand{\beas}{\begin{eqnarray*}}
		\newcommand{\eeas}{\end{eqnarray*}}
	\newcommand{\bas}{\begin{array*}}
		\newcommand{\eas}{\end{array*}}
	\newcommand{\ees}{\end{equation*}}

\numberwithin{equation}{section}
\begin{document}
\onehalfspacing
\noindent
	
\begin{titlepage}
\vspace{10mm}
	
\vspace*{20mm}
\begin{center}
{\Large {\bf Odd Entanglement Entropy and Logarithmic Negativity for Thermofield Double States}\\
}
		
\vspace*{15mm}
\vspace*{1mm}
{Mostafa Ghasemi$^a$, Ali Naseh$^a$ and Reza Pirmoradian$^{a,b}$}

\vspace*{1cm}

{\it 
	${}^a$ School of Particles and Accelerators, Institute for Research in Fundamental Sciences (IPM)
	P.O. Box 19395-5531, Tehran, Iran

${}^b$ Department of physics, Islamic Azad University Central Tehran Branch,Tehran, Iran\\}

\vspace*{0.5cm}
{E-mails: {\tt ghasemi.mg@ipm.ir,  naseh@ipm.ir, rezapirmoradian@ipm.ir}}

\vspace*{1cm}
\end{center}

\date{\today}

\begin{abstract} 
We investigate the time evolution of odd entanglement entropy (OEE) and logarithmic
negativity (LN) for the thermofield double (TFD) states in free scalar quantum field
theories using the covariance matrix approach.
To have mixed states, we choose non-complementary subsystems, either adjacent or disjoint intervals on each side of the TFD. We find that the time evolution pattern of OEE is a linear growth followed by saturation. On a circular lattice, for longer times the finite size effect demonstrates itself as oscillatory behavior.
In the limit of vanishing mass, for a subsystem containing a single degree of freedom on each side of the TFD, we analytically find the effect of zero-mode on the time evolution of OEE which leads to logarithmic growth in the intermediate times. 
Moreover, for adjacent intervals we find that the LN is zero for times $t < \beta/2$ (half of the inverse temperature) and after that, it begins to grow linearly. For disjoint intervals at fixed temperature, the vanishing of LN is observed for times $t<d/2$ (half of the distance between intervals). We also find a similar delay to see linear growth of $\Delta S=S_{\text{OEE}}-S_{\text{EE}}$.   
All these results show that the dynamics of these measures are consistent with the quasi-particle picture, of course apart from the logarithmic growth. 


\end{abstract}

\end{titlepage} 

\section{Introduction} \label{intro}
Understanding the non-equilibrium dynamics of the isolated many-body quantum systems has been one of the major research avenues, both theoretical
and experimental, in the last few decades\cite{Polkovnikov:2010yn,Kaufman:2016hyn,Gogolin:2016hwy,Essler:2016ufo,Vidmar:2016jtx}. One of the protocols for preparing the out of equilibrium systems is the global quench setting that enables us to study thermalization in isolated quantum systems and can be best understood by phenomenological quasi-particle picture\cite{Calabrese:2005in,Calabrese:2006rx,Calabrese:2007rg,Alba:2016lvc,Alba:2017lvc}. In this setup, an isolated system is initially prepared at $t=0$ in a given ground state $|\psi_{0}\rangle$ of a certain Hamiltonian $H_{0}$ and undergoes a sudden change in a parameter of the Hamiltonian at $t=0$. The system evolves with a new Hamiltonian $H$ in such a way that $ [ H,H_{0}] \neq 0 $ and the time evolved state at time $t$ is given by $\vert \psi(t) \rangle =e^{-iHt} |\psi_{0}\rangle$. Since the evolution is unitary, the final state of the system is described by a pure state and one of the appropriate concepts for understanding the thermalization is quantum entanglement, $S_{\text{EE}}$ \cite{Horodecki:2009zz}. When considering this time evolution,  there is a regime in which the initial growth is linear and it occurs when the lattice spacing $\delta$ is much
smaller than the correlation length, $\beta$, which itself is much smaller than the subsystem size $l$, i.e.,
\bea
\delta \ll \beta \ll l.
\eea
In the quasi-particle picture, the quench creates independent entangled pairs which propagate on a circle (with circumference $\mathcal{L}$) in a ballistic fashion with an effective velocity 
\bea\label{vn}
v_{n} = \frac{\mathcal{L}}{2\pi}\partial_{n}\omega_{n},
\eea
where $n$ is an integer that runs from $0$ to the total number of lattice sites and $\omega_n$ is given by (\ref{omegak}). The upper bound of velocity is known as Lieb-Robinson
bound \cite{Lieb:1972zax}. When the quasi-particles and their partners laying in the interval and it is complementary, respectively(or vise versa), we observe the linear growth in the entanglement entropy where
Alba and Calabrese  \cite{Alba:2016lvc,Alba:2017lvc} provided a formula for it. By  carefully tracking quasi-particles on a circle leaving and
re-entering the interval, a phenomenological
relation for  $ S^{\text{TFD}}_{\text{EE}}(t)-S^{\text{TFD}}_{\text{EE}}(t=0)$ in TFD state\footnote{In the following we will define this state.}, is proposed \cite{Alba:2016lvc,Alba:2017lvc,Chapman:2018hou} 
\bea\label{crucial}
&&\hspace{-3.0cm} S^{\text{TFD}}_{\text{EE}}(t)-S^{\text{TFD}}_{\text{EE}}(t=0) =
\nonumber \\
&& \sum_{n} \mathbf{s}_{n}^{\text{TFD}} \mathcal{L}\hspace{.1cm} \text{frac} \big(\frac{v_n t}{\mathcal{L}}\big) \hspace{2.5cm} \text{if}\hspace{.2cm} \mathcal{L}\hspace{.1cm} \text{frac}\big(\frac{v_n t}{\mathcal{L}}\big) <l,
\cr \nonumber\\
&& \sum_{n} \mathbf{s}_{n}^{\text{TFD}} l \hspace{4.25cm} \text{if}\hspace{.2cm} l\leq \mathcal{L}\hspace{.1cm} \text{frac}\big(\frac{v_n t}{\mathcal{L}}\big)< \mathcal{L}-l,
\cr \nonumber\\
&& \sum_n \mathbf{s}_{n}^{\text{TFD}} \mathcal{L}\left(1- \hspace{.1cm}\text{frac}\big(\frac{v_n t}{\mathcal{L}}\big)\right)\hspace{1cm} \text{if}\hspace{.2cm}\mathcal{L}-l\leq \mathcal{L}\hspace{.1cm}\text{frac}\big(\frac{v_n t}{\mathcal{L}}\big),\label{Pheno}
\eea
where frac denotes the fractional part and $\mathbf{s}^{\text{TFD}}_{n}$ is\footnote{The part within the large parentheses represents a contribution to
the thermodynamic entropy of the free boson system at the inverse temperature $\beta$ from the mode $n$.}
\bea
\mathbf{s}_{n}^{\text{TFD}} =\frac{2}{\mathcal{L} } \left(\frac{\beta \omega_n}{e^{\beta\omega_n}-1}-\log(1-e^{-\beta \omega_n})\right).\label{sthn}
\eea
However, the entanglement entropy is only a proper measure to capture the dynamics of entanglement for pure quantum states and to have a deeper insight into the spreading of information in the out of equilibrium situations, we need to investigate the dynamics of the mixed states correlation measures. One of those measures is mutual information (MI) which is defined  for disjoint subregions $A$ and $ B$. It is given as a linear combination of entanglement entropies of that regions,
\begin{equation}
I(A,B)= S_{\text{EE},A}+ S_{\text{EE},B}- S_{\text{EE},A\cup B}.
\end{equation}
However, MI includes both classical and quantum correlations between
the subregions $A$ and $B$ and since it does not vanish for separable states it would not be a good measure of entanglement \cite{Plenio:2007zz}. 
Besides the MI, in the quantum information context, there are various measures for capturing the entanglement dynamics of mixed states. But, these measures are usually based on optimization procedures \cite{Horodecki:2009zz}, which are intractable approaches and we need a way to have explicit computational prescriptions. Among all these proposed  entanglement measures for mixed states, the logarithmic negativity (LN) is the best computable measure which is expected to capture only quantum correlations \cite{Eisert:1998pz,Plenio:2005cwa,Vidal:2002zz}. Recently, another information theoretic quantity is introduced which is called the odd entanglement entropy (OEE) \cite{Tamaoka:2018ned}. In the context of the AdS/CFT duality\cite{Aharony:1999ti}, the LN \cite{Kudler-Flam:2018qjo,Kusuki:2019zsp} and OEE\cite{Tamaoka:2018ned} are dual to the same geometric object which is called the entanglement wedge cross-section (EWC)\cite{Takayanagi:2017knl,Nguyen:2017yqw}. The dynamics of EWC has been recently studied in \cite{Kusuki:2019rbk, Kusuki:2019evw,BabaeiVelni:2020wfl,Sahraei:2021wqn} and the authors of \cite{Kudler-Flam:2020xqu,Kudler-Flam:2020url} have denoted that these two measures are proportional in the case of integrable systems.\\

The LN and OEE are defined by taking partial transposing of the state which it is obtained as follows. Let $ \vert e_{i}^{(1)} \rangle$ and $ \vert e_{i}^{(2)} \rangle$ denote the orthogonal basis for states of $\mathcal{H}_{A_{1}}$ and $\mathcal{H}_{A_{2}}$, respectively. A density matrix acting on the bipartite Hilbert space $\mathcal{H}=\mathcal{H}_{A_{1}}\otimes \mathcal{H}_{A_{2}}$, with the basis $| e_{i}^{(1)} e_{i}^{(2)} \rangle= |e_{i}^{(1)}\rangle\otimes |e_{i}^{(2)}\rangle$,  is denoted by $\rho_{A_{1}A_{2}}$. The $ \rho_{A_{1} A_{2}}$ can be expanded in a basis $\vert e_{i}^{(1)}  e_{j}^{(2)} \rangle$ of $\mathcal{H}$ as follows
\begin{equation}
\rho_{A_{1}A_{2}}=\sum_{ijkl} \big{\langle}  e_{i}^{(1)}  e_{j}^{(2)} \big{\vert} \rho_{A_{1}A_{2}} \big{\vert} e_{k}^{(1)}  e_{l}^{(2)} \big{\rangle}\hspace{.1cm}  \big{\vert} e_{i}^{(1)}  e_{j}^{(2)} \big{\rangle} \big{\langle}  e_{k}^{(1)}  e_{l}^{(2)} \big{\vert}.
\end{equation}
The partial transposition of the density matrix $\rho_{A_{1}A_{2}}$ with respect to the subsystem $A_{2}$ is given by swapping the matrix elements in the subsystem $A_2$,
\begin{equation}
\big{\langle}  e_{i}^{(1)}  e_{j}^{(2)} \big{\vert} \rho_{A_{1}A_{2}}^{T_{A_{2}}} \big{\vert} e_{k}^{(1)}  e_{l}^{(2)} \big{\rangle}=\big{\langle}  e_{i}^{(1)}  e_{l}^{(2)} \big{\vert} \rho_{A_{1}A_{2}}  \big{\vert} e_{k}^{(1)}  e_{j}^{(2)} \big{\rangle}.
\end{equation}
Accordingly, with respect to the original basis of $\mathcal{H}$ one has
\begin{equation}
\rho_{A_{1}A_{2}}^{T_{A_{2}}} =\sum_{ijkl} \big{\langle}  e_{i}^{(1)}  e_{l}^{(2)} \big{\vert} \rho_{A_{1}A_{2}} \big{\vert} e_{k}^{(1)}  e_{j}^{(2)} \big{\rangle} \hspace{.1cm} \big{\vert} e_{i}^{(1)}  e_{j}^{(2)} \big{\rangle} \big{\langle}  e_{k}^{(1)}  e_{l}^{(2)} \big{\vert}.
\end{equation}
By defining  $S_{\text{OEE}}^{(n_{\circ})}(\rho_{A_{1}A_{2}})$ as follows
\bea
S_{\text{OEE}}^{(n_{\circ})}(\rho_{A_{1}A_{2}})=\frac{1}{1-n_{\circ}} \Big( \text{Tr}(\rho_{A_{1}A_{2}}^{T_{A_{2}}})^{n_{\circ}}-1\Big),
\eea
where $n_{\circ}$ is an odd positive integer, the $S_{\text{OEE}}$ is given by\cite{Tamaoka:2018ned}
\begin{equation}
S_{\text{OEE}}(\rho_{A_{1}A_{2}})=\lim_{n_{\circ}\rightarrow 1}S_{\text{OEE}}^{(n_{\circ})}.
\end{equation}
Since, $\rho_{A_{1}A_{2}}^{T_{A_{2}}}$ is a Hermitian operator and the partial transposition is not a completely positive map, then $\rho_{A_{1}A_{2}}^{T_{A_{2}}}$, in general, can have negative eigenvalues.  Accordingly, the $S_{\text{OEE}}$ can be  written as
\begin{equation}
S_{\text{OEE}}(\rho_{A_{1}A_{2}})=-\sum_{\lambda_{i}>0} \lambda_{i}  \log  \lambda_{i} +\sum_{\lambda_{i}<0}\vert \lambda_{i} \vert \log \vert \lambda_{i} \vert,
\end{equation}
where $\lambda_{i}$'s are the eigenvalues of the $\rho_{A_{1}A_{2}}^{T_{A_{2}}}$ matrix. It is worth mentioning that the OEE reduces to the entanglement entropy for pure states. Moreover, if we subtract the von Neumann entropy $S_{\text{EE}}(\rho_{A_{1}A_{2}})$ from OEE, it has been suggested that the obtained quantity might be dual to the entanglement wedge cross-section for holographic theories\footnote{Although, it has been recently argued that there might be counterexamples for this duality; for more details see \cite{Dong:2021clv}.}\cite{Tamaoka:2018ned} 
\begin{equation}
E_{W}(\rho_{A_{1}A_{2}})=S_{\text{OEE}}(\rho_{A_{1}A_{2}})-S_{\text{EE}}(\rho_{A_{1}A_{2}}).\label{EWCS}
\end{equation}
For holographic theories this quantity is positive but, in general, it can be negative\footnote{Recently, OEE for Lifshitz scalar theories has been studied in \cite{Mollabashi:2020ifv} where it has been shown that it can be negative.}. The LN can also be defined along the same lines:
\begin{equation}
\mathcal{E}_{A_{1}A_{2}} \equiv \log \vert \vert \rho_{A_{1}A_{2}}^{T_{A_{2}}} \vert \vert=\log \text{Tr} \vert \rho_{A_{1}A_{2}}^{T_{A_{2}}} \vert,
\end{equation}
where the trace norm $\vert\vert \mathcal{O} \vert \vert \equiv \text{Tr}\sqrt{\mathcal{O}^{\dagger}\mathcal{O}}$. According to the eigenvalues $\lambda_{i}$ of the operator $\rho_{A_{1}A_{2}}^{T_{A_{2}}}$ one has 
\begin{equation}
\mathcal{E}_{A_{1}A_{2}} = -\sum_{\lambda_{i}>0} \lambda_{i} +\sum_{\lambda_{i}<0}\vert \lambda_{i} \vert
\end{equation}
It is easy to see that for pure states, the LN reduces to $n=\frac{1}{2}$ Rényi entropy\cite{Vidal:2002zz}. It is worth mentioning again that LN is expected to capture only quantum correlations and has been studied previously in several works\cite{Peres:1996dw,Horodecki:1996nc,Simon:2000zz,Audenaert:2002xfl,Calabrese:2012ew,Calabrese:2012nk,Calabrese:2014yza,DeNobili:2015dla, Eisler:2016wfo, MohammadiMozaffar:2017chk,Shapourian:2018lsz,Angel-Ramelli:2020wfo,Alba:2018hie}.  
Interestingly, it is  argued that in the limit of long times and
large subsystems with a fix ratio, the LN equals half of the Rényi mutual information
\bea \label{LN-2}
\mathcal{E}(A_{1},A_{2})=\frac{I^{(1/2)}(A_{1},A_{2})}{2}.  
\eea
Accordingly, the condition $\mathcal{E}(A_{1},A_{2}) = 0$ is necessary (but not sufficient) for
the absence of mutual entanglement.
\\  
                             
One particular fruitful situation for studying thermalization and global quench in the context of holography has been to study these phenomena for the case of thermofield double states (TFD)\cite{Balasubramanian:2010ce,Balasubramanian:2011ur,Liu:2013iza}. Using the dictionary of the AdS/CFT duality, the TFD state is proposed to be dual to the eternal AdS black hole\cite{Maldacena:2001kr} and provides a setup to probe various aspects of black holes from a quantum information theoretic perspective. For example, the TFD state can be used to study scrambling and quantum chaos\cite{Shenker:2013pqa,Shenker:2013yza,Roberts:2014isa}, dynamics of entanglement entropy\cite{Hartman:2013qma}, quantum computational complexity\cite{Susskind:2014yaa}, and so on. This state belongs to the product Hilbert space 
\begin{equation}
\mathcal{H}=\mathcal{H}_{\text{L}}\otimes \mathcal{H}_{\text{R}}.
\end{equation}
By choosing the following Hamiltonian
\begin{equation}
H_{\text{TFD}}=H_{L}\otimes I_{R}-I_{L}\otimes H_{R}
\end{equation}
the time-dependent TFD state is given  by 
\begin{equation}\label{tltr}
|\text{TFD}(t_{L},t_{R})\rangle=\frac{1}{\sqrt{Z(\beta)}}\sum_{n}e^{-\beta E_{n}/2}e^{-i E_{n}(t_{L}+t_{R})}|E_{n}\rangle_{L}|E_{n}\rangle_{R},
\end{equation}
where 
$|E_{n}\rangle_{L,R}$ are the energy eigenstates of the left and right theories (for example two CFTs), respectively, with corresponding  times $t_{L,R}$. Moreover, tracing out either copy leads to a thermal state at the inverse temperature $\beta$ for the other, $\rho_{\text{th}}=\frac{1}{Z(\beta)} e^{-\beta H_{i}}$, $i\equiv L,R$ and $Z(\beta)$ denotes the partition function. According to (\ref{tltr}) and in the spirit of \cite{Hartman:2013qma}, our setup to consider the  dynamics of LN and OEE will be in a rather unusual quantum quench scenario\cite{Chapman:2018hou} in which two decoupled subsystems are entangled via their initial conditions. Moreover, to have a mixed state, we will consider two spatial non-complementary regions. Since the underlying TFD state is a Gaussian state, we can use the covariance matrix approach to calculate LN and OEE. We will observe that their behaviors under the time evolution can be summarized as a linear growth followed by saturation which is similar to the expectations from the quasi-particle picture. Also, we will observe the oscillatory behavior due to the finite size effect as well as a  logarithmic contribution in the intermediate regime due to the existence of the zero-mode.
\\

This paper is organized as follows: In section \ref{Sec-2}, we will briefly review the reconstruction of the TFD state for two harmonic oscillators and then study its time dependency. Then, we will generalize this to a $1+1$ dimensional free real scalar QFT. We will also introduce the covariance matrix formalism and by using this method in section \ref{Sec-3}, we will explain how to evaluate the OEE and LN for subsystems involving an equal number of sites on each spatial side of the TFD state. The numerical results are presented in section \ref{Sec-4}. In section {\ref{Sec-4}}, some previously proposed inequalities for OEE are also checked. In section \ref{Sec-5}, we will choose a special dividing for subsystems and give an analytical formula  for evaluating the zero-mode contribution. In section \ref{Sec-6}, we will conclude and discuss our results and  future directions. In appendix \ref{Ap-2}, we will study the effect of temperature on the dynamics of entanglement. To do so, we will consider two cases: In the first case, the two intervals are adjacent to each other and in the second case, the two intervals are separated by a distance $d$.  Further details for evaluation of zero-mode contribution and logarithmic growth are provided in appendix \ref{Ap-3}.
\section{Covariance matrix for Gaussian TFD state}\label{Sec-2}
As we have mentioned in the introduction, we would like to study the dynamics of OEE and LN. Accordingly, we will focus on the TFD state of a free real scalar QFT. This TFD state is a Gaussian state and therefore we can use the power of the covariance matrix to calculate the OEE and LN\footnote{The Covariance matrix approach has been previously used to probe the dynamics of the entanglement entropy of bosonic and fermionic Gaussian states\cite{Chapman:2018hou, Bianchi:2017kgb,Cotler:2016acd, Hackl:2017ndi, Vidmar:2017uux, Hackl:2018ndi,DiGiulio:2020hlz}.}. Since, in general, the OEE and LN are divergent quantities in the continuum, we will regularize the theory by putting it on a lattice. In the normal mode decomposition, the discretized QFT takes the form of $N$ decoupled simple harmonic oscillators. Accordingly, we will first consider the construction of time-independent as well as time-dependent TFD states for two copies of simple harmonic oscillators. Then, the OEE and LN for the TFD state of discretized QFT will be constructed by a sum on the contribution of each normal mode (simple harmonic oscillator).   
\subsection{Free Real Scalar QFT on a Lattice}
The Hamiltonian of a 1+1 dimensional free real scalar QFT on a circle with a circumference $\mathcal{L}$ is given by
\begin{align}\label{H-1}
H =\frac{1}{2} \int_{-\frac{\mathcal{L}}{2}}^{\frac{\mathcal{L}}{2}} dx\hspace{.5mm}\Big( \Pi^{2}+(\partial_{x}\hspace{.5mm}\Phi)^{2}+m^2\Phi^{2}\Big),
\end{align}
where $\Pi = \frac{\partial L}{\partial{\dot{\Phi}}} = \dot{\Phi}$ is the conjugate momentum. The regularized Hamiltonian on a circular lattice with $N$ sites and lattice spacing $\delta= \frac{\mathcal{L}}{N}$ becomes
\begin{align}
\label{Hcomplex3.3}
H = \sum_{a=1}^{N}\bigg(\frac{\delta}{2}\hspace{.5mm} \mathbf{P}_a^{2}+\frac{m^2}{2\delta}\hspace{.5mm}\mathbf{Q}_a^{2} +\frac{1}{2\delta^3}\left(\mathbf{Q}_{a+1}-\mathbf{Q}_a\right)^{2}\bigg),
\end{align}
where redefined canonical variables are
\bea\label{delta}\label{CO-1}
\mathbf{Q}_a = \Phi(x_a)\delta, \hspace{1cm} \mathbf{P}_a = \Pi(x_a),\label{pq}
\eea
and we have imposed the periodic boundary conditions $\mathbf{Q}_{N+1} = \mathbf{Q}_1$ and $\mathbf{P}_{N+1} = \mathbf{P}_1$. By applying the discrete Fourier transformation\footnote{The canonical commutation relations are given by $[\tilde{Q}_{k},\tilde{P}^{\dagger}_{l}]= i \delta_{k,l}.$}
\begin{align}
\label{Fourier}
\tilde{\mathbf{Q}}_k = \frac{1}{\sqrt{N}}\sum_{a=1}^{N}e^{\frac{2\pi ika}{N}}\hspace{.5mm}\mathbf{Q}_a,\hspace{1cm}\tilde{\mathbf{P}}_k = \frac{1}{\sqrt{N}}\sum_{a=1}^{N}e^{-\frac{2\pi ika}{N}}\hspace{.5mm}\mathbf{P}_a,
\end{align}
where $\tilde{\mathbf{Q}}^{\dagger}_{k} =\tilde{\mathbf{Q}}_{N-k}$ and $\tilde{\mathbf{P}}^{\dagger}_{k} =\tilde{\mathbf{P}}_{N-k}$, the Hamiltonian (\ref{Hcomplex3.3}) reduces to
\begin{align}\label{H-3}
H = \sum_{k=0}^{N-1}\bigg(\frac{\delta}{2}\hspace{.5mm} |\tilde{\mathbf{P}}_k|^2+\frac{1}{2\delta}\omega_k^{2}\hspace{.5mm}|\tilde{\mathbf{Q}}_k|^2\bigg).
\end{align}
In the above formula, the frequencies $\omega_k$ are given by
\bea
\label{omegak}
\omega_k^2 = m^2+\frac{4}{\delta^2}\sin^2(\frac{\pi k}{N}).
\eea 
To construct the corresponding TFD state, one first needs to quantize the Hamiltonian (\ref{H-3}) by defining two sets of creation and annihilation operators
\begin{align}
\label{aad}
\tilde{\mathbf{Q}}_k = \sqrt{\frac{\delta}{2\omega_k}}\left(\hat{a}_{N-k}+\hat{a}_k^{\dagger}\right),\hspace{1cm}\tilde{\mathbf{P}}_k = i \sqrt{\frac{\omega_k}{2\delta}}\left(\hat{a}_k^{\dagger}-\hat{a}_{N-k}\right),
\end{align}
where $[\hat{a}_k,\hat{a}^{\dagger}_k]=1$.  By substituting (\ref{aad}) in (\ref{H-3}) one gets
\begin{align}
\label{H-4}
\hat{H}=\sum_{k=0}^{N-1} \omega_k \left(\hat{a}_k^{\dagger}\hspace{.5mm}\hat{a}_k+\frac{1}{2}\right).
\end{align}
 This is the Hamiltonian of $N$ decoupled simple harmonic oscillators with equal mass  $M=\delta^{-1}$ (not to be confused with the physical mass $m$) and k-dependent frequencies $\omega_{k}$. Ignoring the constant term, the $\hat{a}_k^{\dagger}\hspace{.5mm}\hat{a}_k \omega_{k}\equiv n \omega_{k}$ denotes the total energy of level "$n$" for each of these oscillators with fixed momentum. Since the zero-mode Hamiltonian does not have a normalizable ground state ($\omega_{0}$ vanishes when $m=0$) one can regularize it by introducing a very small dimensionless mass, $m \mathcal{L} \ll 1$.
Considering the decoupled form of
the Hamiltonian (\ref{H-3}),
the corresponding TFD state of the free scalar theory will be a product of TFD states for each of the oscillator modes. Accordingly, in the following we will focus on a single mode (fixed momentum $k$)  and construct its time-dependent TFD state. 
\subsection{TFD state for a simple harmonic oscillator}
In this subsection, we will first construct the TFD state for a simple harmonic oscillator with mass $m$ and frequency $\omega$ at $t=0$ and then will turn to study its time evolution. The creation and annihilation operators for a simple harmonic oscillator are given by 
\bea\label{a}
\hat{a}^\dagger=\sqrt{\frac{m\omega}{2}}\left(\hat{Q}-i\frac{\hat P}{m\omega}\right),  \hspace{.5cm}
\hat{a}=\sqrt{\frac{m\omega}{2}}\left(\hat{Q}+i\frac{\hat P}{m\omega}\right).
\eea
 The $n^{th}$ energy eigenstate is then defined by acting $n$ times with the creation operator on the vacuum state $\mid \hspace{-1mm}0 \rangle$:
\bea\label{staten}
 \mid\hspace{-.5mm}n\rangle= \frac{1}{\sqrt{n!}} \hspace{.5mm}(\hat{a}^\dagger)^n \mid \hspace{-1mm}0 \rangle.
\eea
The action of creation and annihilation operators on this state are given by
\bea \label{leveln}
\hat{a}^\dagger \hspace{-1mm}\mid\hspace{-1mm}n\rangle = \sqrt{n+1}\mid\hspace{-1mm}n+1\rangle,\hspace{.5cm}\hat{a}\hspace{-1mm} \mid\hspace{-1mm}n\rangle = \sqrt{n}\mid\hspace{-1mm}n-1\rangle.
\eea
The TFD state at $t=0$ can be constructed as a superposition of a tensor product of two copies of the energy eigenstate (\ref{staten}), which we label by $L$ and $R$, with special wights\cite{Khanna:2009zz}:
\bea\label{TFD.1}  
\mid\hspace{-1mm}  \text{TFD}\rangle=Z(\beta)^{-\frac{1}{2}}\displaystyle\sum_{n=0}^{\infty}e^{-\frac{\beta}{2} E_{n}}\mid\hspace{-1mm}  n \rangle_L\hspace{1mm}\otimes\mid\hspace{-1mm}  n \rangle_R ,
\eea
where the normalization factor is $Z(\beta)= (1-e^{-\beta\omega})^{-1}$ and $E_{n}$ denotes the energy of those eigenstates.
Considering this normalization factor together with (\ref{staten}), then the state (\ref{TFD.1}) can be alternatively written as
\bea
\label{CTFD.2.1}
\ket{\text{TFD}}= \sqrt{1-e^{-\beta\omega}}\hspace{1mm}\exp{\left(e^{-\frac{\beta }{2}\omega}\hspace{1mm}\hat{a}^\dagger_{L}\hat{a}^\dagger_{R}\right)}\mid\hspace{-1mm}0\rangle_L\hspace{.5mm}\otimes\mid \hspace{-1mm}0\rangle_R.
\eea
It is worth noting that in (\ref{CTFD.2.1}) the operator acting on the total vacuum state is not a unitary operator. It is convenient to re-express  
(\ref{CTFD.2.1}) by acting as a unitary operator on the vacuum state $\mid\hspace{-1mm}0\rangle_L\hspace{.5mm}\otimes\mid \hspace{-1mm}0\rangle_R$. The result is \cite{Klimov}
\bea
\label{TFD.3}
\mid \hspace{-1mm}\text{TFD}\rangle = e^{\alpha\left(\hat{a}_L^{\dagger}\hat{a}_R^{\dagger}-\hat{a}_L\hat{a}_R\right)}\ket{0}_L\otimes\ket{0}_R,
\eea
with
\bea\label{alpha}
\tanh{\alpha} = e^{-\frac{\beta}{2}\omega}.
\eea
The time evolution of the state (\ref{TFD.3}) is given by 
\bea
\label{TTFD.2.1}
\mid\hspace{-1mm}\text{TFD}(t)\rangle = e^{-i\hat{H}_{L}t_{L}} e^{-i\hat{H}_{R}t_{R}} \mid\hspace{-.5mm}\text{TFD} \rangle,
\eea
in which the operators $\hat{H}_{L}$ and $\hat{H}_{R}$ are the corresponding Hamiltonians for the left and right simple harmonic oscillators. These quantities are defined as
\bea
\hat{H}_{L} = \hat{a}^{\dagger}_{L}\hspace{.5mm}\hat{a}_{L}+\frac{1}{2},\hspace{.5cm}\hat{H}_{R} =
\hat{a}^{\dagger}_{R}\hspace{.5mm}\hat{a}_{R}+\frac{1}{2}.
\eea
By choosing $t_{L}=t_{R} = t/2$, which is the common convention in holography \cite{Jefferson:2017sdb}, the time dependent
TFD state (\ref{TTFD.2.1}) takes the following form
\bea
\label{TFDt}
\mid\hspace{-.5mm}\text{TFD}(t)\rangle 
=e^{-\frac{i}{2}\omega t}\hspace{.5mm}\sqrt{1-e^{-\beta\omega}}\hspace{1mm}\exp\hspace{-.5mm}\bigg[e^{-\frac{\beta}{2}\omega}e^{-i\omega t}\hspace{1mm}a_L^{\dagger}a_R^{\dagger}\bigg]\ket{0}_L\otimes\ket{0}_R.
\eea
One can write the state (\ref{TFDt}) in a more compact form by acting as a unitary operator on the vacuum state $\ket{0}_L\otimes\ket{0}_R$ as follows\cite{Klimov, Chapman:2018hou}\footnote{Here, we have dropped the global time-dependent phase, since this does not change the physical state.}
\bea\label{TTFD.2.2}
\mid\hspace{-1mm} \text{TFD}(t)\rangle=\exp\hspace{-.5mm}\bigg[z\hspace{.5mm}\hat{a}^\dagger_{L}\hspace{.5mm}\hat{a}^\dagger_{R}-z^*\hspace{.5mm}\hat{a}_{L}\hspace{.5mm}\hat{a}_{R}\bigg]\hspace{-1mm}\mid\hspace{-1mm}0\rangle_L\otimes\mid\hspace{-1mm}0\rangle_R,
\eea
where 
\bea
z = \alpha\hspace{.5mm}e^{-i\omega t},
\eea
and $\alpha$ is the same as (\ref{alpha}). 
Now by having the time-dependent TFD state (\ref{TTFD.2.2}), one needs to find the covariance matrix associated with it. That is the subject of the next subsection.
\subsection{Covariance matrix formalism}
The system (\ref{H-3}) is described by $2N$ linear observables $\hat{\xi} = (\hat{q}_1, \hat{q}_2,...,\hat{q}_N,\hat{p}_1 ,..., \hat{p}_N)$. The $(\hat{q}_i,\hat{p}_i)$ are canonical operators where $[\hat{q}_i,\hat{p}_j]=i\delta_{ij}$. The two-point functions of these observables in an arbitrary state $\mid\hspace{-1mm}\Psi\rangle$ can be decomposed as
\bea
\label{Gab}
\langle\Psi\hspace{-1mm}\mid \hspace{-1mm}\hat{\xi}^a\hspace{.5mm}\hat{\xi}^b\hspace{-1mm}\mid\hspace{-1mm} \Psi\rangle=\frac{1}{2}\left(G^{a b}+i\Omega^{a b}\right),
\eea
where $G^{a b}=G^{(a b)}$ and  $\Omega^{ab}=\Omega^{[a b]}$ are the symmetric and the antisymmetric parts of the correlation functions, respectively. For a bosonic state, $\Omega^{ab}$ is completely fixed by the commutation relations of $\hat{q}_{i}$ and $\hat{p}_{i}$,
\bea\label{Km}
\Omega^{ab}=
\begin{pmatrix}
	0 & 1\\
	-1 & 0
\end{pmatrix}.
\eea
For a pure Gaussian state with vanishing first moment $\langle\psi|\hat{\xi}^{a}|\psi\rangle=0$, the covariance matrix is given by the symmetric part of the two-point function 
\begin{equation}\label{G}
G^{ab}=\langle \Psi\hspace{-1mm}\mid \hat{\xi}^{a}\hat{\xi}^{b}+\hat{\xi}^{b}\hat{\xi}^{a}\mid\hspace{-1mm} \Psi\rangle.
\end{equation}
By using Wick's  theorem, one can compute all of the $n$-point functions from $G^{ab}$. Hence, it can be used to label the Gaussian states. For a mixed state $\rho$, when $\langle\Psi|\hat{\xi}^{a}|\Psi\rangle=0$, the covariance matrix is defined by \cite{Eisert:2003hpa, Weedbrook:2012zau}
\begin{equation}
G^{ab}=\text{Tr}\bigg(\rho\hspace{.5mm}\big( \hat{\xi}^{a}\hat{\xi}^{b}+\hat{\xi}^{b}\hat{\xi}^{a}\big)\bigg).
\end{equation}
To find the covariance matrix associated with the TFD state (\ref{TFD.3}), one can first find the covariance matrix for the vacuum state and then by a unitary transformation modify it for the TFD state. To do so, we restrict ourselves to the space of Gaussian states. In this space, 
the general unitary operator $\hat{U}(s)$ can be expressed by Hermitian operators which are quadratic in the canonical operators $\hat{\xi}$,
\bea
\label{hatK}
\hat{U}(s) = e^{-i s \hat{K}},\hspace{1cm}\text{with}\hspace{.5cm}\hat{K} =\frac{1}{2}\hat{\xi}^{a}\hspace{.5mm} k_{(a,b)}\hspace{.5mm}\hat{\xi}^{b} \equiv \frac{1}{2}\hat{\xi}\hspace{.5mm} k\hspace{.5mm} \hat{\xi}^{T}.
\eea
Accordingly, the transformed state is
\begin{equation}\label{Gs}
|G_{s}\rangle=\hat{U}(s)|G_{0}\rangle,
\end{equation}
where the subscript ``$0$'' indicates the vacuum state (for the time-dependent state (\ref{TTFD.2.2}) it refers to the TFD state (\ref{TFD.3})). To find the corresponding covariance matrix, one needs the operation of $\hat{U}(s)$ on $\hat{\xi}^{a}$ which can be obtained as follows
\bea\label{ext1}
\hat{U}^{\dagger}(s)\hspace{.5mm}\hat{\xi}^{a}\hspace{.5mm}\hat{U}(s) = \sum_{n=0}^{\infty} \frac{s^n}{n!} [i \hat{K},\hat{\xi}^{a}]_{(n)}.
\eea
In the above formula, $[i \hat{K},\hat{\xi}^{a}]_{(n)}$ is defined recursively by $[i \hat{K},\hat{\xi}^{a}]_{(n)}=[i\hat{K}, [i \hat{K},\hat{\xi}^{a}]_{(n-1)}]$, and $[i \hat{K},\hat{\xi}^{a}]_{(0)}=[i \hat{K},\hat{\xi}^{a}]$. With respect to the (\ref{hatK}) and the commutation relation $[\hat{\xi}^{a},\hat{\xi}^{b}] =i\Omega^{a b}$, one can find that
\bea
[i \hat{K},\hat{\xi}^{a}] = \Omega^{a b}\hspace{.5mm} k_{(b, c)}\hspace{.5mm}\hat{\xi}^{c}.
\eea
By defining $K^{a}_{b}= \Omega^{a c}\hspace{.5mm} k_{(c,b)}$, the above formula can be rewritten as 
\bea\label{Kxi}
[i \hat{K},\hat{\xi}^{a}] = K^{a}_{b}\hspace{.5mm}\hat{\xi}^{b}.
\eea
Hence, the operation of $\hat{U}(s)$ on $\hat{\xi}^{a}$, (\ref{ext1}), can be expressed as follows
\bea
\label{UdxiU}
\hat{U}^{\dagger}(s)\hspace{.5mm}\hat{\xi}^{a}\hspace{.5mm}\hat{U}(s) = (e^{s K})^{a}_{b}\hspace{1mm}\hat{\xi}^{b}\equiv U(s)^{a}_{b}\hspace{1mm}\hat{\xi}^{b}.
\eea
Now, the relations (\ref{hatK}) together with the relation (\ref{UdxiU}) implies that the covariance matrix associated with transformed state $|G_{s}\rangle$ becomes
\bea\label{GsG0}
&& G_{s}^{(a,b)} =  \langle G_{s}\hspace{-1mm}\mid \left(\hat{\xi}^{a}\hat{\xi}^{b}+\hat{\xi}^{b}\hat{\xi}^a\right) |G_{s}\rangle
=
U(s)^{a}_{c} \hspace{1mm}G_{0}^{(c,d)} \hspace{1mm}U(s)^{b}_{d},
\eea
where $G_{0}^{(c,d)}$ is its counterpart for the vacuum. 
Accordingly, in the compact notation, we have
\begin{equation}
|G_{s}\rangle =\hat{U}(s) |G_{0}\rangle =|U(s)\hspace{1mm}G_{0} \hspace{1mm}U^{\top}(s)\rangle ,  \qquad G_{s}=U(s)\hspace{1mm}G_{0} \hspace{1mm}U^{\top}(s). 
\end{equation}
In the following, we will demonstrate how to evaluate the covariance matrix associated with the time-independent as well as time-dependent TFD states. This is easily achieved by using the above formalism. 
\subsubsection{Covariance matrix for TFD state of the harmonic oscillator}
Let us begin with constructing the covariance matrix associated with the time-independent TFD state (\ref{TFD.3}) corresponding to $N=1$ and then extend it to the time-dependent case (\ref{TTFD.2.2}). The Hamiltonian of the system is described by
\bea\label{H}
\hat{H}=\frac{1}{2m}\Big(\hat{p}^{2}_{L}+\hat{p}^{2}_{R}+m^{2}\omega^{2}(\hat{q}^{2}_{L}+\hat{q}^{2}_{R})\Big).
\eea
By rewritting (\ref{TFD.3}) in terms of the $(q_L, p_L, q_R, p_R)$ coordinates and reading the corresponding $\hat{K}$ and $U(s)$, the covariance matrix of TFD state (\ref{TFD.3}) for a single-mode is given by
\bea\label{rotation} 
G^{ab}_{\alpha} = \begin{pmatrix}
	\frac{\cosh(2\alpha)}{m \omega}& 0 & -\frac{\sinh(2\alpha)}{m \omega} & 0 \vspace{.5cm}
	
	\\
	0& m\omega \cosh(2\alpha) & 0 &m\omega \sinh(2\alpha)\vspace{.5cm}
	
	\\
	-\frac{\sinh(2\alpha)}{m \omega} & 0& \frac{\cosh(2\alpha)}{m \omega}& 0 \vspace{.5cm}
	
	\\
	0 &m\omega \sinh(2\alpha)& 0& m\omega \cosh(2\alpha)
\end{pmatrix}.
\eea
The time-dependent covariance matrix associated with time-dependent TFD state (\ref{TTFD.2.2}) can be derived as follows. The Hamiltonian (\ref{H})
has the form $\hat{H}=\frac{1}{2}\hat{\xi^{a}}k_{ab}\hat{\xi^{b}}$, where the matrix representation of $k_{ab}$ with respect to  $\hat{\xi} = (\hat{q}_L, \hat{p}_L,\hat{q}_R,\hat{p}_R)$ is given by
\bea 
\label{rotationn}
k_{ab} = \begin{pmatrix}
	m \omega^{2}& 0 & 0& 0\\
	0& \frac{1}{m} & 0 &0 \\
	0& 0& m \omega^{2}& 0 \\
	0 &0 & 0& \frac{1}{m}
\end{pmatrix}.
\eea
The simplectic generator $K^{a}_{b}=\Omega^{ac} k_{(c,b)}$
becomes
\bea
K_{b}^{a} = \begin{pmatrix}
	0& \frac{1}{m} & 0& 0\\
	-m \omega^{2}& 0 & 0 &0 \\
	0& 0& 0& \frac{1}{m} \\
	0 &0 & -m \omega^{2}& 0
\end{pmatrix},
\eea
and therefore
\bea
U(t)=\exp(tK) = \begin{pmatrix}
	m\cos(\omega t)& \frac{\sin(\omega t)}{m} & 0& 0 \vspace{.5cm}
	
	\\
	-\frac{\sin(\omega t)}{m}& m\cos(\omega t) & 0 &0 \vspace{.5cm}
	
	\\
	0& 0& m\cos(\omega t)& \frac{\sin(\omega t)}{m} \vspace{.5cm}
	
	\\
	0 &0 & -\frac{\sin(\omega t)}{m}& m\cos(\omega t)
\end{pmatrix}.
\eea
Using this and noting to (\ref{rotation}), the covariance matrix $G_{\alpha}^{ab}(t) = U(t)\hspace{.5mm} G_{\alpha}^{ab} \hspace{.5mm}U^{\top}(t)$ becomes
\bea\label{rotationnn}
G^{ab}_{\alpha}(t) = \begin{pmatrix}
\frac{\cosh(2\alpha)}{m \omega}& 0 & -\frac{\sinh(2\alpha)\cos(\omega t)}{m \omega} & \sin(\omega t)\sinh(2\alpha) \vspace{.5cm}
	
\\
0& m\omega \cosh(2\alpha) & \sin(\omega t)\sinh(2\alpha) &m\omega \sinh(2\alpha)\cos(\omega t) \vspace{.5cm}
	
\\
-\frac{\sinh(2\alpha)\cos(\omega t)}{m \omega} & \sin(\omega t)\sinh(2\alpha)& \frac{\cosh(2\alpha)}{m \omega}& 0 \vspace{.5cm}
	
\\ 
\sin(\omega t)\sinh(2\alpha) &m\omega \sinh(2\alpha)\cos(\omega t) & 0& m\omega \cosh(2\alpha)
\end{pmatrix}.
\eea
\subsubsection{Covariance matrix for TFD state of real scalar QFT}
As we explained previously, according to the discretized Hamiltonian (\ref{H-3}) on a lattice with $N$ sites, the  TFD state (\ref{TTFD.2.2}) is described by $2N$ degrees of freedom on each side. Again, one can choose the  following coordinates 
\begin{equation}
\xi^{a}=\bigg(q_{m,L},p_{m,L},q_{m,R},p_{m,R}\bigg),
\end{equation}
where $q_{m}=\frac{1}{\sqrt{N}}\Phi_{L}(x_{m})$, $p_{m}=\frac{\mathcal{L}}{\sqrt{N}}\Pi_{L}(x_{m})$, and $x_{m}=m \delta$ which $\delta$ is the distance between the two sites. Using the Fourier space coordinates
\begin{equation}
\tilde{q}_{k}=\frac{1}{\sqrt{N}}\sum_{m=1}^{N}e^{\frac{2\pi i k m}{N}}q_{m},
\hspace{.5cm}
\tilde{p}_{k}=\frac{1}{\sqrt{N}}\sum_{m=1}^{N}e^{-\frac{2\pi i k m}{N}}p_{m},
\end{equation}
and the extension of (\ref{rotationnn}), the covariance matrix associated with the time-dependent TFD state for a real scalar QFT (on the lattice) is real \cite{Chapman:2018hou,Hackl:2018-andi,DiGiulio:2020hlz} and is given by
\bea\label{GA}
G(t)=\begin{pmatrix}
	G_{LL}^{\Phi\Phi}&G_{LL}^{\Phi\Pi}&G_{LR}^{\Phi\Phi}&G_{LR}^{\Phi\Pi} \vspace{.3cm}
	
	\\ 
	G_{LL}^{\Pi\Phi}&G_{LL}^{\Pi\Pi}&G_{LR}^{\Pi\Phi}&G_{LR}^{\Pi\Pi} \vspace{.3cm}
	
	\\
	G_{RL}^{\Phi\Phi}&G_{RL}^{\Phi\Pi}&G_{RR}^{\Phi\Phi}&G_{RR}^{\Phi\Pi} \vspace{.3cm}
	
	\\
	G_{RL}^{\Pi\Phi}&G_{RL}^{\Pi\Pi}&G_{RR}^{\Pi\Phi}&G_{RR}^{\Pi\Pi}
\end{pmatrix},
\eea
where, 
\bea\label{EQ-1}
&&G_{LL}^{\Phi\Phi}=G_{RR}^{\Phi\Phi} =\frac{1}{N}\sum_{k=0}^{N-1}e^{\frac{2\pi i k(m-n)}{N}}\hspace{1mm}\frac{\cosh(2\alpha_{k})}{\lambda_k},
\cr \nonumber\\
&&
G_{LL}^{\Pi\Pi}=G_{RR}^{\Pi\Pi}=\frac{1}{N}\sum_{k=0}^{N-1}e^{\frac{2\pi i k(m-n)}{N}}\hspace{1mm}\lambda_{k}\cosh (2\alpha_{k}),
\cr\nonumber\\
&&
G_{LL}^{\Phi\Pi}=G_{LL}^{\Pi\Phi}=G_{RR}^{\Phi\Pi}=G_{RR}^{\Pi\Phi}=0,
\cr\nonumber\\
&&G_{LR}^{\Phi\Phi}=G_{RL}^{\Phi\Phi}=-\frac{1}{N}\sum_{k=0}^{N-1}e^{\frac{2\pi i k(m-n)}{N}}\hspace{1mm}\frac{\cos\left(\omega_{k}t\right) \sinh(2\alpha_{k})}{\lambda_{k}},
\cr\nonumber\\
&&
G_{LR}^{\Pi\Pi}=G_{RL}^{\Pi\Pi}=\frac{1}{N}\sum_{k=0}^{N-1}e^{\frac{2\pi i k(m-n)}{N}}\hspace{1mm}\lambda_{k}\cos\left(\omega_{k}t\right)\sinh(2\alpha_{k}),
\cr\nonumber\\
&&
G_{LR}^{\Phi\Pi}=G_{LR}^{\Pi\Phi}=G_{RL}^{\Phi\Pi}=G_{RL}^{\Pi\Phi}=\frac{1}{N}\sum_{k=0}^{N-1}e^{\frac{2\pi i k(m-n)}{N}}\hspace{1mm}\sin\left(\omega_{k}t\right)\sinh(2\alpha_{k}),
\eea
with $\alpha_k=\frac{1}{2}\log \coth(\frac{\beta\hspace{.1mm}\omega_{k}}{4})$, $\lambda_{k}=\mathcal{L}\omega_{k}$ and $\mathcal{L}$ denote the circumference of a circle. Note that $L$ or $R$ are separately described by a thermal state so the associated blocks are time-independent. But the crossing blocks, $LR$ and $RL$, correspond to correlations between the left and right sides and are time-dependent. In the next section, we will see that the time dependence of OEE and LN comes from these blocks.
	
\section{Time evolution of $S_{\text{OEE}}$ and $\mathcal{E}$}\label{Sec-3}
\begin{figure}[H]
\centering \includegraphics[scale=.38]{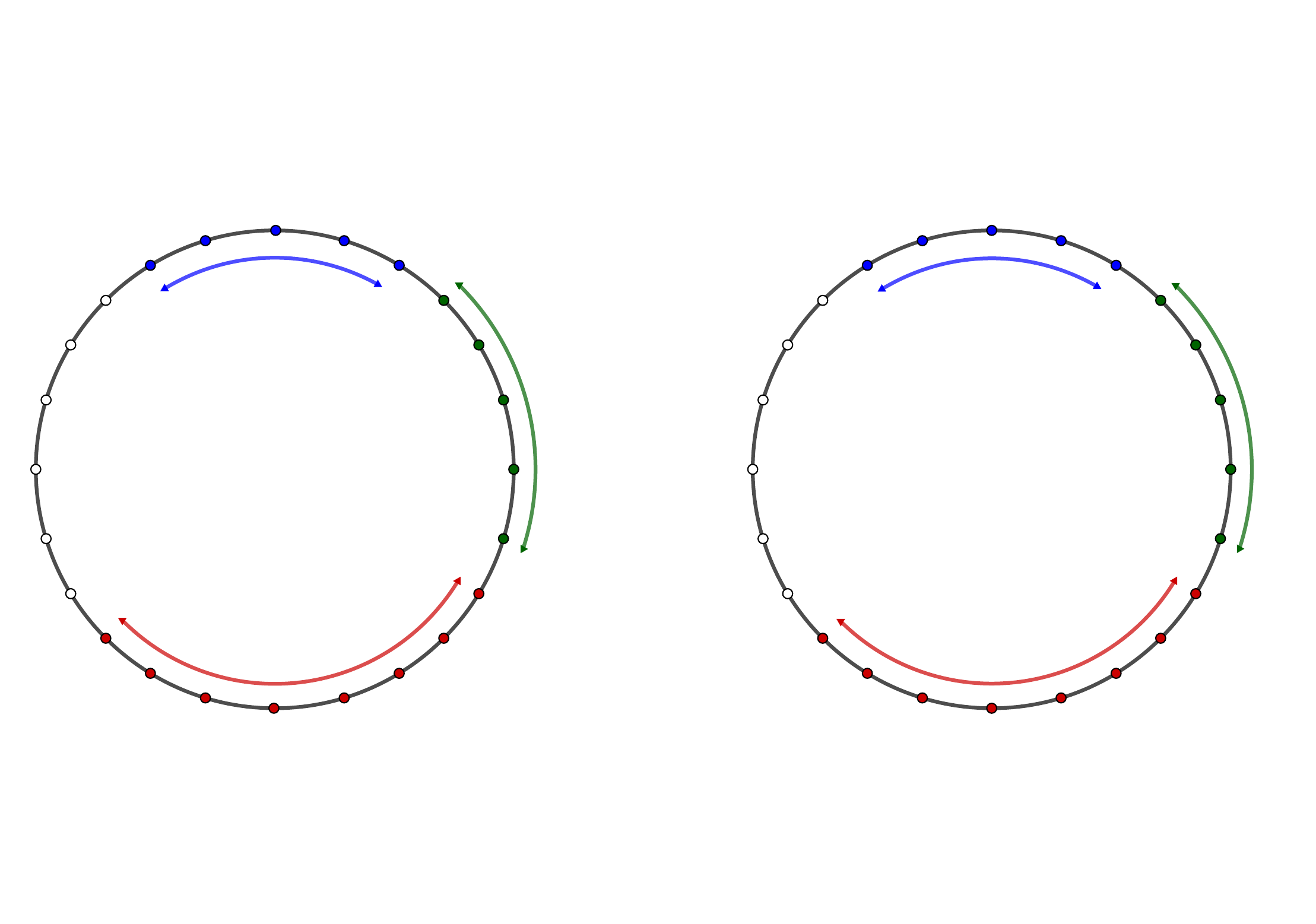}
\put(-330,228){\rotatebox{360}{\fontsize{14}{14}\selectfont $A_{L_1}$}}
\put(-110,228){\rotatebox{360}{\fontsize{14}{14}\selectfont $A_{R_1}$}}
\put(-330,188){\rotatebox{360}{\fontsize{14}{14}\selectfont $l_{1}$}}
\put(-110,188){\rotatebox{360}{\fontsize{14}{14}\selectfont $l_{1}$}}
\put(-330,86){\rotatebox{360}{\fontsize{14}{14}\selectfont $l_{2}$}}	
	\put(-110,86){\rotatebox{360}{\fontsize{14}{14}\selectfont $l_{2}$}}	
	\put(-330,46){\rotatebox{360}{\fontsize{14}{14}\selectfont $A_{L_2}$}}	
	\put(-110,46){\rotatebox{360}{\fontsize{14}{14}\selectfont $A_{R_2}$}}
	\put(-240,160){\rotatebox{360}{\fontsize{14}{14}\selectfont $d$}}
	\put(-20,160){\rotatebox{360}{\fontsize{14}{14}\selectfont $d$}}
	\vspace{-1.5cm}
	
	\caption{Our setup for the decomposition of the system is depicted in this figure. We have taken $N$ sites on each side and the total subsystem consists of four spatially
		disconnected regions: two intervals on the left ($A_{L}$) with the separation $d$ and the corresponding (identical) intervals and the same separation
		on the right ($A_{R}$) with $N_{A}$ sites on each side.}\label{SUB-1}	
\end{figure}
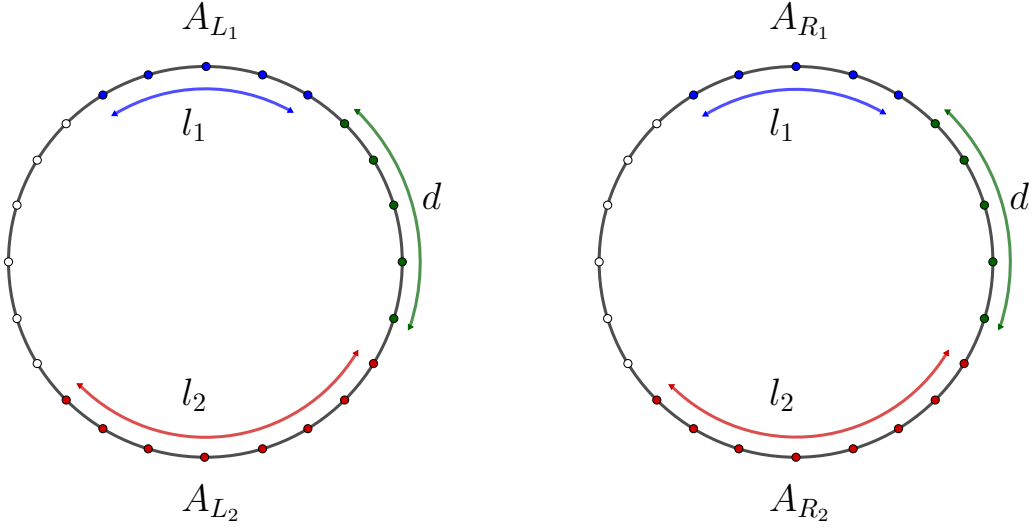

In this section, using the mentioned covariance matrix formalism, we will investigate the time evolution of OEE and LN for the time-dependent $1+1$ TFD state of a real scalar QFT  regularized on a circular lattice. In above it is shown that this state is just a product of TFD states for each of the oscillator modes, (\ref{TTFD.2.2}).  In order to evaluate these quantum information quantities we will take a subregion of the entire quantum system that contains a part in both the left and the right QFTs, see figure \ref{SUB-1}.
Let us describe the whole system as
\bea
\bigg(A_{L} \cup B_{L}\bigg)\cup\bigg(A_{R}\cup B_{R}\bigg),
\eea
in the corresponding TFD state. According to  figure \ref{SUB-1}, the $B_{L(R)}$ are the complement of $A_{L(R)}$. Moreover, we decompose the region $A_{L(R)}$ into two subregions $A_{L_{1}(R_{1})}$ and $A_{L_{2}(R_{2})}$ where they describe two (in general disconnected) subregions on each side 
\bea\label{decom}
\bigg((A_{L_1}\cup A_{L_2}) \cup B_{L}\bigg)\cup\bigg((A_{R_1}\cup A_{R_2})\cup B_{R}\bigg).
\eea
We would like to study  $S_{\text{OEE}}(\rho_{12}(t))$ and $\mathcal{E}(\rho_{12}(t))$ between two non-complementary regions $A_1= (A_{L_1}\cup A_{R_1})$ and $A_2=(A_{L_2}\cup A_{R_2})$.
The reduced covariance matrix $G_{12}$ can be obtained from the covariance matrix (\ref{GA}) where each of the regions $(A_{L_{1}}\cup A_{L_{2}})$ and $(A_{R_{1}}\cup A_{R_{2}})$ contain $N_{A}$ sites.  Accordingly, $G_{12}$ is decomposed into four blocks corresponding to $LR$ decomposition in position space such that each block is a $2N_A\times 2N_A$ matrix. Two of these blocks, $G_{12}^{LL}$ and $G_{12}^{RR}$, are time-independent and the time dependence of the covariance matrix comes from mixed blocks, $G_{12}^{LR}(t)$ and $G_{12}^{RL}(t)$:
\begin{equation}\label{G12}
G_{12}^{ab} (t)= \left(
\begin{array}{cc}
G_{12}^{LL}  & G_{12}^{LR}(t) \vspace{.4cm}

\\
G_{12}^{RL}(t) & G_{12}^{RR}\\
\end{array} \right)_{4 N_A\times 4 N_A},
\end{equation}
with
\bea
\label{GA11}
G_{12}^{LL}= \left(
\begin{array}{cc}
	G_{LL}^{\Phi\Phi} & 0\vspace{.4cm}
	
	\\
	0 & G_{LL}^{\Pi\Pi}\\
\end{array} \right)_{2 N_A\times 2 N_A}
\hspace{1cm}
G_{12}^{LR}(t)= \left(
\begin{array}{cc}
	G_{LR}^{\Phi\Phi} &G_{LR}^{\Phi\Pi}\vspace{.4cm}
	
	\\
	G_{LR}^{\Pi\Phi} &  G_{LR}^{\Pi\Pi}\\
\end{array} \right)_{2 N_A\times 2 N_A}.
\eea
In the above expression, $\left\{ G_{LL}^{\Phi\Phi}, G_{LL}^{\Pi\Pi},\dots \right\}$ are given by the equation (\ref{EQ-1}) with $m,n$ restricted to the entangling region. In addition,  $G_{12}^{RR}$, $G_{12}^{RL}(t)$ blocks can be obtained by changing $L$ by $R$ in (\ref{GA11}). Once the covariance matrix $G_{12}^{ab}(t)$, (\ref{G12}), is found one can express the $S_{\text{EE}}(\rho_{12}(t))$ in terms of eigenvalues $(\nu_{i})$ of the symplectic form
\bea\label{J} J=i\hspace{.5mm}\Omega_A^{-1}G_{12}(t).
\eea  
Doing so, one obtains \cite{Eisert:2003hpa, Weedbrook:2012zau} \footnote{We are taking the absolute value of the eigenvalues of the symplectic form. Since these eigenvalues come in pairs, we should include a factor of $\frac{1}{2}$ in evaluating the entanglement entropy.}:
\begin{equation}\label{SOEEE}
S_\text{EE}\left(\rho_{12}(t)\right)=\frac{1}{2}\sum_{i=1}^{4N_{A}}s_{\text{EE}}(|\nu_i|),
\end{equation}
where,
\begin{equation}\label{SEE}
s_{\text{EE}}(\nu_{i})=\left(\frac{\nu_i+1}{2}\right)\log\left(\frac{\nu_i+1}{2}\right)- \left(\frac{\nu_i-1}{2}\right) \log\left(\hspace{.5mm}\frac{ \nu_i-1}{2}\hspace{.5mm}\right).
\end{equation}
It is useful to mention that the Rényi entropies $S_{n}$ for $n > 0$ can be computed by replacing the $s_{\text{EE}}(\nu_{i})$ in \eqref{SOEEE} with $s_{n}(\nu_{i})$ defined as
\begin{equation}\label{SEE-2}
s_{n}(\nu_{i})=\frac{1}{n-1}\log\left[\frac{\left(\nu_i+1\right)^{n}-\left(\nu_i-1\right)^{n}}{2^{n}}\right].
\end{equation}
The entanglement entropy \eqref{SOEEE} can be recovered in the limit
$n\rightarrow 1$. In addition, the second Rényi entropy is $S_{2}=\frac{1}{2}\log \det (G)$. Now, in order to compute $S_{\text{OEE}}$, we must take a partial transpose with respect to momentum degrees of freedom in the $A_2 $ subregion \cite{Simon:2000zz,Audenaert:2002xfl}. This can be accomplished by acting with the time-reversal operator $\mathcal{R}_{A_{2}}$ on each block of $G_{12}^{ab}$. For example  
\begin{equation}
\tilde{G}^{LL}_{12}=\mathcal{R}_{A_{2}}.G_{12}^{LL}.\mathcal{R}_{A_{2}},
\end{equation}
with $\mathcal{R}_{A_{2}}$ given by a $2N_A \times 2N_A$ square matrix:
\begin{equation}
\mathcal{R}_{A_{2}}=\text{diag}\{1,1,\cdots,1,-1,\cdots,-1\}.
\end{equation}
The number of ``$-1$'' elements in the above expression is equal to the length of the subregion $A_{L_{2}}$. It is worth mentioning that the operator $\mathcal{R}_{A_{2}}$ is the same for the other blocks, $G_{12}^{LR}(t),G_{12}^{RL}(t)$ and $G_{12}^{RR}(t)$. This is because we are considering the simplest case here in which the $L$ and $R$ blocks are exactly the same. Now, by having $\tilde{G}^{ab}_{12}$, one can compute eigenvalues $(\tilde{\nu}_{i})$ of partial transposed symplectic form 
\bea\label{Jt}
\tilde{J}=i\hspace{.5mm} \Omega_A^{-1}\tilde{G}_{A_{1}A_{2}}.
\eea
Having these, the odd entanglement entropy, $S_{\text{OEE}}$, becomes \cite{Audenaert:2002xfl}
\begin{equation}\label{SOEE}
S_\text{OEE}\left(\rho_{12}(t)\right)=\frac{1}{2}\sum_{i=1}^{4N_{A}}\tilde{s}_\text{odd}(|\tilde{\nu}_{i}|),
\end{equation}
with
\begin{equation}\label{Stodd}
\tilde{s}_\text{odd}(\tilde{\nu}_{i})=\left(\frac{\tilde{\nu}_{i}+1}{2}\right)\log\left(\frac{\tilde{\nu}_{i}+1}{2}\right)-\text{sgn} \left(\frac{\tilde{\nu}_{i}-1}{2}\right) \bigg{\vert}\frac{ \tilde{\nu}_{i}-1}{2}\bigg{\vert}\log\left(\hspace{.5mm}\bigg{\vert}\frac{ \tilde{\nu}_{i}-1}{2}\bigg{\vert}\hspace{.5mm}\right).
\end{equation}
It is worth emphasizing that the partial transposition is not a completely positive map and the appearance of negative eigenvalues is a sign of quantum entanglement\cite{Peres:1996dw}. This is the main reason to consider the absolute value in (\ref{Stodd}) in comparison with (\ref{SEE}). Moreover, by knowing the eigenvalues $\tilde{\nu_{i}}$, the trace norm of the reduced density matrix $\rho_{12}^{T_{2}}$ becomes \cite{Audenaert:2002xfl,Calabrese:2012nk}\footnote{The out-of-equilibrium dynamics of the negativity after a \emph{different}
quench has been studied previously in several works\cite{Coser:2014gsa,ViktorEisler:2014wfo,Hoogeveen:2014bqa,Wen:2015qwa,Alba:2018hie,Fujita:2018lfj,Kudler-Flam:2020url,Kudler-Flam:2020xqu}.}
\bea
\text{Tr}\vert \rho_{12}^{T_{2}} \vert=\prod_{i=1}^{4N_{A}}\left[\vert \frac{\tilde{\nu}_{i}+1}{2}\vert-\vert \frac{\tilde{\nu}_{i}-1}{2}\vert \right]^{-1}=\prod_{i=1}^{4N_{A}}\text{max}\left( 1,\frac{1}{\tilde{\nu}_{i}}\right),
\eea
which implies that the LN becomes\footnote{The factor $1/2$ comes from the fact that we are taking the absolute values of eigenvalues which come in pairs.}
\bea \label{LN-1}
\mathcal{E}(A_{1},A_{2})=-\frac{1}{2}\sum_{i=1}^{4N_{A}}\log\left[ \text{min}\left( 1,\tilde{\nu}_{i}\right) \right].  
\eea
According to the above relation, only the symplectic eigenvalues $\tilde{\nu} < 1$ contribute to the LN. Hence, in order to have quantum correlation we must have at least one symplectic eigenvalue which is less than one. Note that the LN is a relative entanglement measure, hence, it is symmetric with respect to the exchange of subsystems.\\

Before closing this section, let us
clarify some points that will be useful for interpreting the numerical results presented in the next section. The entanglement entropy enjoys several properties. One of them is known as subadditivity. With respect to the decomposition 
\begin{equation}\label{BO-1}
	A=(A_{L_{1}}\cup A_{L_{2}})\cup(A_{R_{1}}\cup A_{R_{2}}),
\end{equation}
the entanglement entropy satisfies
\begin{equation}\label{BO-1}
S_{\text{EE}}(\rho_A(t))\leqslant S_{\text{EE}}(\rho_{A_{L}}(t))+S_{\text{EE}}(\rho_{A_{R}}(t)),
\end{equation}
in which, $A_{L(R)}=A_{L_{1}(R_{1})}\cup A_{L_{2}(R_{2})}$. 
The subadditivity provides a time-independent upper bound for the entanglement entropy. The aforementioned decomposition results in a block structure in the reduced covariance matrix $G_{A}$ which consists of four blocks corresponding to $LR$ decomposition in the position space.
The time-independent blocks, $G_{12}^{LL}$ and $G_{12}^{RR}$, are related to thermal entropy corresponding to reduced thermal density matrix of each individual side, $(A_{L_{1}}\cup A_{L_{2}})$ and $(A_{R_{1}}\cup A_{R_{2}})$. Since intervals are equal and symmetric, these thermal entropies are equal. Therefore, the time-independent  upper bound is twice the thermal entropy. This means that with respect to the bound (\ref{BO-1}), the growing behavior of the entanglement entropy over a large range of times ultimately terminates in twice the thermal entropy.
\\

There is also another inequality that is specific to the Gaussian states \cite{Bianchi:2017kgb},
\begin{equation}\label{BO-2}
S_{\text{EE}}(\rho_A(t))\leqslant S_{2}(\rho_{A}(t))+2N_{A}(1-\log 2).
\end{equation}
In the above inequality, $N_{A}$ denotes the number of bosonic degrees of freedom associated with each side of the TFD state
in the subregion $A$  and $S_{2}(\rho_{A_{L}}(t))=\frac{1}{2}\log \det G_{A}(t)$.
The inequality (\ref{BO-2}) can be derived using the below expressions for entanglement entropy and second Rényi entropy,
\begin{equation}
S_{\text{EE}}(\rho_A(t))=\frac{1}{2}\sum_{i=1}^{4 N_{A}}s(\nu_i), \qquad S_{2}=\frac{1}{2}\sum_{i=1}^{4N_{A}}\log(\nu_{i}),
\end{equation}
where $s({\nu}_i)$ is given by (\ref{SEE}).
According to (\ref{SEE}), for all $\nu\geqslant 1$, we have
\begin{equation}
\log(\nu)\leqslant s(\nu)\leqslant \log(\nu)+(1-\log 2).
\end{equation}
Therefore, by summing over all eigenvalues we can conclude that the second Rényi entropy provides an  upper bound for entanglement entropy of Gaussian states. We will especially use the inequality (\ref{BO-2}) in the analysis of zero-mode in section {\ref{Sec-5}}.

\section{Numerical results}\label{Sec-4}
In this section, we will present our numerical results for the time dependency of OEE and LN based on equations (\ref{SOEE}) and (\ref{LN-1}). These results are expressed in terms of two dimensionless parameters $m\mathcal{L}$ and $\beta/\mathcal{L}$ where $m$ refers to the mass parameter in (\ref{H-1}), $\mathcal{L}$ is the length of the circular lattice with lattice spacing $\delta$, $\mathcal{L} = N\delta$, and $\beta$ is the inverse temperature. Let us remind that we are decomposing the circular lattice according to (\ref{decom}). The subsystem $A$ is further decomposed into two subsystems $A_1$ and $A_2$ with length $l_1$ and $l_2$ respectively: 
\bea
N_{A} =N_{A_1}+N_{A_2} = (l_1 + l_2)/\delta= l/\delta.
\eea 
Since we are interested in describing the TFD state (for single-mode see (\ref{TTFD.2.2})), one has two copies of this decomposition: one for the left QFT and one for the right QFT where on each side we have $N_{A}$ sites (See figure \ref{SUB-1}).
In what follows, we will study different cases for massless as well as massive real scalar QFTs. We consider the behavior of OEE and LN by changing $N_{A}$ ($N_{A_1}$ or $N_{A_2}$ or both of them), the separation distance $d$ (between $A_1$ and $A_2$ on each side), lattice spacing $\delta$ and inverse temperature $\beta$. The obtained results will also be compared with their counterparts for the entanglement entropy. To do so, we will discretize  the circular lattice with circumference $\mathcal{L}$ into $N=2n+1$ sites (almost $1501$) and the length of
intervals $l_{1}$ and $l_{2}$ on each side vary from $0.1 \mathcal{L}$ to $0.5 \mathcal{L}$. Before going to the details of numerical results, it is worth to explain why we choose the lattice cites an odd number. It is clear from (\ref{pq}) that $\mathbf{P}_{a}$ and $\mathbf{Q}_{a}$ are real degrees of freedom., i.e. $\mathbf{P}_{a}^{\dagger}=\mathbf{P}_{a}$ and $\mathbf{Q}_{a}^{\dagger}=\mathbf{Q}_{a}$. But when we pass to the normal mode basis by Fourier transforming, the $\tilde{\mathbf{P}}_{k}$ and $\tilde{\mathbf{Q}}_{k}$ are no longer real with the exceptions of $k=0$ and $k=N/2$ (for even $N$). To restore (each) two real degrees of freedom, we have imposed the constraints $\tilde{\mathbf{P}}_{k}^{\dagger}=\tilde{\mathbf{P}}_{-k}$ and $\tilde{\mathbf{Q}}_{k}^{\dagger}=\tilde{\mathbf{Q}}_{-k}$ which imply that the negative and positive momentum modes will be mixed, $[\tilde{\mathbf{Q}}_{k},\tilde{\mathbf{P}}_{-l}] = i \delta_{k,l}$. Accordingly, to avoid over-counting (constraining) those two excepted modes, we choose $N = 2n+1$ for all calculations. 
\\

\begin{figure}[H]	\includegraphics[scale=.39]{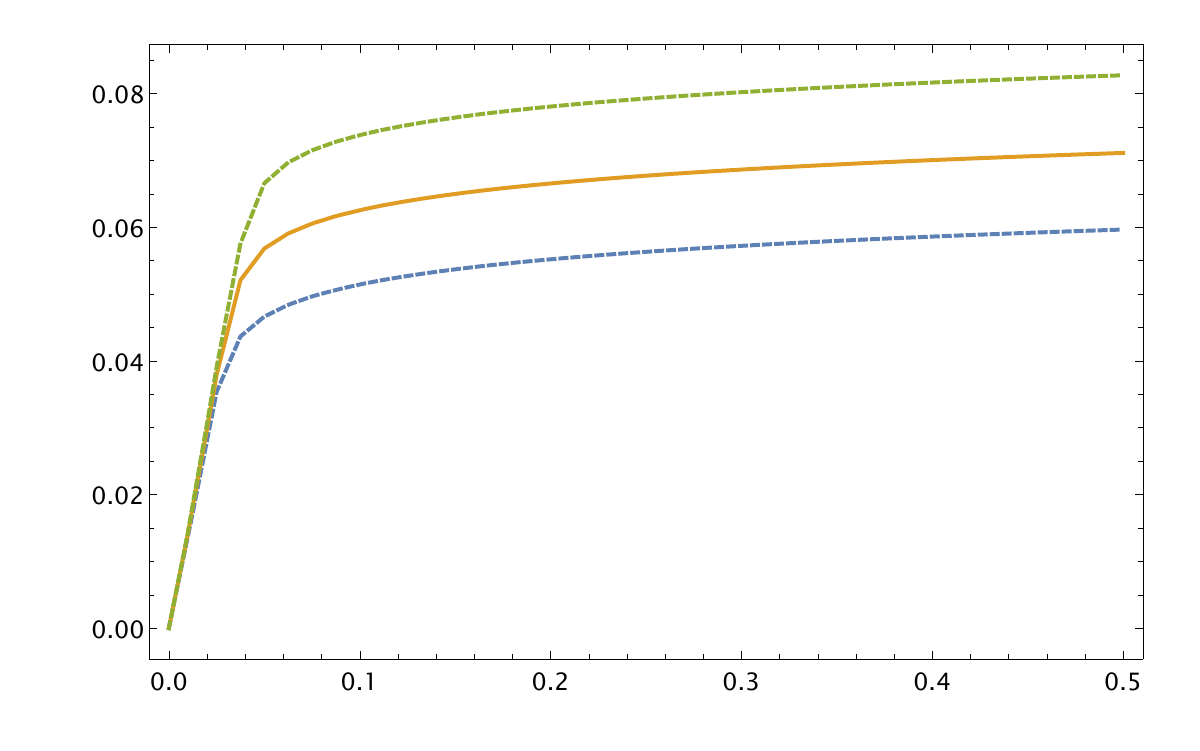}\put(-232,45){\rotatebox{-270}{\fontsize{13}{13}\selectfont $\frac{S_{\text{EE}}(t)-S_{\text{EE}}(0)}{S_{\text{th}}}$}}		\put(-110,-5){{\fontsize{11}{11}\selectfont $t/\mathcal{L}$}}\hspace{.7cm}\includegraphics[scale=.39]{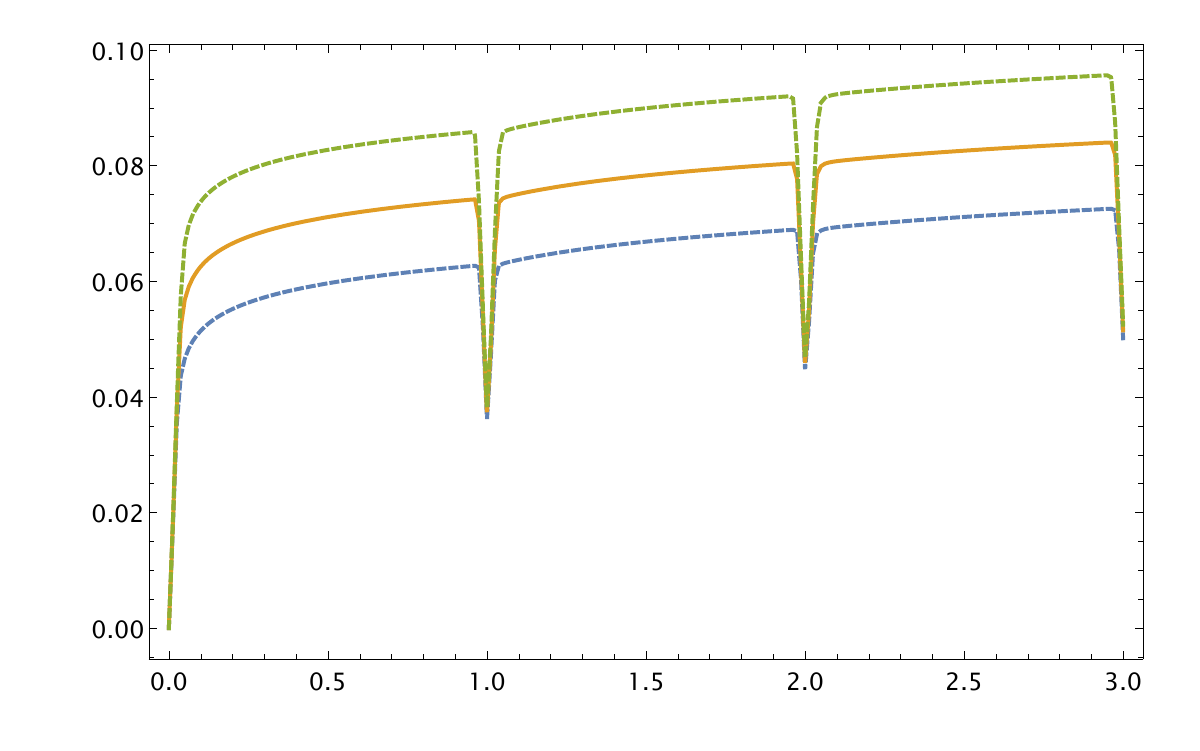}\put(-232,45){\rotatebox{-270}{\fontsize{13}{13}\selectfont $\frac{S_{\text{EE}}(t)-S_{\text{EE}}(0)}{S_{\text{th}}}$}}		\put(-110,-5){{\fontsize{11}{11}\selectfont $t/\mathcal{L}$}}
\vspace{.4cm}
\includegraphics[scale=.39]{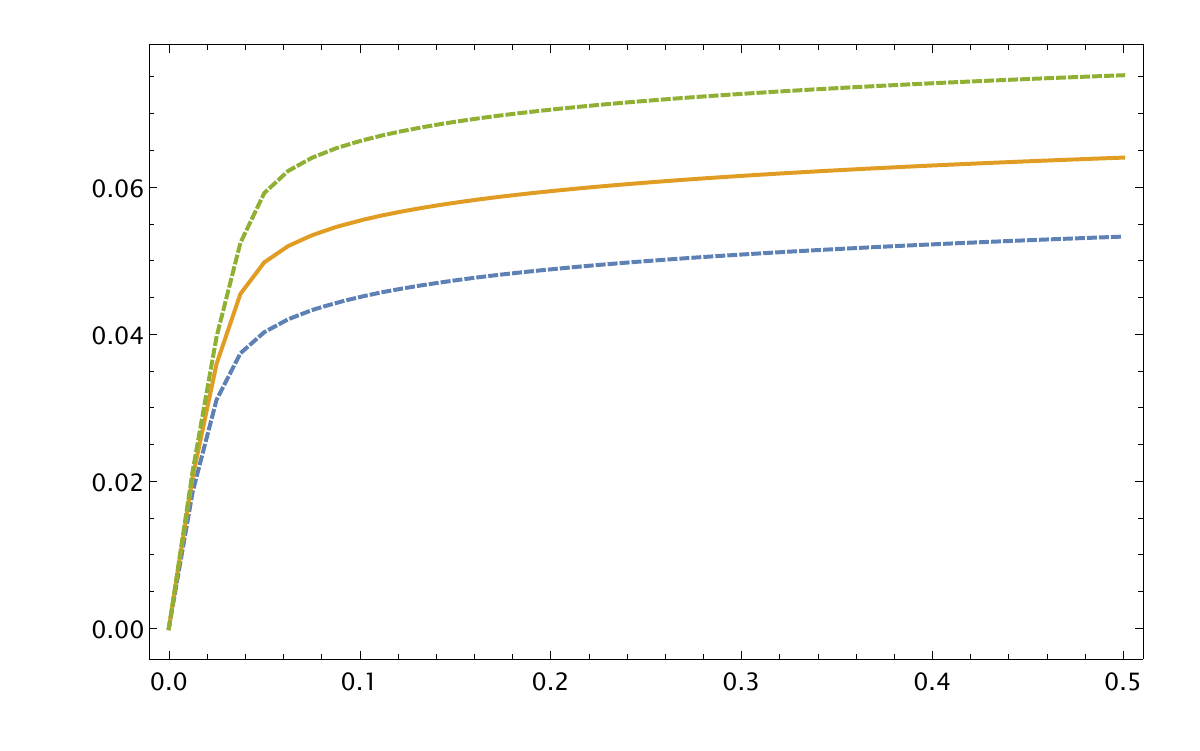}\put(-232,35){\rotatebox{-270}{\fontsize{13}{13}\selectfont $\frac{S_{\text{OEE}}(t)-S_{\text{OEE}}(0)}{S_{\text{th}}}$}}		\put(-110,-5){{\fontsize{11}{11}\selectfont $t/\mathcal{L}$}}\hspace{.85cm}\includegraphics[scale=.39]{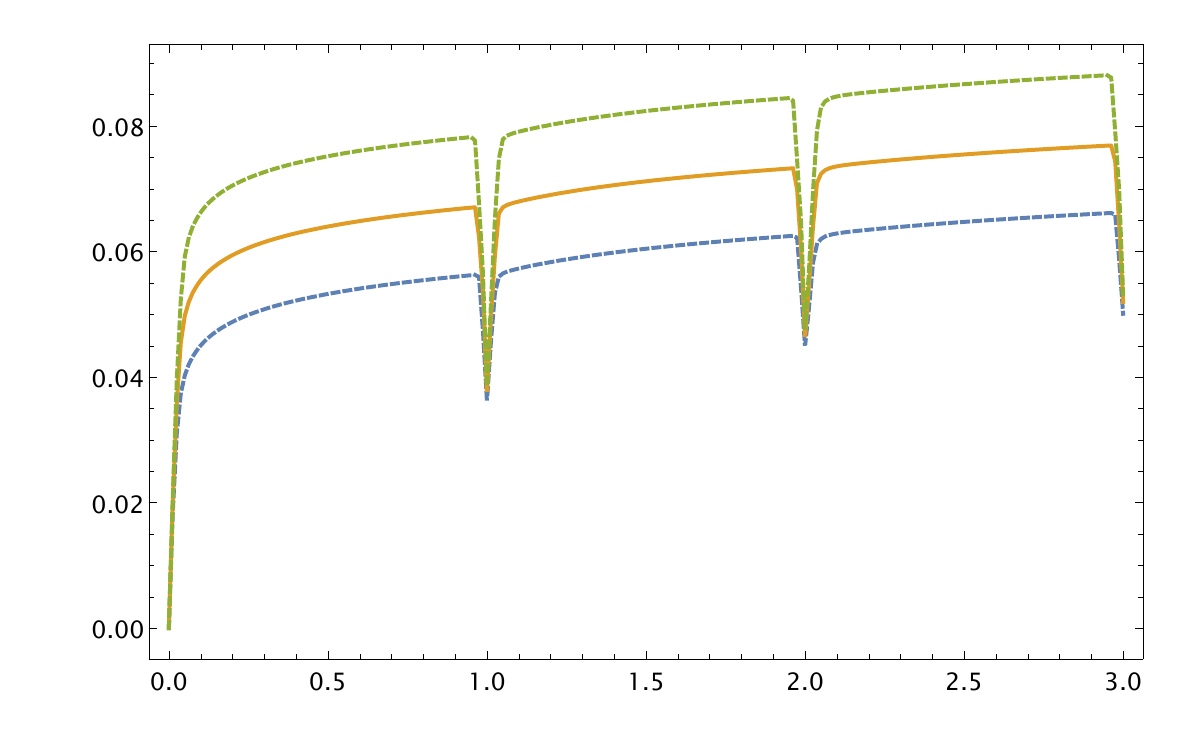}\put(-232,35){\rotatebox{-270}{\fontsize{13}{13}\selectfont $\frac{S_{\text{OEE}}(t)-S_{\text{OEE}}(0)}{S_{\text{th}}}$}}		\put(-110,-5){{\fontsize{11}{11}\selectfont $t/\mathcal{L}$}}
\vspace{.4cm}
\includegraphics[scale=.39]{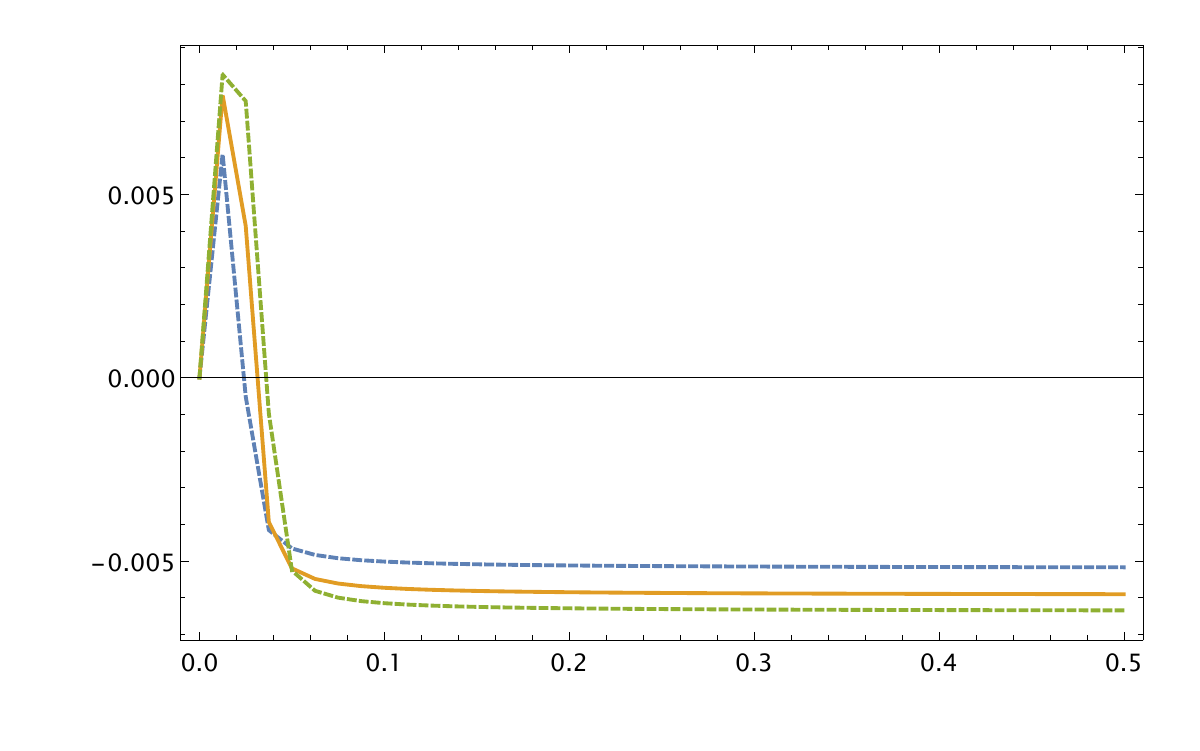}\put(-230,55){\rotatebox{-270}{\fontsize{13}{13}\selectfont $\frac{	\mathcal{E}(t)-	\mathcal{E}(0)}{S_{\text{th}}}$}}		\put(-110,0){{\fontsize{11}{11}\selectfont $t/\mathcal{L}$}}\hspace{.85cm}\includegraphics[scale=.39]{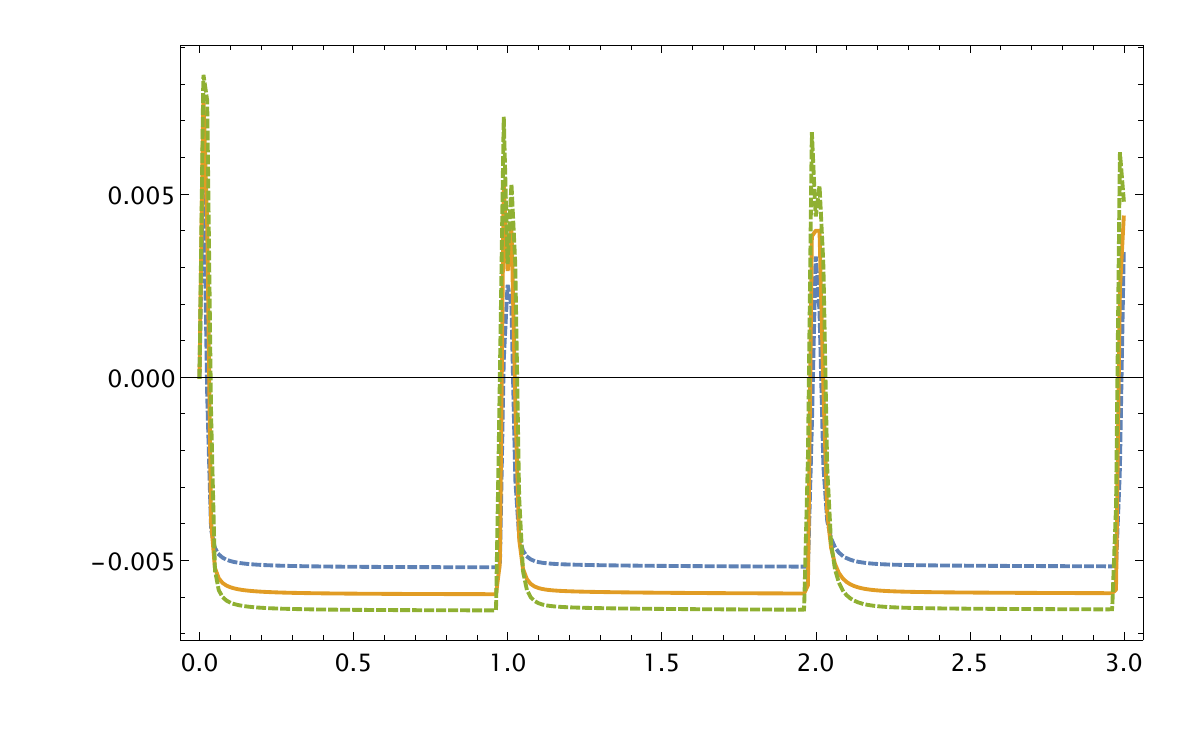}\put(-230,55){\rotatebox{-270}{\fontsize{13}{13}\selectfont $\frac{	\mathcal{E}(t)-	\mathcal{E}(0)}{S_{\text{th}}}$}}		\put(-110,0){{\fontsize{11}{11}\selectfont $t/\mathcal{L}$}}
\caption{The time evolution of $S_{\text{EE}}$(upper plot), $S_{\text{OEE}}$ (middle plot) and $\mathcal{E}$ (bottom plot) for various regimes in which the initial values are subtracted and normalized with respect to the thermal entropy $S_{\text{th}}$. We take $N=1501$, $N_{A_{L(R)}}\hspace{-.1cm}=\hspace{-.1cm}N_{A_{L_{1}(R_{1})}}\hspace{-.1cm}+\hspace{-.05cm}N_{A_{L_{2}(R_{2})}}\hspace{-.1cm}=\hspace{-.1cm}20+a$ where $a=21,31,41$ for dashed blue, orange, dashed green curves respectively and $m \mathcal{L}=10^{-3}$, $\beta=10^{-2} \mathcal{L}$. The upper-left plot denotes the growth of $S_{\text{EE}}$ which is linear at times approximately equal to the size of an interval. The upper-right plot is for longer periods of time. One can see the finite size
effects which induce a periodic behavior with periodicity $\mathcal{L}$. We see a logarithmic growth due to the presence of a zero-mode which we study further analytically in the next section. The middle-left and right panels denote the time dependency of  $S_{\text{OEE}}$. The effective time evolution of  $S_{\text{EE}}$ and $S_{\text{OEE}}$ are the same. The bottom-left plot denotes the short-time behavior of $\mathcal{E}$ for $a=21$ (dashed blue), $31$(orange), $41$(dashed green). In early times we see growth and then reduction with respect to the initial value of LN.  The bottom-right plot is for a long time regime, in which the finite size effect exhibit itself as periodic behavior.}\label{MasslessSSLN-adj1}
\end{figure}
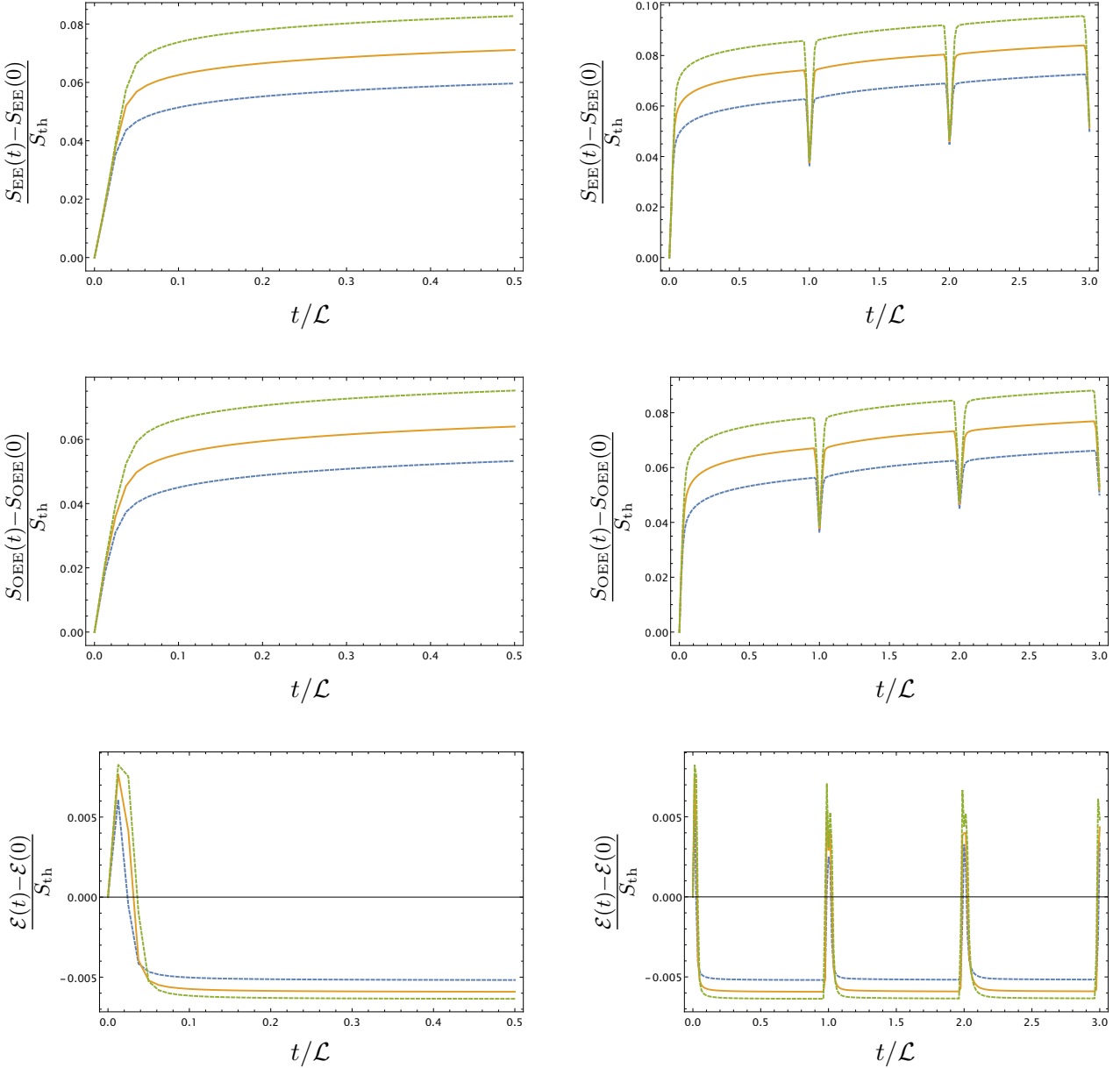

Let us first consider the case in which the two intervals $A_1$ and $A_2$ on each side are adjacent to each other, i.e., the separation distance vanishes, $d=0$. In figure  \ref{MasslessSSLN-adj1}, we investigate the time dependence of entanglement entropy $S_{\text{EE}}$(upper plots), odd entanglement entropy $S_{\text{OEE}}$(middle plots) and logarithmic negativity $\mathcal{E}$ (bottom plots) for short-time (left plots) and long-time (right plots) scales in which the initial value is subtracted and normalized with respect to the thermal entropy $S_{\text{th}}$\footnote{For a real scalar QFT, the thermal entropy in the continuum limit is given by
\bea\nonumber
S_{\text{th}} = \mathrm{vol} \int \frac{\mathrm{d}^{d-1}k}{(2\pi)^{d-1}}\bigg[\frac{\beta(\omega_k)}{e^{\beta(\omega_k)}-1}-\log(1-e^{-\beta(\omega_k)})\bigg].
\eea
}. 
We take  $N=1501$, $N_{A_{L(R)}}\hspace{-.1cm}=\hspace{-.1cm}N_{A_{L_{1}(R_{1})}}\hspace{-.1cm}+\hspace{-.05cm}N_{A_{L_{2}(R_{2})}}=20+a$, where $a=21,31,41$ (dashed blue, orange, dashed green, respectively), and $m \mathcal{L}=10^{-3}$, $\beta=10^{-2} \mathcal{L}$.  For short-times (i.e. times smaller than $\mathcal{L}/2$  where the system is not sensitive to finite size effects) the growth of $S_{\text{EE}}$ as well as $S_{\text{OEE}}$ is linear and lasts until approximately $t\sim l$; this linear growth is then followed by saturation. For long-times (i.e. times larger than $\mathcal{L}- l$ where the finite size effects can be visible) the time dependency of $S_{\text{EE}}$ (upper-right panel) as well as $S_{\text{OEE}}$ (middle-right panel) is periodic with periodicity $\mathcal{L}$. To be more precise, the pattern of the time evolution of $S_{\text{EE}}$, as well as $S_{\text{OEE}}$, is consisting of linear growth for early times, $t\sim l$,  a quasi-plateau of width approximately $\mathcal{L}-2l$ in the intermediate times and a linear decreasing up to order $t\sim l$ and then repeating the same structure\footnote{ Similar revivals of quantum states after \emph{different} quantum quench has been already seen
\cite{Cardy:2014rqa,Najafi:2017esc,Modak:2020faf}.}. The slope of linear growth, for both of these quantities, is equal with good accuracy to two times of thermal entropy density of each copy at inverse temperature $\beta$.
This matches with (\ref{Pheno}), since in massless theory the group velocity is the same for all species of quasi-particles and the slope becomes the sum of entropy densities (\ref{sthn}) over all species. We have said the quasi-plateau since for these flat regimes we observe a logarithmic growth. In section \ref{Sec-5}, we explicitly show that this non-trivial growth is due to the zero momentum mode. It is also worth mentioning that, upon a closer examination of figure \ref{MasslessSSLN-adj1}, the linear regime does not start right away and a different behavior can be seen around $t = 0$ which we believe is a manifestation of an expected quadratic growth at early times following a quantum quench.
The time dependence of  LN is depicted in the bottom plots for $a=21,31,41$
(dashed blue, orange, dashed green curves, respectively). 
In early times (the bottom-left panel), we observe growth and then decrease followed by saturation. The oscillatory behavior can be observed for long times (the bottom-right panel) due to the finite size effects. This can also be understood by a quasi-particle picture where meeting quasi-particles on the 
opposite side of the circle leads to reducing the correlations between the subsystem and its complement.\\

\begin{figure}[H]
\includegraphics[scale=.39]{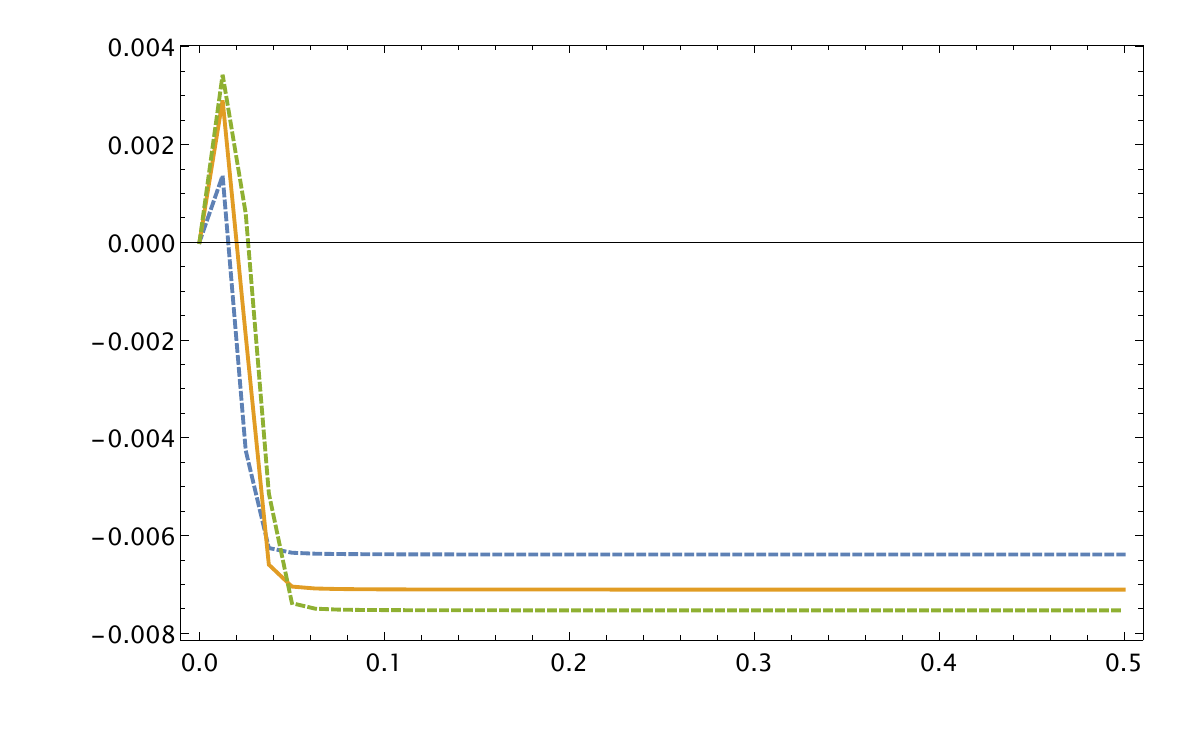}\put(-228,47){\rotatebox{-270}{\fontsize{13}{13}\selectfont $\frac{	\Delta S(t)-	\Delta S(0)}{S_{\text{th}}}$}}		\put(-110,0){{\fontsize{11}{11}\selectfont $t/\mathcal{L}$}}\hspace{.7cm}\includegraphics[scale=.39]{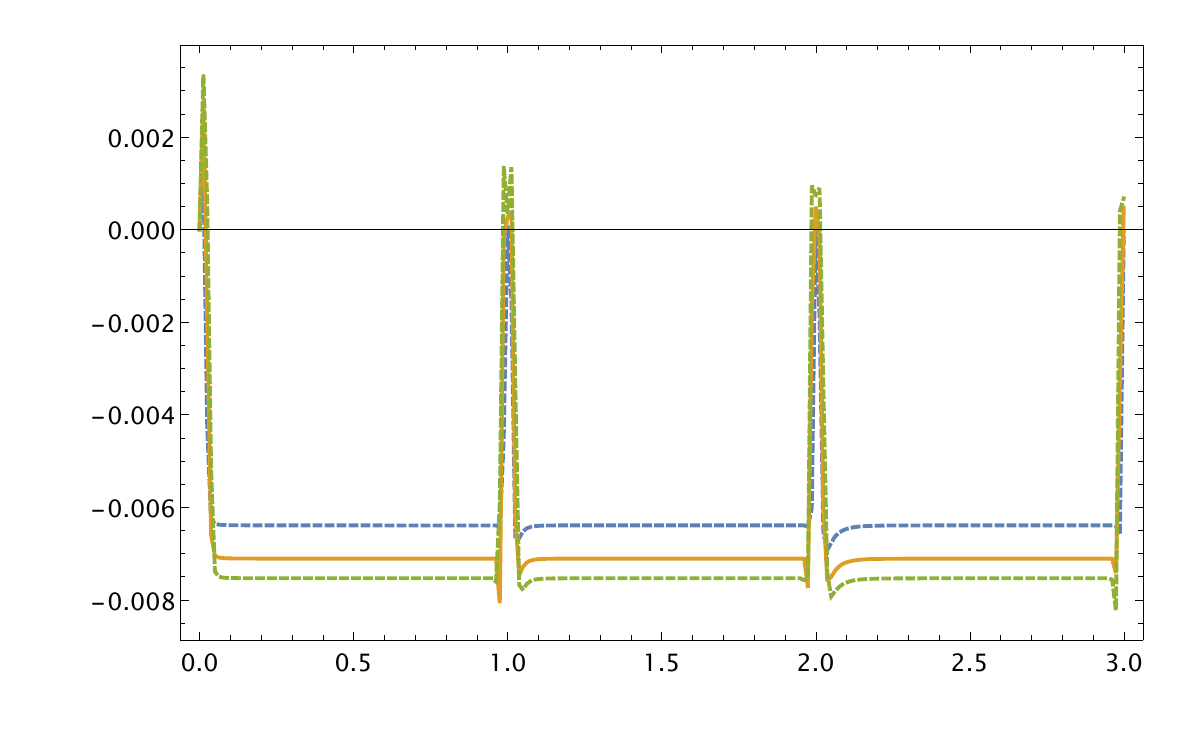}\put(-228,47){\rotatebox{-270}{\fontsize{13}{13}\selectfont $\frac{	\Delta S(t)-	\Delta S(0)}{S_{\text{th}}}$}}		\put(-110,0){{\fontsize{11}{11}\selectfont $t/\mathcal{L}$}}
\caption{Time dependence of  $\Delta S=S_{\text{OEE}}-S_{\text{EE}}$  normalized by the thermodynamic entropy, $S_{\text{th}}$, for the free scalar theory where $N=1501$, $N_{A_{L(R)}}\hspace{-.1cm}=\hspace{-.1cm}N_{A_{L_{1}(R_{1})}}\hspace{-.1cm}+\hspace{-.05cm}N_{A_{L_{2}(R_{2})}}\hspace{-.1cm}=\hspace{-.1cm}20+a$, with $a=21$(dashed blue), $a =31$(orange), and  $a=41$(dashed green) and $m \mathcal{L}=10^{-3}$, $\beta=10^{-2} \mathcal{L}$. Left panels correspond to short-time scales and the right panels are for long-time scales.}
\label{MassiveLN-adj1}
\end{figure}

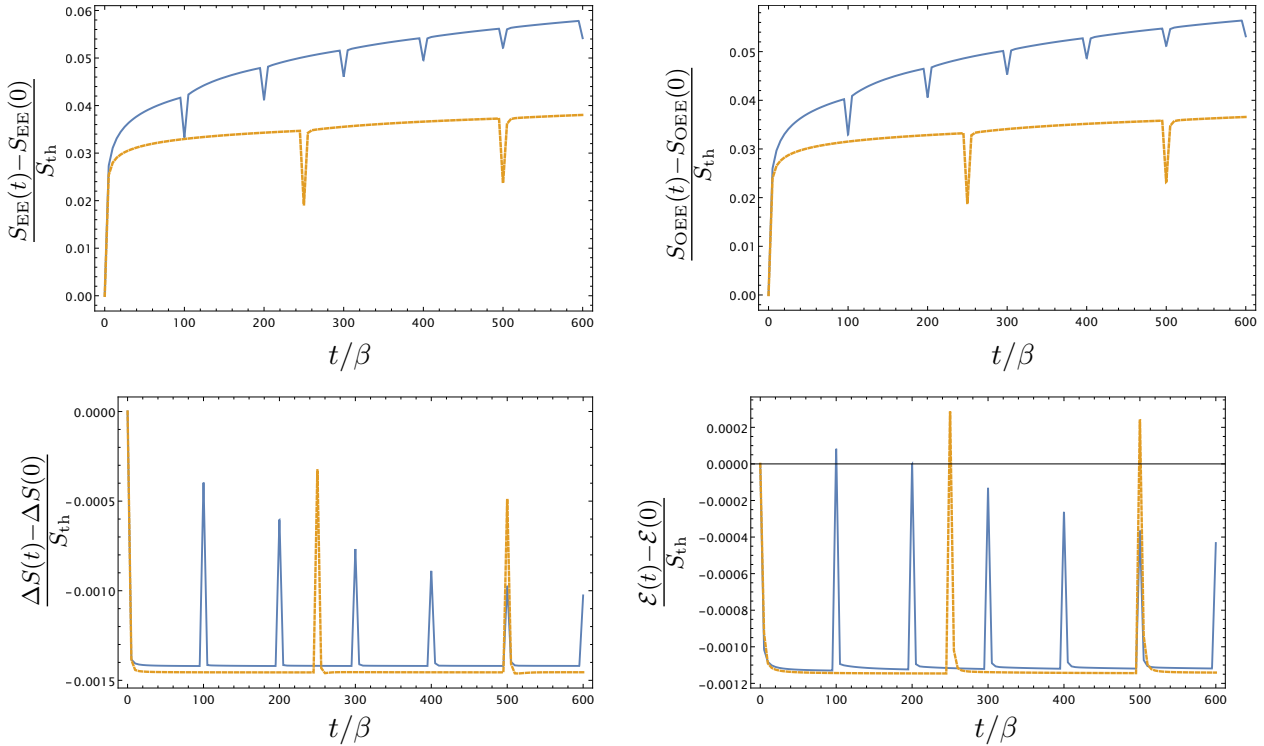
\begin{figure}[H]	
\includegraphics[scale=.39]{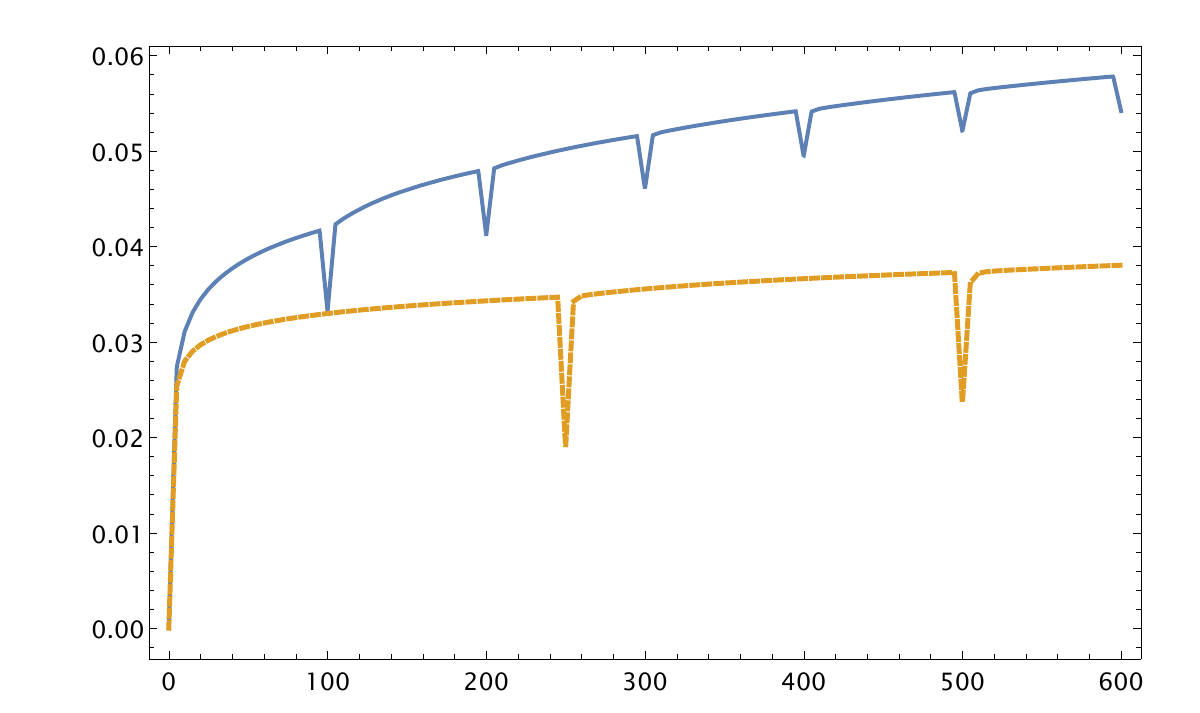}\put(-230,35){\rotatebox{-270}{\fontsize{13}{13}\selectfont $\frac{S_{\text{EE}}(t)-S_{\text{EE}}(0)}{S_{\text{th}}}$}}		\put(-110,-10){{\fontsize{11}{11}\selectfont $t/\beta$}}
\hspace{.8cm}\includegraphics[scale=.39]{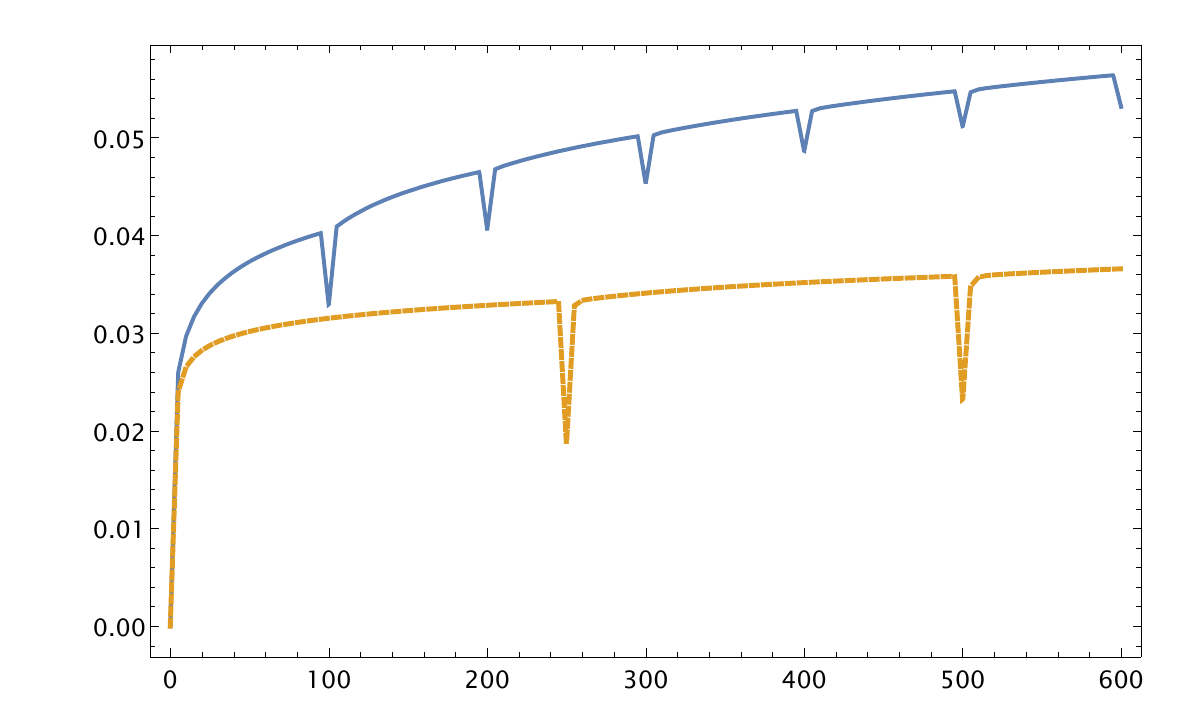}\put(-232,28){\rotatebox{-270}{\fontsize{13}{13}\selectfont $\frac{S_{\text{OEE}}(t)-S_{\text{OEE}}(0)}{S_{\text{th}}}$}}		\put(-110,-10){{\fontsize{11}{11}\selectfont $t/\beta$}}

\includegraphics[scale=.39]{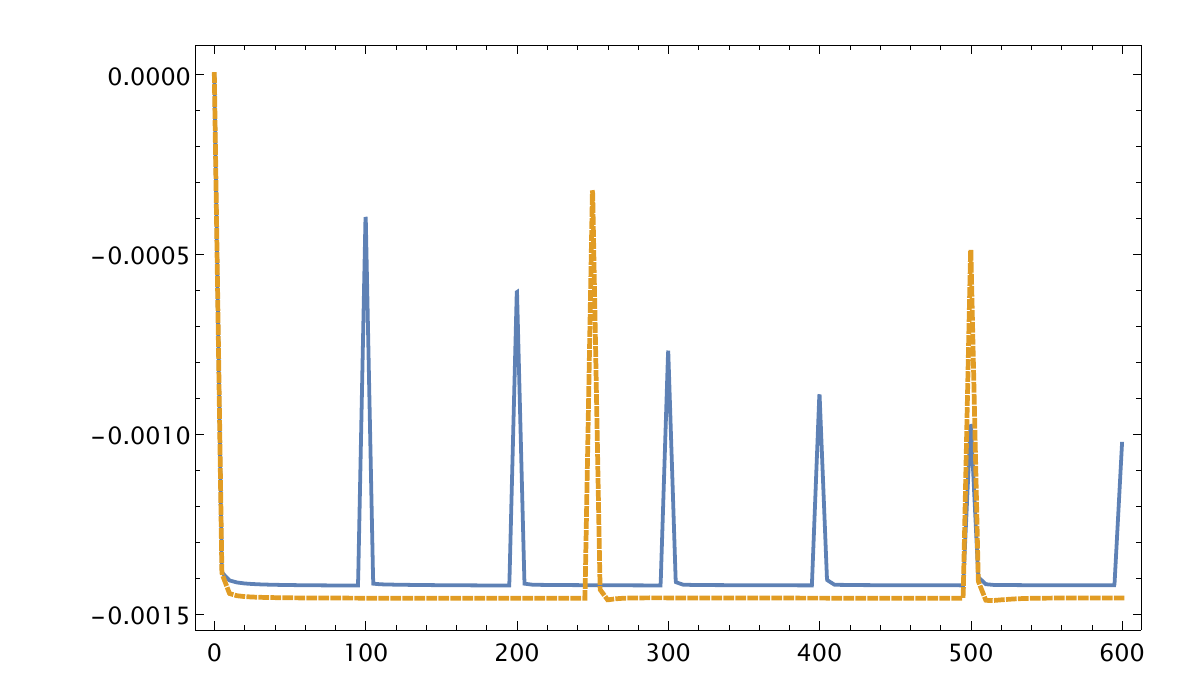}\put(-225,38){\rotatebox{-270}{\fontsize{13}{13}\selectfont $\frac{	\Delta S(t)-	\Delta S(0)}{S_{\text{th}}}$}}		\put(-110,-10){{\fontsize{11}{11}\selectfont $t/\beta$}}
\hspace{.4cm}\includegraphics[scale=.39]{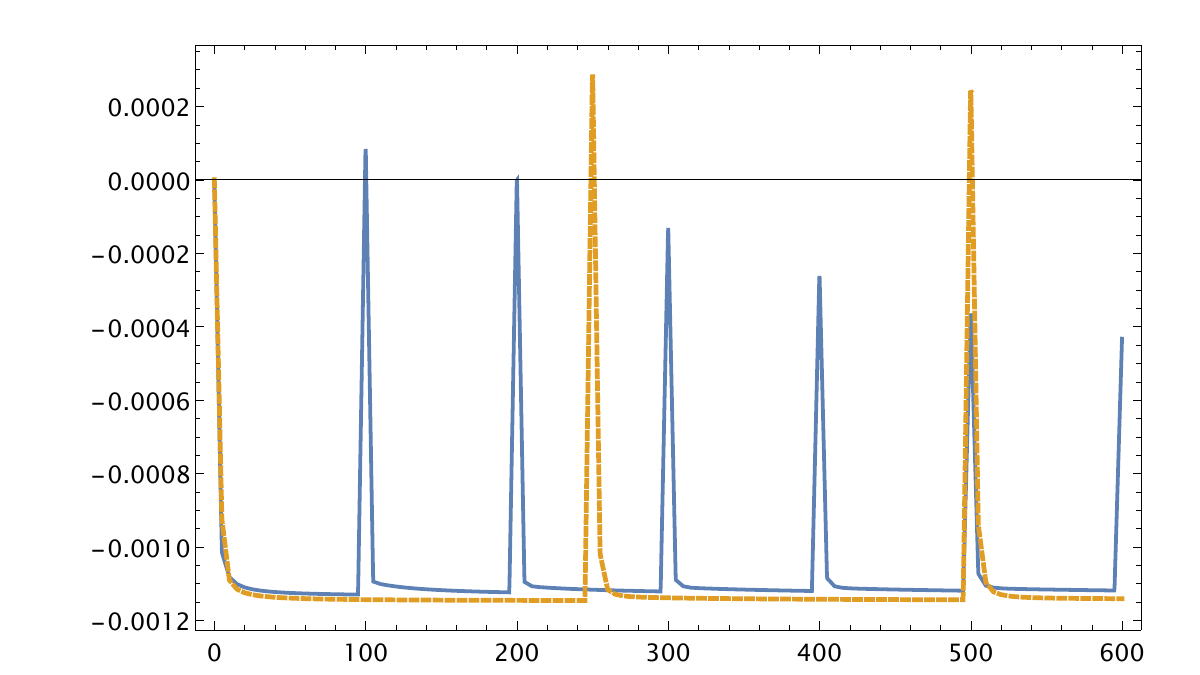}\put(-231,38){\rotatebox{-270}{\fontsize{13}{13}\selectfont $\frac{		\mathcal{E}(t)-	\mathcal{E}(0)}{S_{\text{th}}}$}}	\put(-105,-10){{\fontsize{11}{11}\selectfont $t/\beta$}}
\caption{The dynamics of $S_{\text{EE}}$(upper-left), $S_{\text{OEE}}$(upper-right), $\Delta S=S_{\text{OEE}}-S_{\text{EE}}$(bottom-left), and $\mathcal{E}$(bottom-right) (subtracted from the initial values and normalized with thermal entropy) extrapolated from circle to a line with fixed lattice spacing $\delta$ and $m\beta=10^{-5}$.  
The blue curve denotes the theory on a circle with the total number of sites $N=1501$ and $N_{A_{L(R)}}=N_{A_{L_{1}(R_{1})}}\hspace{-.1cm}+\hspace{-.05cm}N_{A_{L_{2}(R_{2})}}\hspace{-.1cm}=\hspace{-.1cm}2+20$.
The dashed-orange curve represents the same theory on a circle with fixing $\delta$ and increasing the total number of sites (and accordingly subsystem sites) to $N=3751$. 
Increasing $N$ causes the effects of finite size to be transferred to larger times. We also observe the logarithmic growth at the intermediate times.}
\label{QFT-SEEOL}
\end{figure}

\begin{figure}[H]
\centering
\includegraphics[scale=.38]{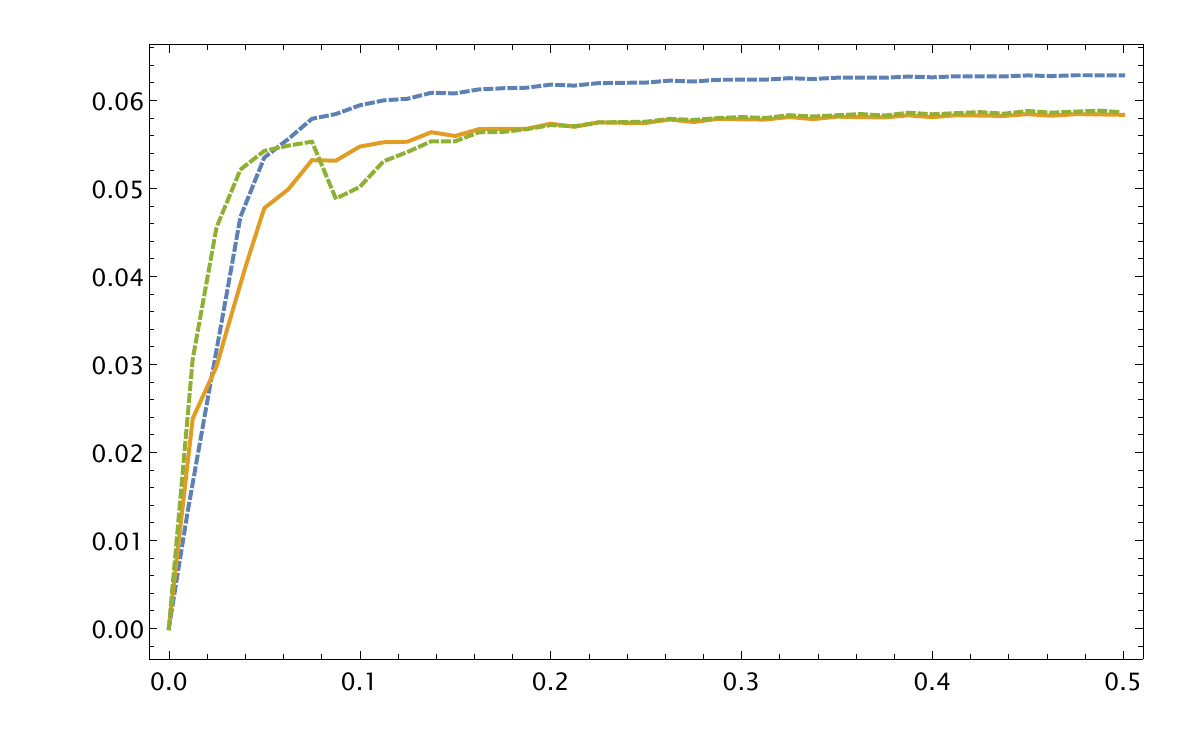}\put(-224,35){\rotatebox{-270}{\fontsize{13}{13}\selectfont $\frac{S_{\text{EE}}(t)-S_{\text{EE}}(0)}{S_{\text{th}}}$}}		\put(-110,-5){{\fontsize{11}{11}\selectfont $t/\mathcal{L}$}}
\hspace{.8cm}\includegraphics[scale=.38]{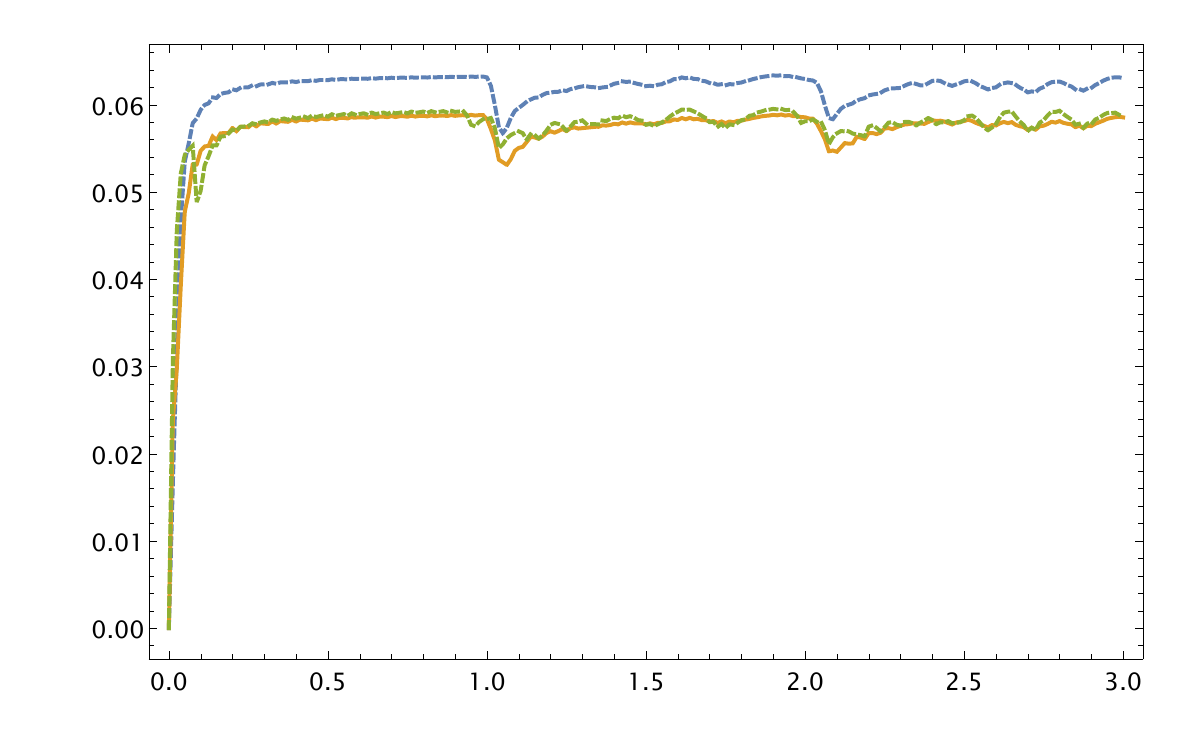}\put(-224,35){\rotatebox{-270}{\fontsize{13}{13}\selectfont $\frac{S_{\text{EE}}(t)-S_{\text{EE}}(0)}{S_{\text{th}}}$}}		\put(-110,-5){{\fontsize{11}{11}\selectfont $t/\mathcal{L}$}}
\vspace{0cm}

\includegraphics[scale=.37]{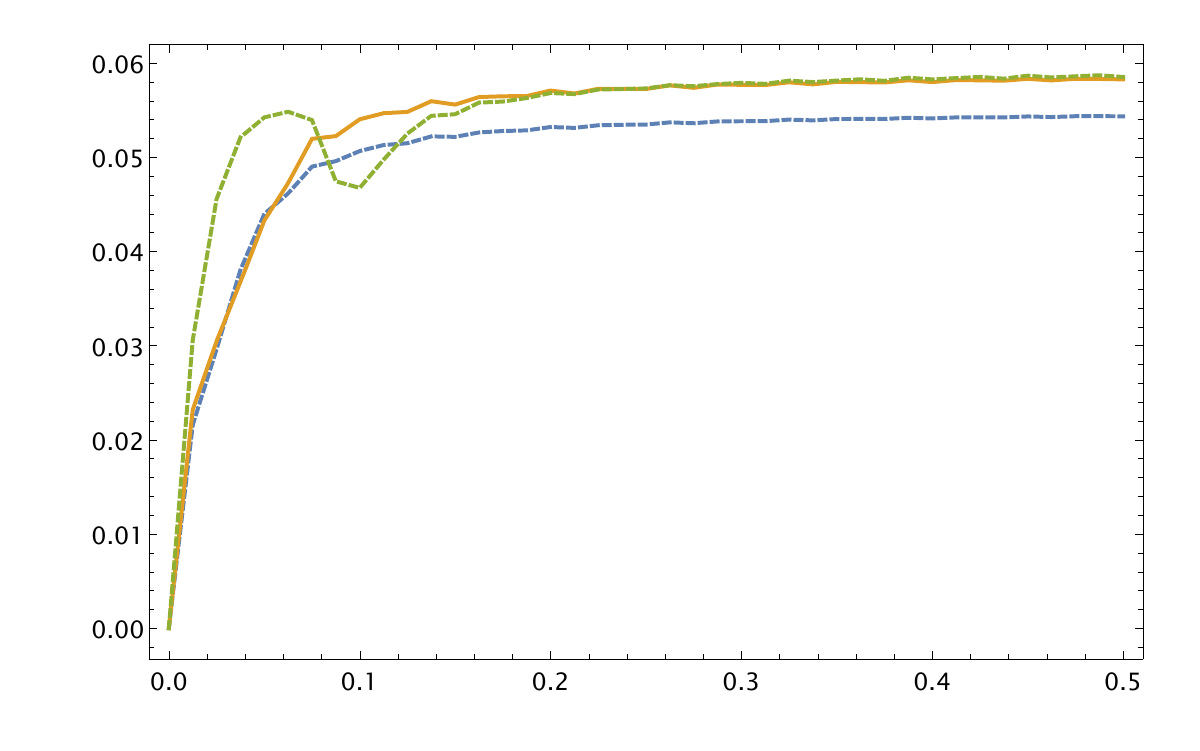}\put(-220,35){\rotatebox{-270}{\fontsize{13}{13}\selectfont $\frac{S_{\text{OEE}}(t)-S_{\text{OEE}}(0)}{S_{\text{th}}}$}}		\put(-110,-5){{\fontsize{11}{11}\selectfont $t/\mathcal{L}$}}
\hspace{.8cm}\includegraphics[scale=.38]{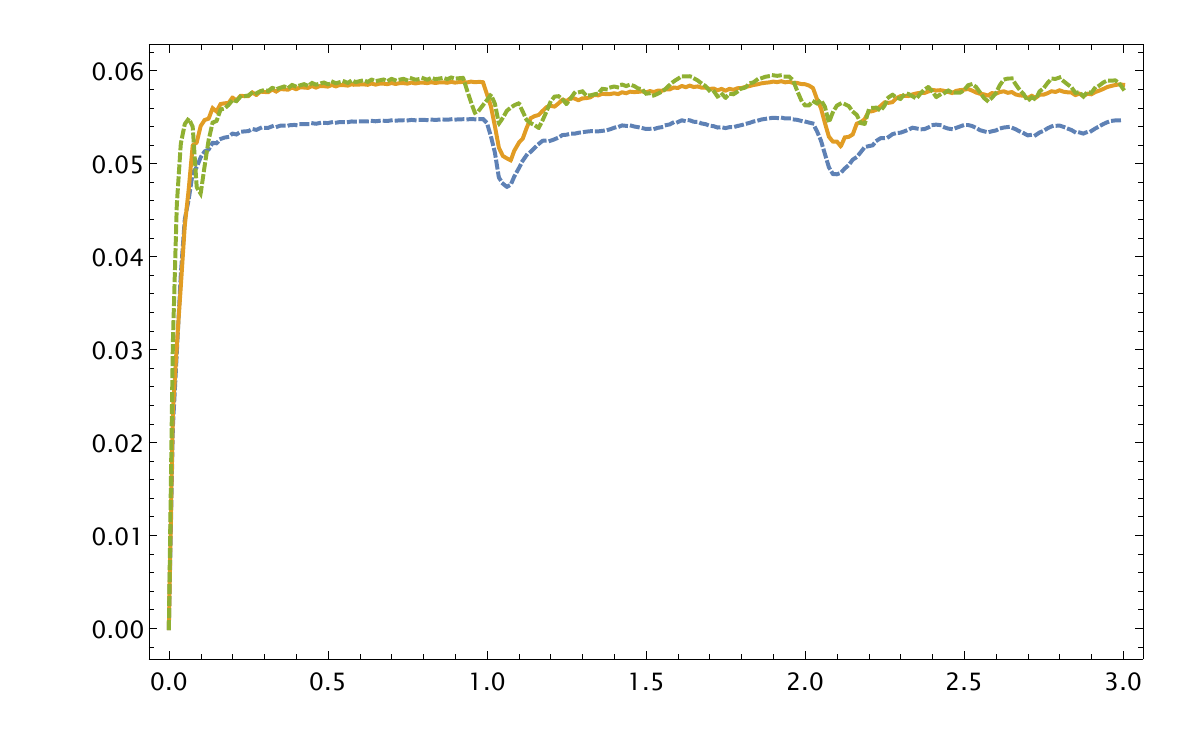}\put(-225,35){\rotatebox{-270}{\fontsize{13}{13}\selectfont $\frac{S_{\text{OEE}}(t)-S_{\text{OEE}}(0)}{S_{\text{th}}}$}}		\put(-110,-5){{\fontsize{11}{11}\selectfont $t/\mathcal{L}$}}
\vspace{.2cm}
	
\caption{Time dependence of $S_{\text{EE}}$ (upper panels), $S_{\text{OEE}}$ (bottom panels) is plotted for the free massive scalar theory when 
$N=1501$ , $N_{A_{L(R)}}\hspace{-.1cm}=\hspace{-.1cm}N_{A_{L_{1}(R_{1})}}\hspace{-.1cm}+\hspace{-.05cm}N_{A_{L_{2}(R_{2})}}\hspace{-.1cm}=\hspace{-.1cm}20+31$ and $m \mathcal{L}=10^{2}$, $\beta=10^{-2} \mathcal{L}$. The curves correspond to the $d=0$(dashed blue), $d=10$(orange), and  $d=100$(dashed green). The upper-left plot denotes the linear regime for the growth of entanglement entropy. The linear growth is for times approximately equal to the size of an interval. The upper-right plot is for a longer period of time for which the finite size effects induce a periodic behavior with periodicity $\mathcal{L}$. The lower-left and right panels denote time behavior of odd entanglement entropy. The effective evolution of both $S_{\text{EE}}$ and $S_{\text{OEE}}$ are the same in different  times. We also see the memory effect in $S_{\text{EE}}$ and $S_{\text{OEE}}$ when $d > l$.} \label{Massive-dist1}
\end{figure}
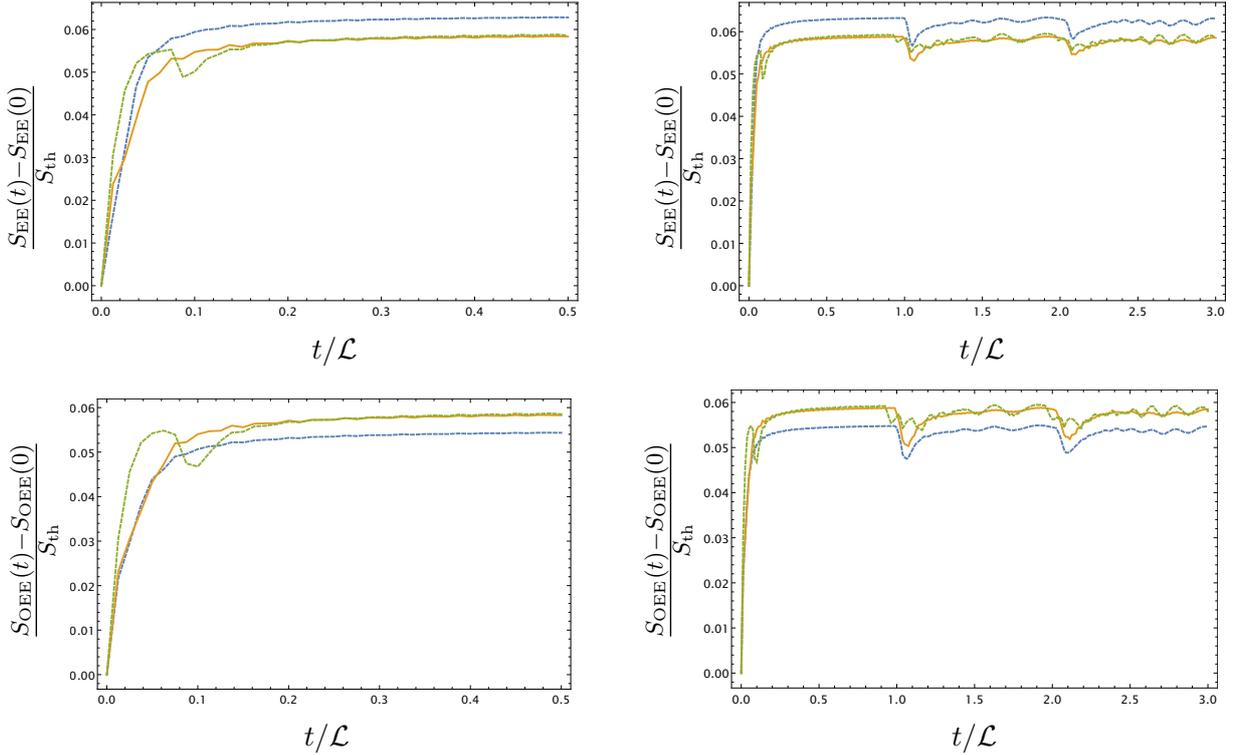

\begin{figure}[H]
\centering
\includegraphics[scale=.39]{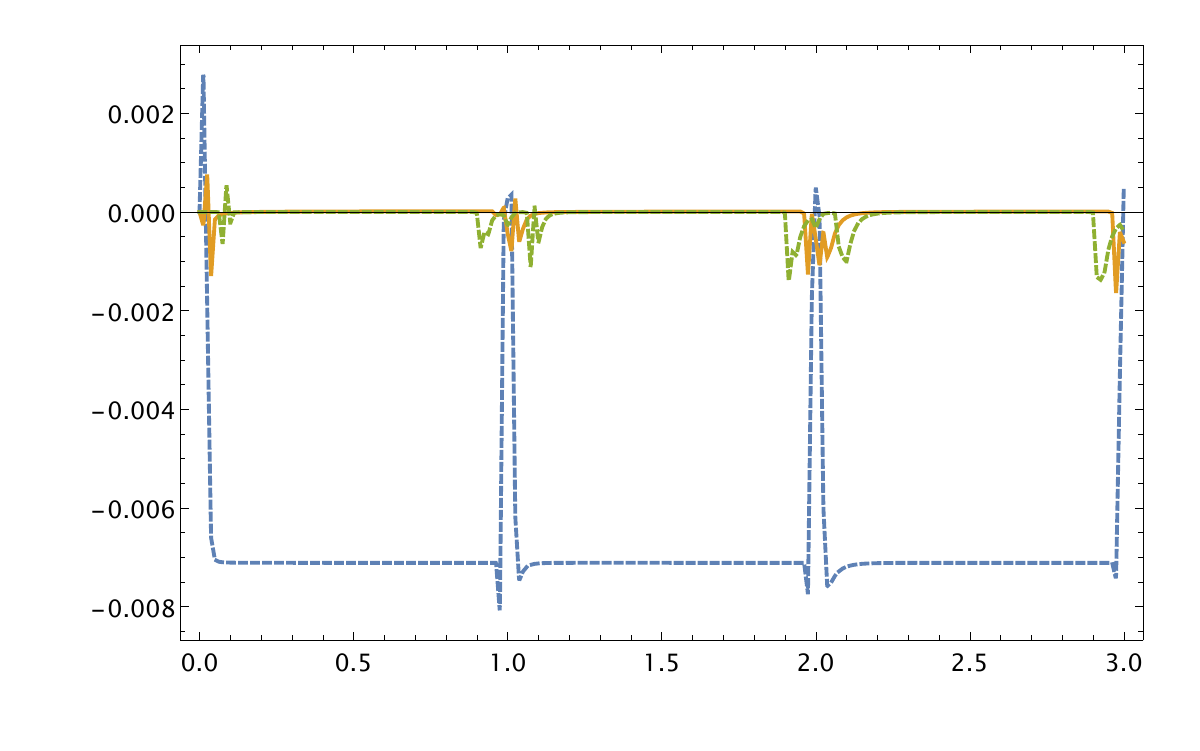}\put(-228,45){\rotatebox{-270}{\fontsize{13}{13}\selectfont $\frac{	\Delta S(t)-	\Delta S(0)}{S_{\text{th}}}$}}		\put(-110,0){{\fontsize{11}{11}\selectfont $t/\mathcal{L}$}}
\hspace{.8cm}
\includegraphics[scale=.39]{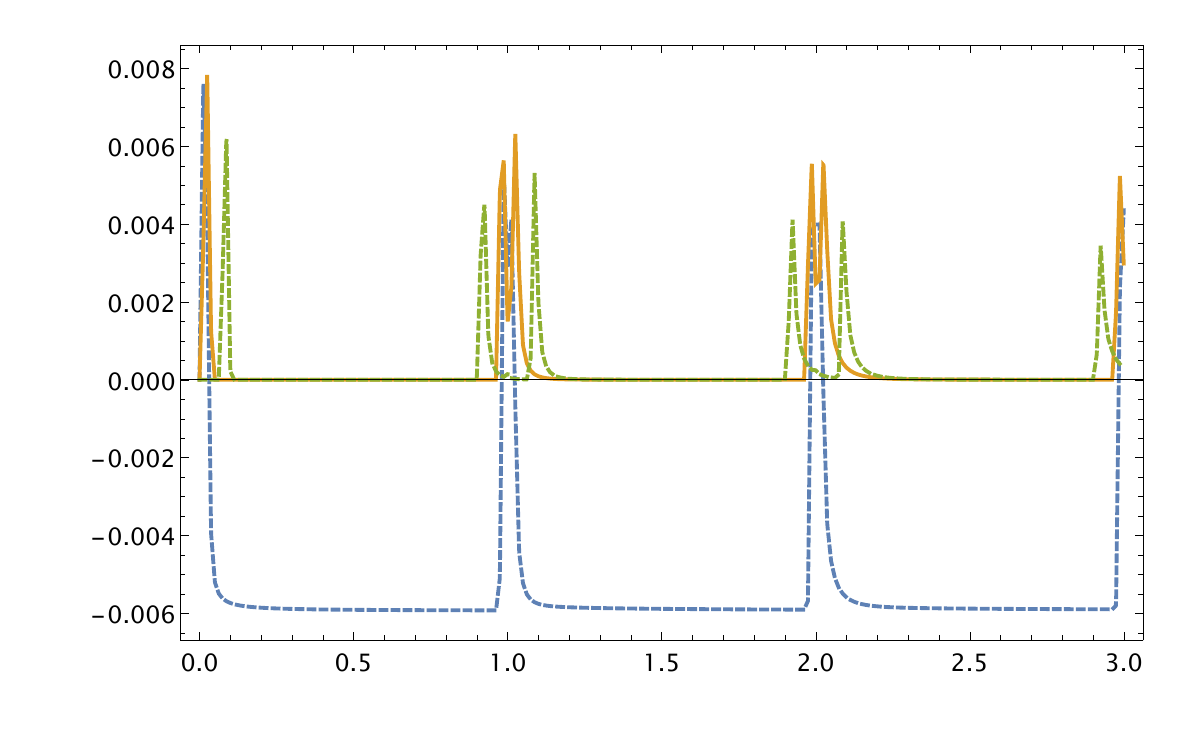}\put(-228,55){\rotatebox{-270}{\fontsize{13}{13}\selectfont $\frac{\mathcal{E}(t)-\mathcal{E}(0)}{S_{\text{th}}}  $}}		\put(-110,0){{\fontsize{11}{11}\selectfont $t/\mathcal{L}$}}
\vspace{.4cm}
\includegraphics[scale=.39]{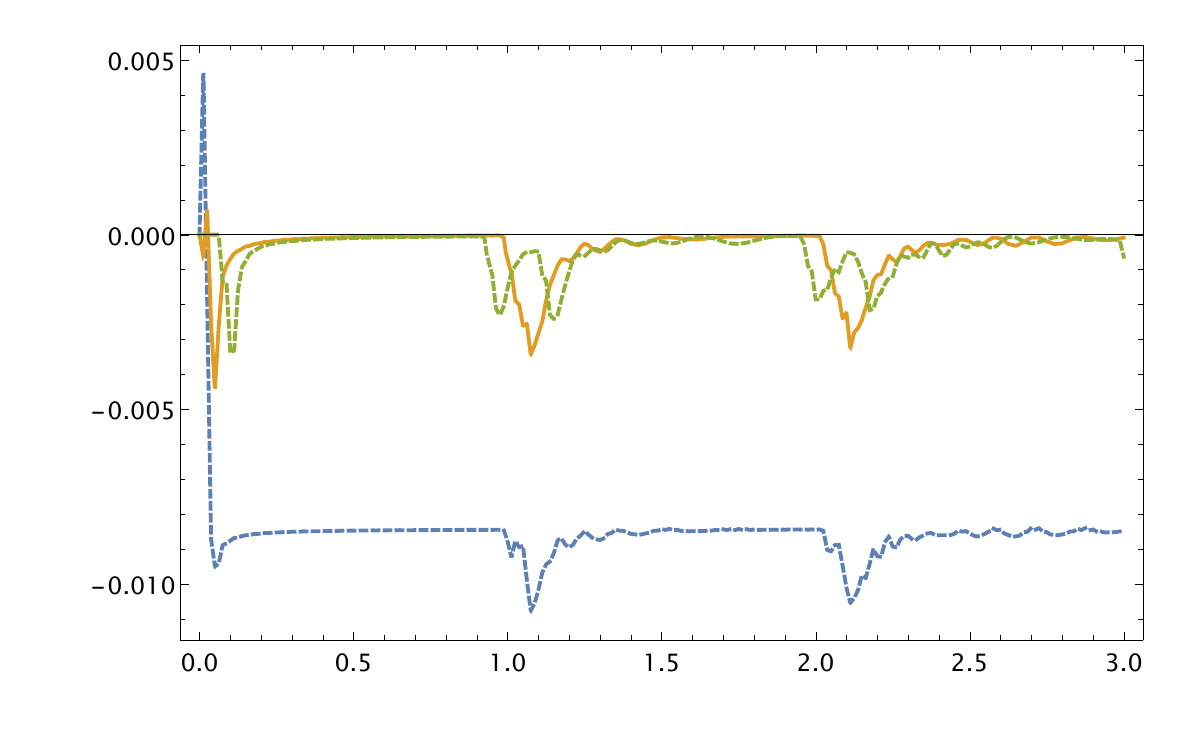}\put(-228,55){\rotatebox{-270}{\fontsize{13}{13}\selectfont $\frac{\Delta S(t)-\Delta S(0)}{S_{\text{th}}}$}}		\put(-110,0){{\fontsize{11}{11}\selectfont $t/\mathcal{L}$}}
\hspace{.8cm}
\includegraphics[scale=.39]{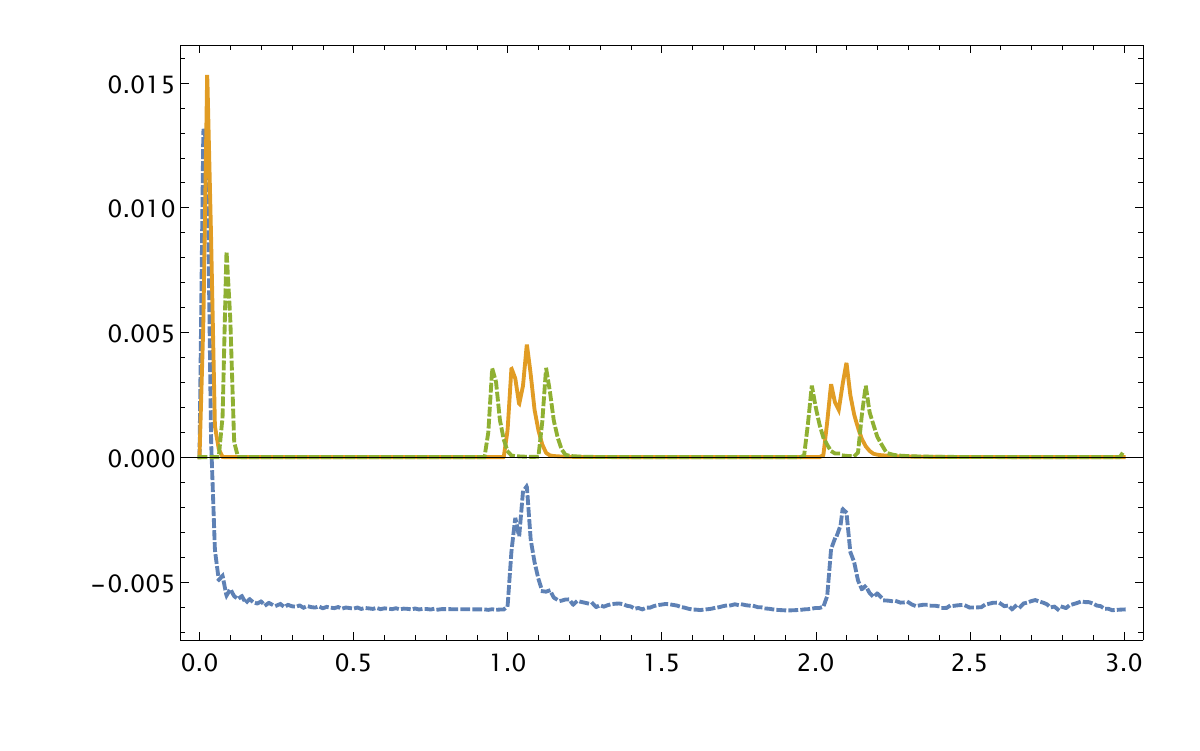}\put(-228,55){\rotatebox{-270}{\fontsize{13}{13}\selectfont $\frac{\mathcal{E}(t)-\mathcal{E}(0)}{S_{\text{th}}}  $}}		\put(-110,0){{\fontsize{11}{11}\selectfont $t/\mathcal{L}$}}
\caption{The up-left and up-right plots denote the long-time behavior of $\Delta S=S_{\text{OEE}}-S_{\text{EE}}$ and $\mathcal{E}(t)$, respectively, for massless free scalar theory. The two figures in bottom are for the massive case. These are normalized by the thermodynamic entropy $S_{\text{th}}$ when $N=1501$, $N_{A_{L(R)}}\hspace{-.1cm}=\hspace{-.1cm}N_{A_{L_{1}(R_{1})}}\hspace{-.1cm}+\hspace{-.05cm}N_{A_{L_{2}(R_{2})}}\hspace{-.1cm}=\hspace{-.1cm}20+31$, $\beta=10^{-2} \mathcal{L}$ and $m \mathcal{L}=10^{-3}$ for massless theory, $m \mathcal{L}=10^{2}$ for massive one. The lines denote $d=0$ (dashed blue), $d =10$ (orange), and  $d=100$ (dashed green).}\label{MassiveLN-dist1}
\end{figure}

An important feature of $\mathcal{E}$ is the sudden decreasing of entanglement. It is worth mentioning that one should not worry about growing behavior over a large range of times for $S_{\text{EE}}$ since according to (\ref{BO-1}) we know that this growth ultimately terminates. The similar behavior of $S_{\text{OEE}}$ over long times is the sign that this quantity also has a subadditivity characteristic. In figure \ref{MassiveLN-adj1}, we consider the time dependence of  $\Delta S= S_{\text{OEE}}- S_{\text{EE}}$ which is normalized with respect to thermodynamic entropy $S_{\text{th}}$ and subtracted from initial value for massless theory $m \mathcal{L}=10^{-3}$. Accordingly, $\Delta  S(t)-\Delta S(0)$ initially increases and then decreases followed by saturation. This is same as the behavior (qualitatively) for logarithmic negativity $\mathcal{E}$ (figure \ref{MasslessSSLN-adj1}).  In the right panel, the long time behavior is plotted and the finite size effect can be seen. 
In the figure \ref{QFT-SEEOL}, we explore the decompactification  limit (equivalently continuous limit) when theory lives on a line instead of a circle\footnote{We would like to thank Erik Tonni for discussion about QFT limit.}. For this purpose, we fixing the lattice spacing $\delta$, mass $m$, and inverse temperature $\beta$ while increasing the total number of lattice sites $N$. We observe a time delay to see the oscillatory behavior for the  larger $N$. In another words, finite size effects are pushed to later and later times by increasing the radius of circle on which theory lives. Moreover, we observe that the coefficient of logarithmic growth becomes smaller by increasing the value of $N$. 
This is consistent with the result of next section since for the fixed mass and fixed lattice spacing and increasing the system size we expect to only see the initial contribution in (\ref{analytic}) where the coefficient of logarithmic growth changes from $1$ to $1/2$. The existence of this long-lived logarithmic growth comes back to the non-local nature of the zero-mode and it is related to the periodic boundary condition on a circle.
This means that the logarithmic growth will be absent where the translational invariance is broken. This happens for example for Dirichlet instead of periodic boundary condition.
In the following, we extend the analysis to the case of two disjoint intervals on each side with a separation $d$. Moreover, the numerical results about changing the inverse temperature $\beta$ are provided in appendix \ref{Ap-2}. In figure \ref{Massless-dist1} we investigate the time evolution of the normalized entanglement entropy $S_{\text{EE}}$ (upper plots) and odd entanglement entropy $S_{\text{OEE}}$ (lower plots) which are subtracted from the initial values for short-times (left plots) and long-times (right plots) for different values of $d$. We take  $N=1501$, $m \mathcal{L}=10^{-3}$, $\beta=10^{-2} \mathcal{L}$, and  $N_{A_{L(R)}}=N_{A_{L_{1}(R_{1})}}+N_{A_{L_{2}(R_{2})}}=20+31$. For short-times (times smaller than $\mathcal{L}/2$ and in the absence of finite size effects), the growth of the $S_{\text{EE}}$, as well as $S_{\text{OEE}}$, is linear and lasts until approximately $t\sim l$; this is followed by saturation. For long-times (times larger than $\mathcal{L}- l$ where the finite size effects can be visible), the time dependence of $S_{\text{EE}}$ (up-right panel), as well as $S_{\text{OEE}}$ (bottom-right panel), is periodic with periodicity $\mathcal{L}$. Clearly, the pattern of evolution for $S_{\text{EE}}$, as well as $S_{\text{OEE}}$, is: linear growth for early times, $t\sim l$, a plateau of width approximately $\mathcal{L}-2l$ in a middle timed  and then a linear decrease up to order $t\sim l$. The oscillatory behavior can be observed for long times due to the finite size effect.
For the massive theory, the similar results are presented in figure \ref{Massive-dist1}. It is clear that for large masses, the oscillations are rather irregular due to the dephasing of the different kinds of quasi-particles with different group velocities. This dephasing also causes that the slope of linear growth at early times becomes significantly less than $2S_{\text{th}}/\mathcal{L}$. Another effect of changing mass can be seen in this figure where  both $\text{OEE}$ and ${\text{EE}}$ fluctuate around the same value albeit we have increased the distance $d$. Of course this can be also understood by the quasi-particle picture where the entanglement increases, in general, by increasing the distance $d$ while it decreases by increasing the IR regulator mass $m$. 
\vspace{1cm}
\begin{figure}[H]
\centering
\includegraphics[scale=.38]{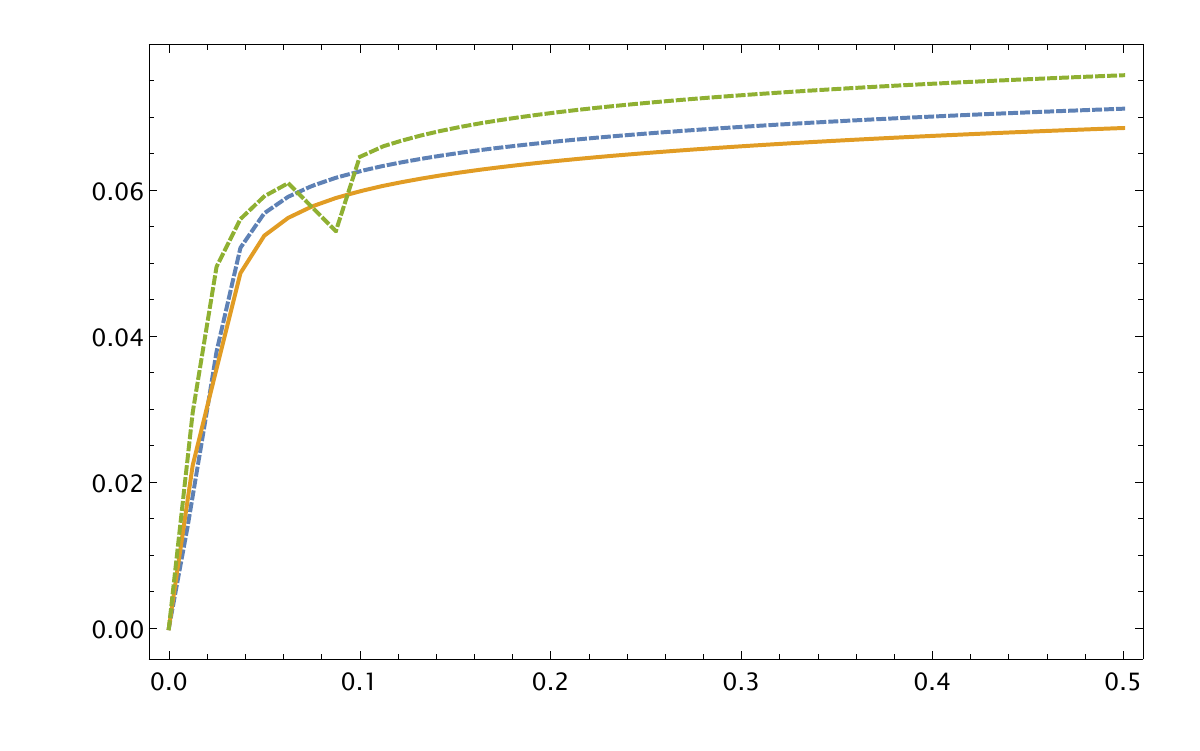}\put(-224,35){\rotatebox{-270}{\fontsize{13}{13}\selectfont $\frac{S_{\text{EE}}(t)-S_{\text{EE}}(0)}{S_{\text{th}}}$}}		\put(-110,-5){{\fontsize{11}{11}\selectfont $t/\mathcal{L}$}}
\hspace{.8cm}\includegraphics[scale=.38]{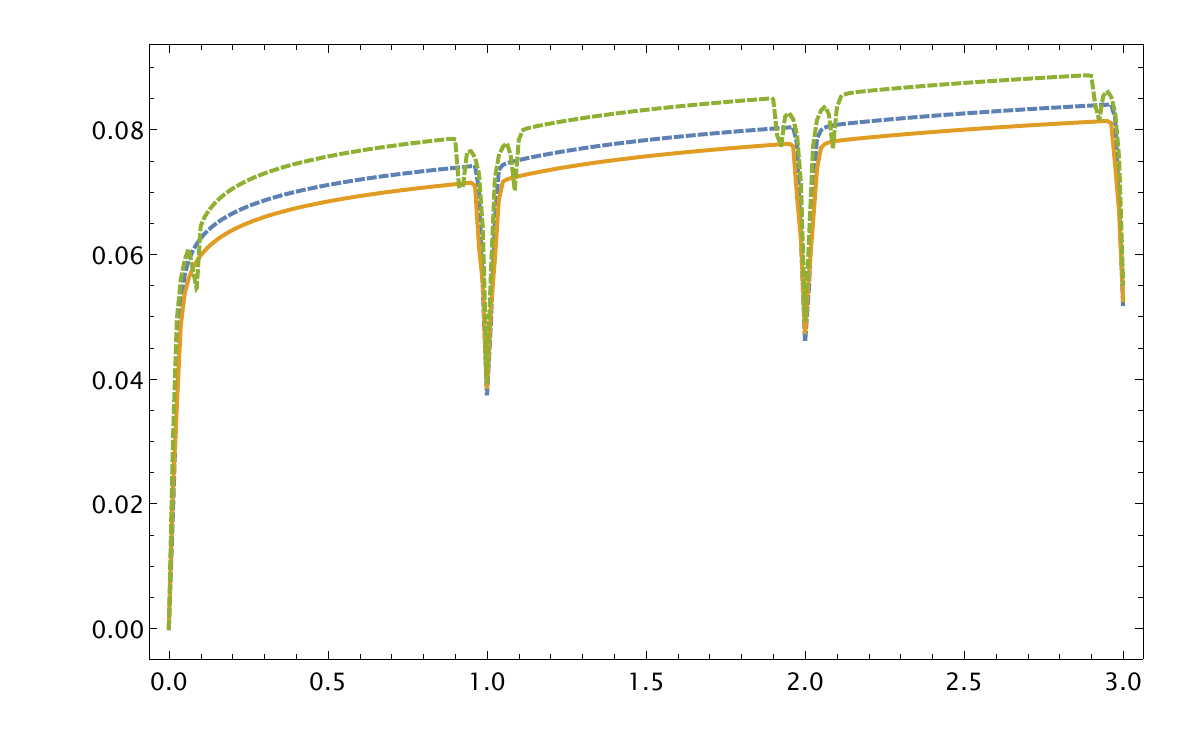}\put(-224,35){\rotatebox{-270}{\fontsize{13}{13}\selectfont $\frac{S_{\text{EE}}(t)-S_{\text{EE}}(0)}{S_{\text{th}}}$}}		\put(-110,-5){{\fontsize{11}{11}\selectfont $t/\mathcal{L}$}}
\vspace{.4cm}
\includegraphics[scale=.37]{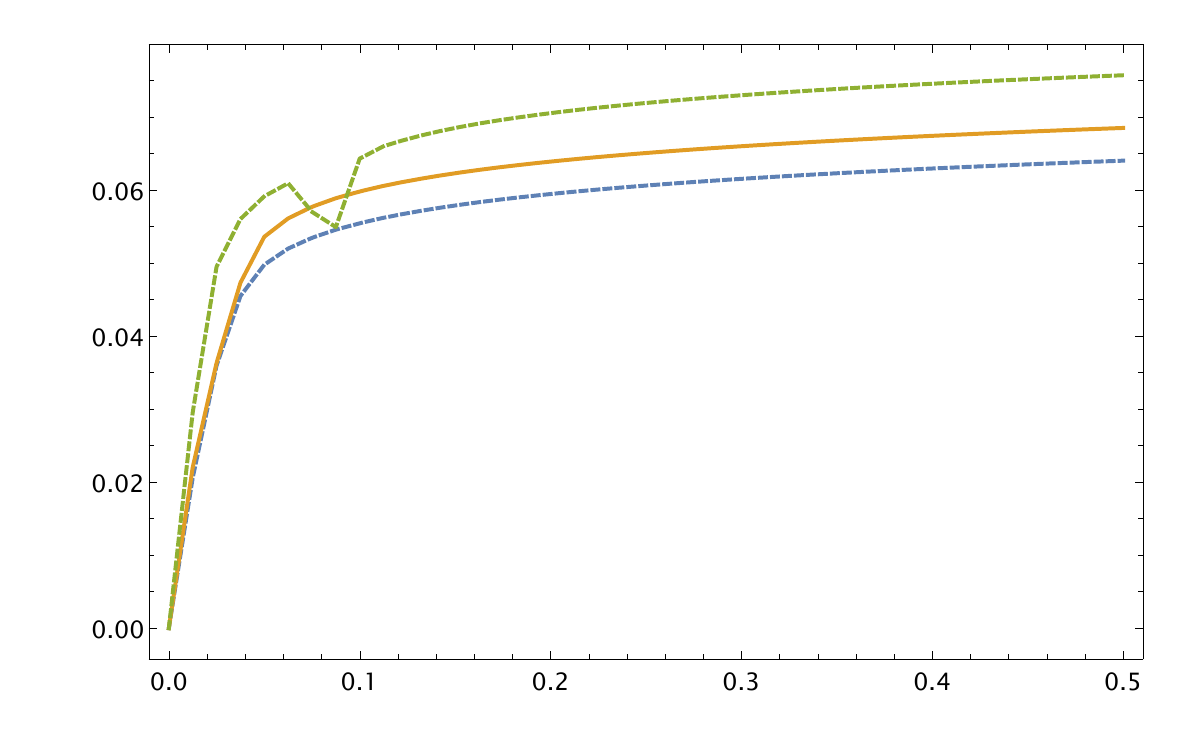}\put(-220,35){\rotatebox{-270}{\fontsize{13}{13}\selectfont $\frac{S_{\text{OEE}}(t)-S_{\text{OEE}}(0)}{S_{\text{th}}}$}}		\put(-110,-5){{\fontsize{11}{11}\selectfont $t/\mathcal{L}$}}
\hspace{.8cm}\includegraphics[scale=.38]{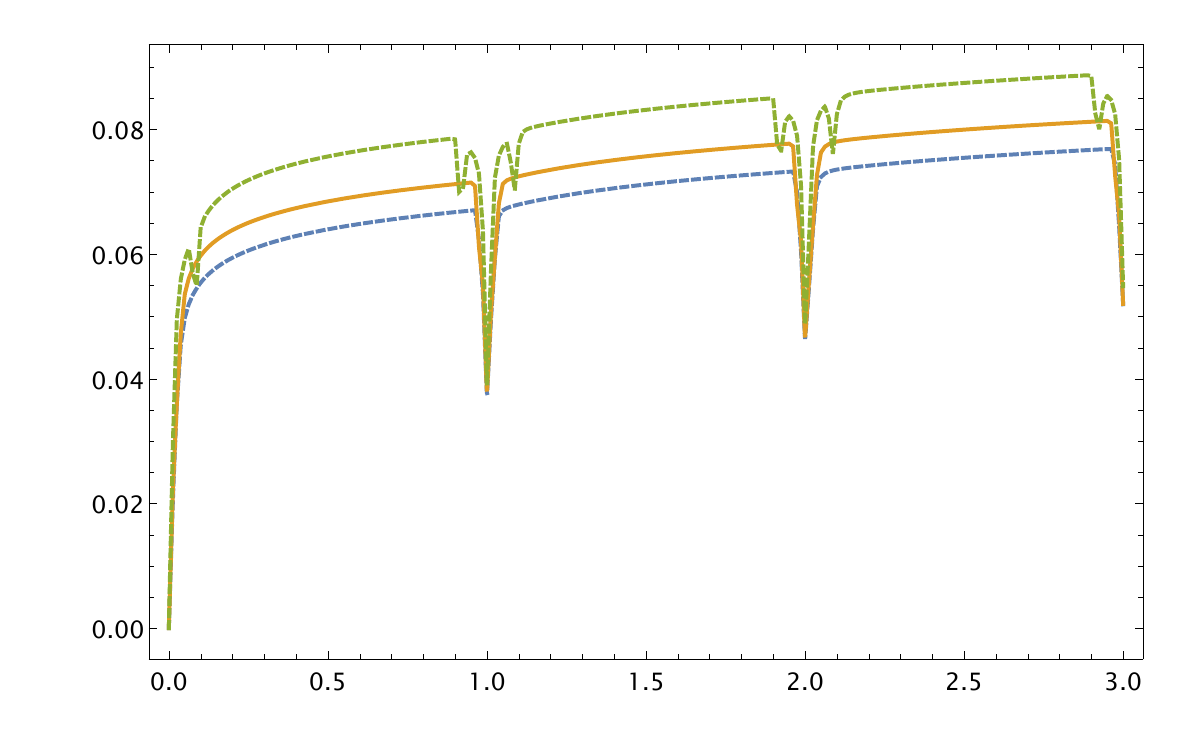}\put(-225,35){\rotatebox{-270}{\fontsize{13}{13}\selectfont $\frac{S_{\text{OEE}}(t)-S_{\text{OEE}}(0)}{S_{\text{th}}}$}}		\put(-110,-5){{\fontsize{11}{11}\selectfont $t/\mathcal{L}$}}
\vspace{.4cm}

\caption{Time dependence of $S_{\text{EE}}$ (upper panels), $S_{\text{OEE}}$ (bottom panels) is plotted for the free massless scalar theory when 
$N=1501$ , $N_{A_{L(R)}}\hspace{-.1cm}=\hspace{-.1cm}N_{A_{L_{1}(R_{1})}}\hspace{-.1cm}+\hspace{-.05cm}N_{A_{L_{2}(R_{2})}}\hspace{-.1cm}=\hspace{-.1cm}20+31$ and $m \mathcal{L}=10^{-3}$, $\beta=10^{-2} \mathcal{L}$. The curves correspond to the $d=0$(dashed blue), $d=10$(orange), and  $d=100$(dashed green). The upper-left plot denotes the linear regime for the growth of entanglement entropy. The linear growth is for times approximately equal to the size of an interval. The upper-right plot is for a longer period of time for which the finite size effects induce a periodic behavior with periodicity $\mathcal{L}$. The lower-left and right panels denote time behavior of odd entanglement entropy. The effective evolution of both $S_{\text{EE}}$ and $S_{\text{OEE}}$ are the same in different  times. Memory effect in $S_{\text{EE}}$ and $S_{\text{OEE}}$ is observed for intervals $d > l$. The dip extends over the range $d/2< t < d/2 +l$, and it is centered at $t= (d+l)/2$.} \label{Massless-dist1}
\end{figure}
Figure \ref{MassiveLN-dist1} illustrates 
the time evolution of the $\Delta S=S_{\text{OEE}}-S_{\text{EE}}$ (left panel) and the logarithmic negativity, $\mathcal{E}(t)$ (right panel) which are normalized with respect to thermodynamic entropy $S_{\text{th}}$ and
subtracted from their initial values. This figure together with figures \ref{MasslessSSLN-adj1}, \ref{MassiveLN-adj1}, \ref{QFT-SEEOL} and figures \ref{TempSS-adjac2} and \ref{TempLN-adjac3} in appendix \ref{Ap-2} contain an interesting result: \emph{At least for free scalar $QFTs$},  the $\Delta S$ and $\mathcal{E}$ are qualitatively the same independent of the mass $m$, lattice spacing $\delta$, separation distance $d$ and inverse temperature $\beta$\footnote{For the disjoint intervals, we also checked the entanglement dynamics in the decompactification limit (or equivalently in the continuum limit): Similar to the adjacent case, the effect of increasing $N$ appears in changing of oscillation period on the circle to larger times. For $S_{\text{EE}}$ and $S_{\text{OEE}}$, the coefficient of logarithmic growth also decreases by increasing the number of total sites $N$.}. These observations has an interesting consequence.
It is well-known that \cite{Coser:2014gsa, Toth:2009} the logarithmic negativity is just a measure of entanglement (quantum correlation) between the sub-systems $A_1$ and $A_2$ and the quasi-particles between $(A_1+A_2)$ and $B$ (figure \ref{SUB-1}) do not contribute to it. To be more precise, all moments of partial transpose quasi-particles between $(A_1+A_2)$ and $B$ matter, but they cancel when taking the replica limit\cite{Elben:2020hpu}\footnote{We would like to thank Pasquale Calabrese for illuminating this point and also discussing our results.}. In contrast, the entanglement entropy is just affected by quasi-particles between $(A_1+A_2)$ and $B$. The time dependency of odd entanglement entropy is \emph{similar} to the entanglement entropy but when we subtract it from entanglement entropy, interestingly, it behaves as logarithmic negativity. This implies that $S_{\text{OEE}}$ not only has a contribution from the quasi-particles between $A_1$ and $A_2$ but also has contribution from quasi-particles between $(A_1+A_2)$ and $B$ in contrast to logarithmic negativity\footnote{Of course this should be checked for a generic quantum state.}.\\

Another interesting witness for this interpretation comes from figures \ref{Massless-dist1} and \ref{Massive-dist1}. Accordingly,
$S_{\text{OEE}}$, in general, increases by increasing the distance $d$ but the situation is different for $S_{\text{EE}}$. By increasing $d$, the $S_{\text{EE}}$ firstly decreases but then increases. According to the above interpretation we can understand these different behaviors in the following way: For each pair of quasi-particles in the region outside the $(A_1+A_2)$, two events are possible: one quasi-particle travels to the region $A_1$ (or $A_2$) and another one remains in the outside region or one quasi-particle travels to the region $A_1$ and another one to the region $A_2$. The occurrence of the first event increases both the $S_{\text{OEE}}$ and the $S_{\text{EE}}$. But the occurrence of the second event increases the $S_{\text{OEE}}$ but it decreases the $S_{\text{EE}}$. Therefore, 
$S_{\text{OEE}}$ is almost increasing but the behavior of $S_{\text{EE}}$ crucially depends on the distance $d$ since the second event is more likely to occur for shorter distances $d$. One more witness comes from the observation of \emph{memory effect} \cite{Asplund:2015eha} in Figures \ref{Massless-dist1} and \ref{Massive-dist1}  when $d>l$. This effect can be understood by noting to the quasi-particles created at the midpoint between two intervals. Actually one quasi-particle entering the region $A_1$ and its partner entering the region $A_2$ around $t \sim d/2$. When the dip exists, it extends over the range $d/2< t < d/2 +l$ and it is centered at $t= (d+l)/2$.\\

\begin{figure}[H]
\centering\includegraphics[scale=.39]{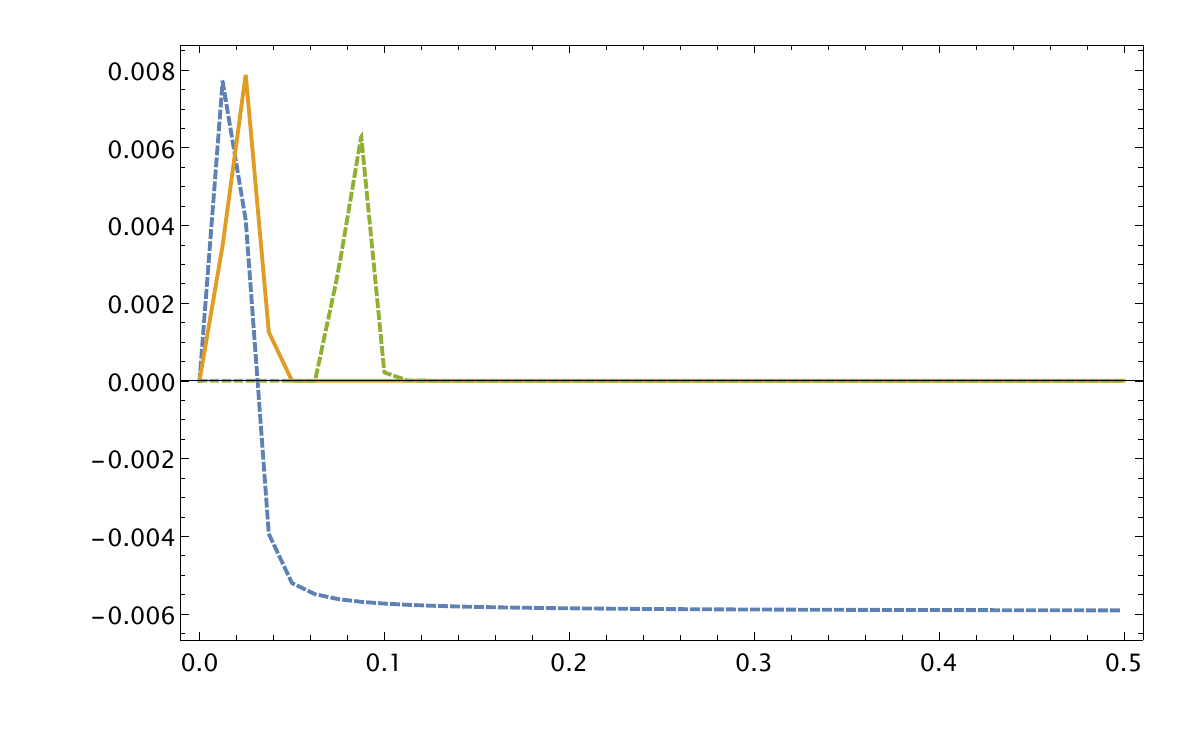}\put(-228,55){\rotatebox{-270}{\fontsize{13}{13}\selectfont $\frac{\mathcal{E}(t)-\mathcal{E}(0)}{S_{\text{th}}}$}}		\put(-110,0){{\fontsize{11}{11}\selectfont $t/\mathcal{L}$}}\hspace{.2cm}\includegraphics[scale=.39]{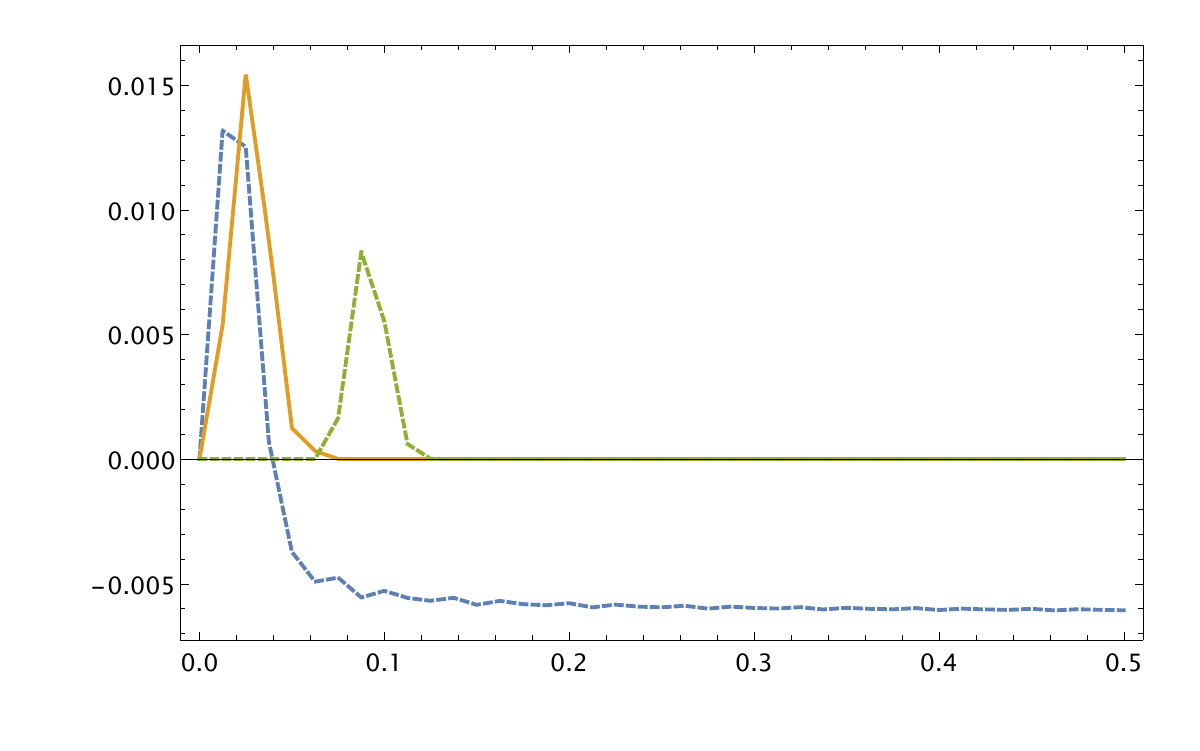}\put(-228,55){\rotatebox{-270}{\fontsize{13}{13}\selectfont $\frac{\mathcal{E}(t)-\mathcal{E}(0)}{S_{\text{th}}}$}}		\put(-110,0){{\fontsize{11}{11}\selectfont $t/\mathcal{L}$}}
\caption{Time dependence of logarithmic negativity normalized by the thermodynamic entropy  $S_{\text{th}}$, for the free scalar theory where $N=1501$, $N_{A_{L(R)}}\hspace{-.1cm}=\hspace{-.1cm}N_{A_{L_{1}(R_{1})}}\hspace{-.1cm}+\hspace{-.05cm}N_{A_{L_{2}(R_{2})}}\hspace{-.1cm}=\hspace{-.1cm}20+31$, and $\beta=10^{-2} \mathcal{L}$ with $d=0$(dashed blue), $d =10$(orange) and $d=100$(dashed green). Left panel corresponds to massless theory with $m \mathcal{L}=10^{-3}$ and the right one is for massive theory with $m \mathcal{L}=10^{2}$.}
\label{shorttime}
\end{figure}

\begin{figure}	\hspace{1.55cm}\includegraphics[scale=.48]{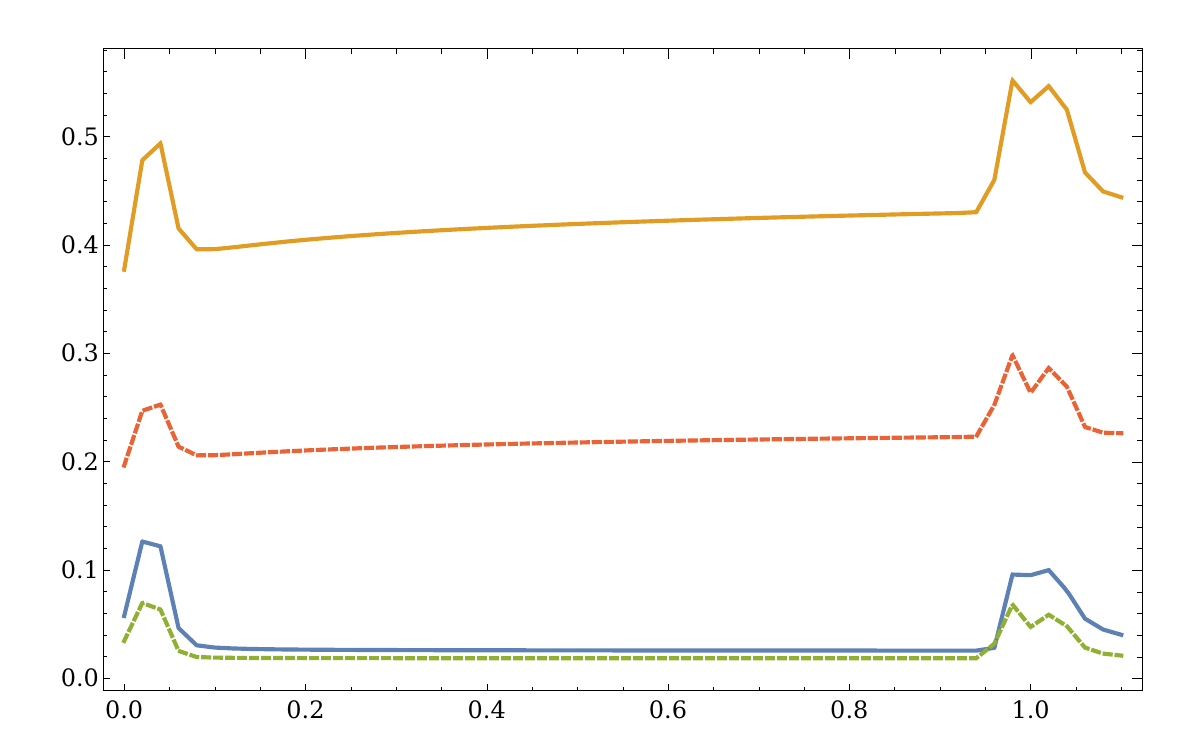}\put(-285,18){\rotatebox{-270}{\fontsize{13}{13}\selectfont $\mathcal{E}(t)/(l_1+l_2), I(t)/(l_1+l_2) $}}	\put(-145,-5){{\fontsize{11}{11}\selectfont $t/\mathcal{L}$}}\hspace{-.2cm}
\includegraphics[scale=.5]{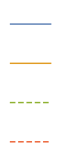}\put(-0,69){\rotatebox{1}{\fontsize{7}{7}\selectfont $  \mathcal{E},N=501,l_1=l_2=15$}}
\put(-0,49){\rotatebox{1}{\fontsize{7}{7}\selectfont $  I,N=501,l_1=l_2=15$}}
\put(-0,29){\rotatebox{1}{\fontsize{7}{7}\selectfont $  \mathcal{E},N=1001,l_1=l_2=30$}}
\put(-0,9){\rotatebox{1}{\fontsize{7}{7}\selectfont $  I,N=1001,l_1=l_2=30$}}
\vspace{.5cm}

\hspace{1.55cm}\includegraphics[scale=.47]{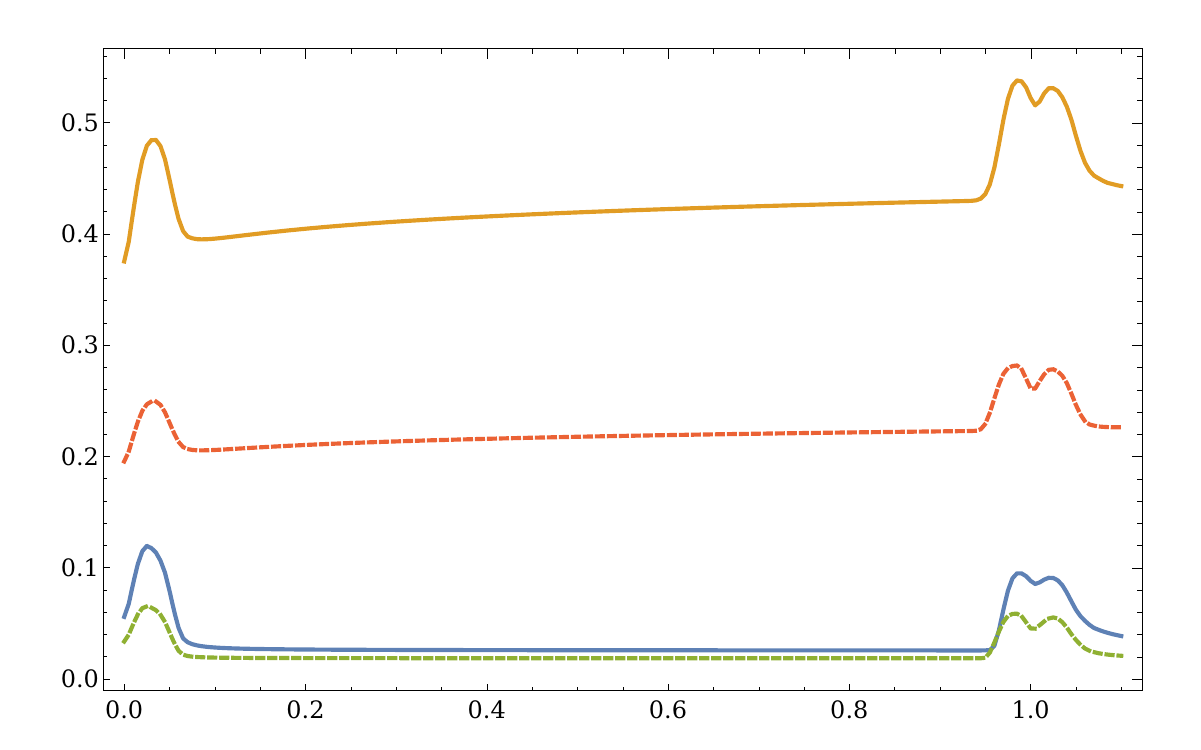}\put(-280,18){\rotatebox{-270}{\fontsize{13}{13}\selectfont $\mathcal{E}(t)/(l_1+l_2), I(t)/(l_1+l_2)$}}		\put(-135,-5){{\fontsize{11}{11}\selectfont $t/\mathcal{L}$}}\hspace{-.1cm}
\includegraphics[scale=.5]{croped1.png}\put(-0,69){\rotatebox{1}{\fontsize{7}{7}\selectfont $  \mathcal{E},N=501,l_1+l_2=10+20$}}
\put(-0,49){\rotatebox{1}{\fontsize{7}{7}\selectfont $  I,N=501,l_1+l_2=10+20$}}
\put(-0,29){\rotatebox{1}{\fontsize{7}{7}\selectfont $  \mathcal{E},N=1001,l_1+l_2=20+40$}}
\put(-0,9){\rotatebox{1}{\fontsize{7}{7}\selectfont $  I,N=1001,l_1+l_2=20+40$}}
\vspace{.5cm}
	
\hspace{1.9cm}	
\includegraphics[scale=.485]{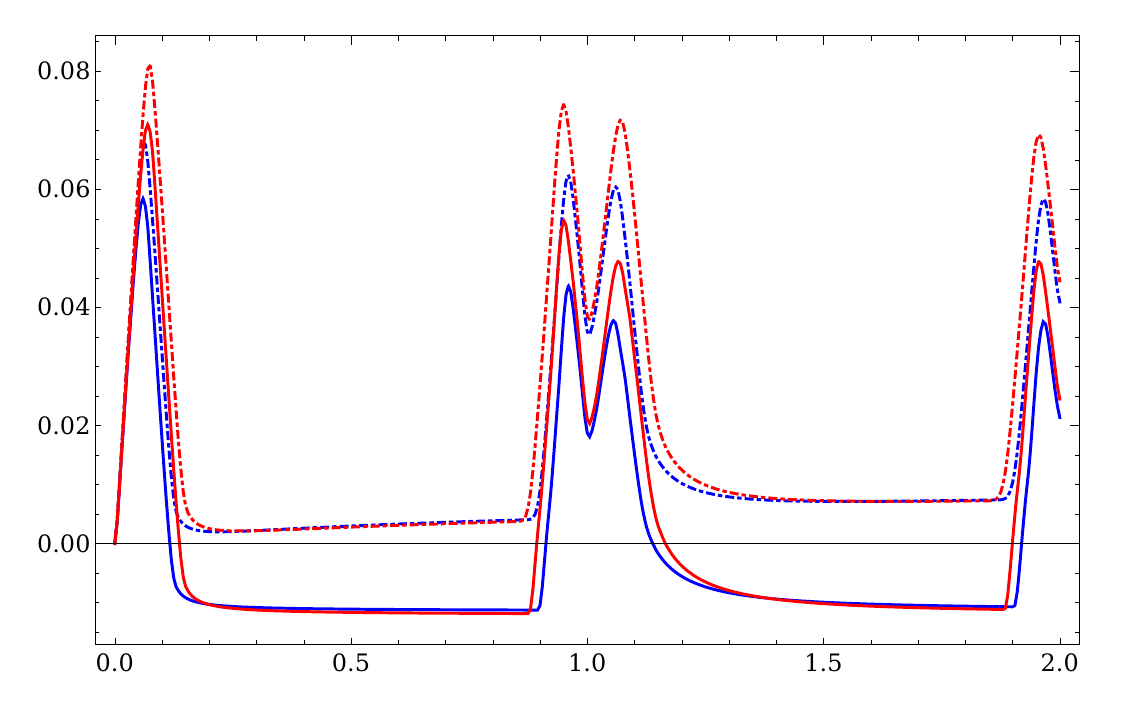}\put(-280,30){\rotatebox{-270}{\fontsize{13}{13}\selectfont $\frac{\mathcal{E}(t)-\mathcal{E}(0)}{S_{\text{th}}} , \frac{I^{1/2}(t)-I^{1/2}(0)}{2 S_{\text{th}}} $}}		\put(-140,-2){{\fontsize{11}{11}\selectfont $t/\mathcal{L}$}}\hspace{-1mm}\includegraphics[scale=.5]{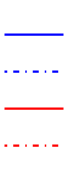}\put(-0,50){\rotatebox{1}{\fontsize{7}{7}\selectfont $ \mathcal{E},l_1=l_2=30$}}
\put(-0,36){\rotatebox{1}{\fontsize{7}{7}\selectfont $  I^{1/2},l_1=l_2=30$}}
\put(-0,22){\rotatebox{1}{\fontsize{7}{7}\selectfont $  \mathcal{E},l_1=l_2=35$}}
\put(-0,8){\rotatebox{1}{\fontsize{7}{7}\selectfont $  I^{1/2},l_1=l_2=35$}}
\caption{
Time dependency of $\mathcal{E}$ and mutual information $I$ are presented in top and middle panels. The top one is for symmetric subsystem configuration and the middle one is for non-symmetric case. The bottom panel demonstrates the time evolution of $\mathcal{E}$(solid curves) and  $1/2$-Rényi mutual information $I^{(1/2)}$ (dotted-dashed curves) for $N=501$,   $N_{A_{L(R)}}\hspace{-.1cm}=\hspace{-.1cm}N_{A_{L_{1}(R_{1})}}\hspace{-.1cm}+\hspace{-.05cm}N_{A_{L_{2}(R_{2})}}\hspace{-.1cm}=\hspace{-.1cm}30+30$ (blue) and $N_{A_{L(R)}}\hspace{-.1cm}=\hspace{-.1cm}N_{A_{L_{1}(R_{1})}}\hspace{-.1cm}+\hspace{-.05cm}N_{A_{L_{2}(R_{2})}}\hspace{-.1cm}=\hspace{-.1cm}35+35$ (red) which are subtracted from zero value and normalized by the thermal entropy $S_{\text{th}}$. }\label{MassiveLN-adj3}	
\end{figure}

It is worth to mention that  figure \ref{MassiveLN-dist1} contains an interesting information in very short times which is provided by figure \ref{shorttime}. It is clear that the LN is zero for times $t < d/2$ and after that, it begins to grow linearly. This is in agreement with \cite{Alba:2018hie,Coser:2014gsa}. A similar delay to start linear growth of $\Delta S=S_{\text{OEE}}-S_{\text{EE}}$ can also be seen in figure \ref{MassiveLN-dist1}.
For further investigation, we would like to study the time evolution of logarithmic negativity $\mathcal{E}$, mutual information $I$, $1/2$-Rényi mutual information $I^{(1/2)}$ and explore their similarities and differences. The numerical results for two adjacent intervals are depicted in figure \ref{MassiveLN-adj3}. We take  $\beta = 10^{-2} \mathcal{L}$ and  $m=10^{-3}/\mathcal{L}$. The top  panel denotes the time evolution of $\mathcal{E}$ and $I$  for long times and symmetric lengths for two subsystems. The middle panel denotes the same quantities but with non-symmetric lengths for subsystems. In the symmetric case, at early times, $\mathcal{E}$ and $I$ have a very similar behavior: initial linear growth followed by an almost linear decreasing up to time $t \sim l_1(\equiv l_{A_1})$. In this case, on time scales of the order of the system's size a difference between these quantities appears. In these time scales, the mutual information reaches a plateau while the logarithmic negativity decreases monotonically until the appearance of finite size effects.
This observation implies that the decreasing behavior is a peculiarity of the entanglement and 
is not reflected by the correlation measures such as mutual information. Moreover, in the decompactification limit (increasing the system and subsystem sizes properly) the rate of decreasing of logarithmic negativity becomes more sharply which it might be the sign of \emph{sudden death of entanglement} before the trivial (finite size effect) revival\footnote{The same phenomenon is explored in different quench setup \cite{Coser:2014gsa} and for other entanglement measures \cite{Yu:2009}. Of course it might be the lattice effect and will absent in true \emph{continuous} QFT \cite{Ferraro:2008zz}.}. Overall, the non-symmetric case has the same characteristics but with a difference which is the appearance of a (narrow) plateau after the first linear growth, for both the logarithmic negativity and mutual information. All these are in agreement with the results of \cite{Coser:2014gsa} which is in a different setup.  
The bottom plot denotes the time evolution of $\mathcal{E}$(solid curves) and $1/2$-Rényi mutual information $I^{(1/2)}$ (dotted-dashed curves). The initial growth which is linear and then decreasing followed by saturation is approximately the same between them which is in agreement with \cite{Alba:2018hie}. This similar behavior implies an interesting unification of these two seemingly different information-theoretic quantities in integrable models which it might break down in chaotic theories.
\\

\begin{figure}[H]	
\includegraphics[scale=.39]{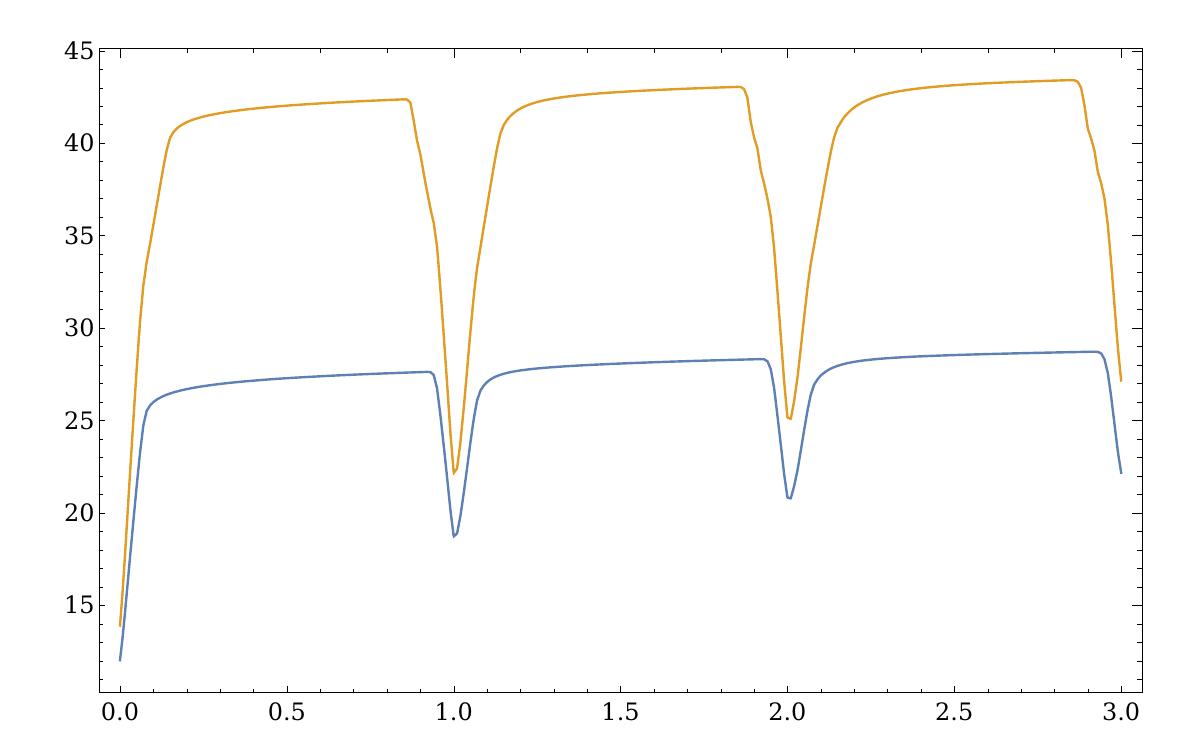}\put(-230,35){\rotatebox{-270}{\fontsize{13}{13}\selectfont $S_{\text{OEE}}(t),S_{\text{EE} }(t)$}}		\put(-110,-10){{\fontsize{11}{11}\selectfont $t/\mathcal{L}$}}
\hspace{.8cm}\includegraphics[scale=.39]{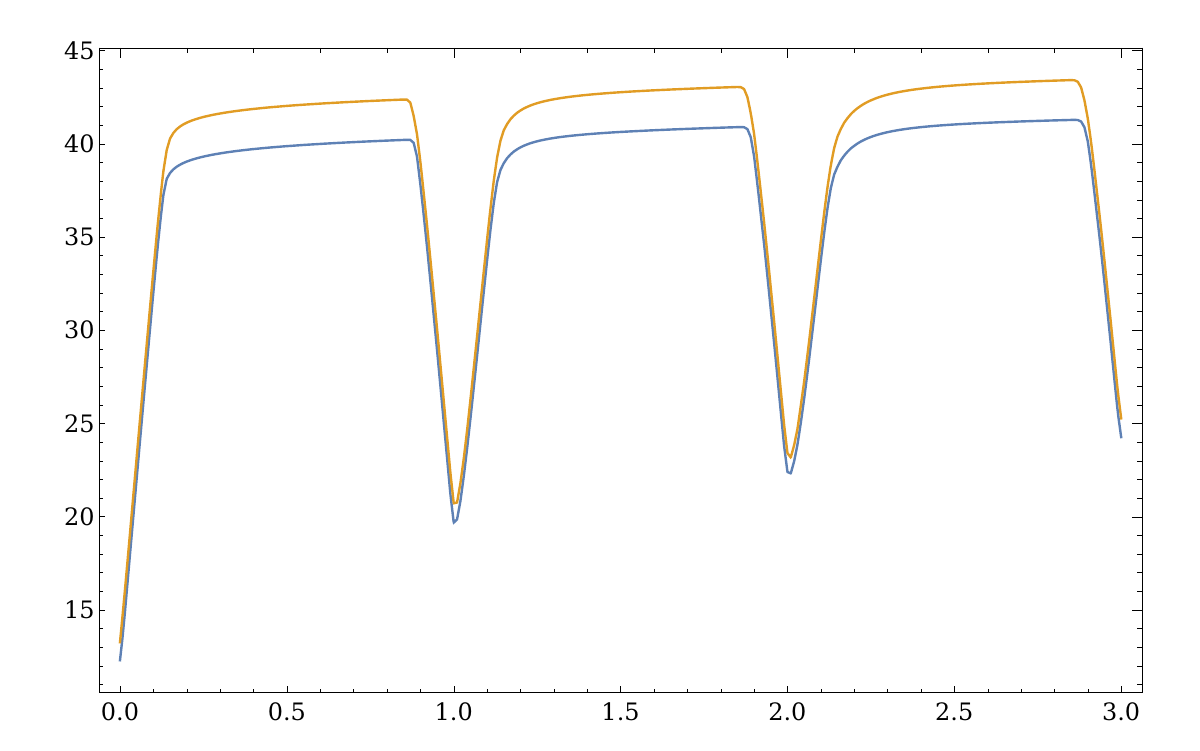}\put(-235,38){\rotatebox{-270}{\fontsize{13}{13}\selectfont$S_{\text{OEE} }(t) , S_{\text{EE} }(t)$}}		\put(-110,-10){{\fontsize{11}{11}\selectfont $t/\mathcal{L}$}}
\caption{The  $S_{\text{EE}}$ (blue) and $S_{\text{OEE}}$ (orange) are presented for $N=501$, $m\mathcal{L}=10^{-3}/ $, $d=0$, $\beta=10^{-2} \mathcal{L}$. In the left plot, $N_{A_{L(R)}}\hspace{-.2cm}=\hspace{-.1cm}N_{A_{L_{1}(R_{1})}}\hspace{-.1cm}+\hspace{-.05cm}N_{A_{L_{2}(R_{2})}}\hspace{-.2cm}=\hspace{-.1cm}35+35$ and in the right plot $N_{A_{L(R)}}=N_{A_{L_{1}(R_{1})}}\hspace{-.1cm}+\hspace{-.05cm}N_{A_{L_{2}(R_{2})}}\hspace{-.1cm}=\hspace{-.1cm}5+65$.} 
\label{IneOEE-3}
\end{figure}

Before closing this section, it is worth studying several proposed inequality for odd entanglement entropy $S_{\text{OEE}}$\cite{Mollabashi:2020ifv}. We can check  numerically some of these inequalities:\footnote{We would like to thank Kotaro Tamaoka for pointing out the last inequality to us.}
\begin{itemize}
\item $S_{\text{OEE}}(A:B)\geq0$ (positive semi-definiteness) \vspace{1.5mm}
\item $S_{\text{OEE}}(A:B_1B_2)\geq S_{\text{OEE}}(A:B_1)$ (monotonicity) \vspace{1.5mm}
\item $S_{\text{OEE}}(A_1:A_2)\geq \text{max}[S_{\text{EE}}(A_1),S_{\text{EE}}(A_2)]$
\end{itemize}
According to the results of this section, the positive semi-definiteness clearly is established. Monotonicity relation means that by enlarging one of two subsystems, the total amount of correlations between the two subsystems increases. According to figure \ref{MasslessSSLN-adj1}, it is clear that both of  $S_{\text{EE}}$ and $S_{{\text{OEE}}}$ satisfies the monotonicity relation. 
In figure \ref{IneOEE-3}, we can also simply see that the $S_{\text{OEE}}$ satisfies the last inequality. In this figure, $S_{\text{EE}}(t)$ refers to the greater one between  $S_{\text{EE}}(A_1)$ and $S_{\text{EE}}(A_2)$.
\section{Logarithmic growth at intermediate times}\label{Sec-5}
In the previous section, we have observed the existence of a logarithmic growth regime for odd entanglement entropy, instead of a sharp saturation, at intermediate times. In the case of entanglement entropy of a pure state, this behavior was observed previously\cite{Alba:2016lvc,Cotler:2016acd,Chapman:2018hou}. This behavior, as we will discuss more in the following, is due to the presence of a zero-mode, namely the momentum mode with $k=0$ in the massless limit. It is worth emphasizing that this gapless zero-mode does not lead to a ballistic propagation as in the quasi-particle picture, but it instead has a diffusive nature\cite{Cotler:2016acd}\footnote{The authors have considered both a boundary state quench and a global mass quench. For other relevant zero-mode effect discussions see refs\cite{Yazdi:2016cxn,DiGiulio:2021oal}.}. The authors of \cite{Chapman:2018hou} have studied the logarithmic contribution based on analyzing the entanglement entropy for a single degree of freedom on each side of TFD in the massless limit, i.e., $m\rightarrow 0$, and found that
\bea \label{S0}
&& \hspace{-1cm} S_{\text{EE}}(t)\hspace{.1cm} \sim  \left\{\begin{array}{rl} \frac{1}{2}\log (t/\mathcal{L}) & \hspace{.5cm}\delta\ll t <\mathcal{L},\\\log (t/\mathcal{L}) & \hspace{.5cm} \mathcal{L} < t\ll m^{-1},\\\log |\sin (mt)| & \hspace{.5cm} \mathcal{L} < t. \end{array} \right.
\eea
In order to proceed for odd entanglement entropy, we will follow \cite{Chapman:2018hou} and determine the consequences of the existence of the zero-mode analytically. We will take a single site in each entangling region, and derive the full asymptotic behavior of odd entanglement entropy in the limit $\beta,m^{-1} \ll \mathcal{L}$, which will turn out to be the same as (\ref{S0}). When we extend the analysis for logarithmic negativity, no logarithmic correction will be observed.
 It is worth noting that in order to make the study of this logarithmic behavior feasible, we will focus on a subsystem consisting of a single lattice site on each side of the TFD state. Accordingly, we will intentionally suppress the
linear regime by choosing a subsystem size that vanishes in the continuum limit. Then, we will
extend the observed asymptotic behavior to the case of larger subsystems due to the fact that the zero-mode is completely non-local and therefore affects local subsystems in a similar way, regardless of their size. We should also emphasize that this expression will fail to describe the entanglement accurately in the regime $t < \delta$ in which the precise form of oscillations of various modes will determine the time evolution.\\

For highly entangled states, the von Neumann entropy $S_{\text{EE}}(\rho_{A_{12}}(t))$ can be approximated as\cite{Bianchi:2017kgb,Hackl:2017ndi}\footnote{The error scales as $\exp(-2S_{2})$. Therefore, it decreases exponentially for highly entangled states\cite{Chapman:2018hou}.} 
\begin{equation}\label{EE-Z}
S_{\text{EE}}(\rho_{A_{12}}(t))\sim S_{2}(\rho_{A_{12}}(t))+2 N_{A_{12}}(1-2 \log 2),
\end{equation}
where $S_{2}(\rho_{A_{12}}(t))$ is the second Rényi entropy,
\begin{equation}\label{RE-Z}
	S_{2}(\rho_{A_{12}}(t))=\frac{1}{2}\log(\det(G_{A_{12}}(t))). 
\end{equation}
Using (\ref{Stodd}), we can find a similar upper bound formula for the odd entanglement entropy. Actually, instead of $G_{A_{12}}(t)$ we use $\tilde{G}_{A_{12}}(t)$ and approximate the $S_{\text{OEE}}$ for a single-mode on each side with the modified second Rényi entropy, $\tilde{S_{2}}$.
As we discussed in section \ref{Sec-3},
in order to find $\tilde{G}^{ab}_{12}$, one first need to find the covariance matrix, $G_{12}^{ab}$, and then take a partial transpose with respect to the momentum degrees of freedom in the $A_2$ subregion. This can be accomplished by acting with the time-reversal operator $\mathcal{R}_{A_{2}}$,
\begin{equation}
	\tilde{G}^{ab}_{12}=\mathcal{R}_{A_{2}}.G_{12}^{ab}.\mathcal{R}_{A_{2}},
\end{equation}
where $\mathcal{R}_{A_{2}}$ is a square matrix of length $2N_A \times 2N_A$,
\begin{equation}
	\mathcal{R}_{A_{2}}=\text{diag}\{1,1,\cdots,1,-1,\cdots,-1\},
\end{equation}
and the number of $-1$ elements is equal to the length of subregion $A_{2}$. Here, $N_{A_1}\equiv N_{A_{R_1}}=1$ and $N_{A_2}\equiv N_{A_{L_1}}=1$, so we have
\begin{equation}
\det(\tilde{G}^{ab}_{12})=\det(\mathcal{R}_{A_{2}}.G_{12}^{ab}.\mathcal{R}_{A_{2}})=\det(G_{12}^{ab}).
\end{equation}
Note that with respect to this specific choice of subsystems, one obtains $\det(\tilde{G}^{ab}_{12})=\det(G_{12}^{ab})$. Therefore, we can use the second Rényi entropy itself for evaluating the zero-mode effect on the evolution of $S_{\text{OEE}}$. In the following, we explicitly compute it.\\

The covariance matrix, $G_{A}(t)$, associated to subregion with a single site at position $x$ on both sides of the TFD, with respect to the dimensionless basis  $\tilde{\xi}_{k}^{a}=(\tilde{q}_{k}^{L},\tilde{q}_{k}^{R},\tilde{p}_{k}^{L},\tilde{p}_{k}^{R})$\footnote{These variables are related to variables of momentum space dual to the discretized field according to $q_{a}=\frac{1}{\sqrt{N}}\Phi(x_a)$ and $p_{a}=\frac{\mathcal{L}}{\sqrt{N}}\Pi(x_a)$.}, is given as follows:
\begin{align}\label{EQ-}
&\tilde{G}^{ab}_k (t) =\langle \text{TFD}(t)\mid \tilde{\xi}^{a}_{k}\tilde{\xi}^{\dagger b}_{k}+\tilde{\xi}^{\dagger b}_{k}\tilde{\xi}^{a}_{k}\mid \text{TFD}(t)\rangle
\nonumber\\ \nonumber\\
&\hspace{-.1cm}= \left(
\begin{array}{cccc}
\frac{\cosh(2\alpha_{ k})}{\lambda_k}    &\frac{\cos(\omega_{ k}t)\sinh(2\alpha_{ k})}{\lambda_{k}} & 0&-\sin(\omega_{ k}t)\sinh(2\alpha_{ k}) \\
\frac{\cos(\omega_{ k}t)\sinh(2\alpha_{ k})}{\lambda_{k}} & \frac{\cosh(2\alpha_{ k})}{\lambda_k} &-\sin(\omega_{ k}t)\sinh(2\alpha_{ k})&  0\\
0 & -\sin(\omega_{ k}t)\sinh(2\alpha_{ k}) & 
\lambda_{k}\cosh (2\alpha_{ k})&-\lambda_{k}\cos(\omega_{ k}t)\sinh(2\alpha_{ k})\\
-\sin(\omega_{ k}t)\sinh(2\alpha_{ k})& 0 & 
-\lambda_{k}\cos(\omega_{ k}t)\sinh(2\alpha_{ k})& \lambda_{k}\cosh (2\alpha_{ k})\\
\end{array} \right),
\end{align}
where,
\bea
\omega_{ k} =\sqrt{ m^2+\frac{4}{\delta^2}\sin^2(\frac{\pi k}{N})},\hspace{.5cm} \alpha_{ k} =\frac{1}{2}\log \coth(\frac{\beta \omega_{ k}}{4}),\hspace{.5cm}  \lambda_{k}=\omega_{k}\mathcal{L}.
\eea
By using the inverse Fourier transformation $\xi_{x}^{a}=\frac{1}{\sqrt{N}}\sum _{k=1}^{N}e^{-\frac{2\pi i k x}{N}}\tilde{\xi}_{k}^{a}$, in the position basis $\xi_{x}^{a}=(q_{x}^{L},q_{x}^{R},p_{x}^{L},p_{x}^{R})$, we have
\begin{align}\label{EQ-Z1}
&G^{ab}_{x,y} =\frac{1}{N}\sum_{k} e^{-\frac{2\pi i k(x-y)}{N}}\tilde{G^{ab}_k} =\frac{1}{N}\sum_{k} e^{-\frac{ 2\pi i k(x-y)}{N}} 
\nonumber\\\nonumber\\
&\hspace{-.2cm}\times\left(
\begin{array}{cccc}
\frac{\cosh(2\alpha_{k})}{\lambda_k} &\frac{\cos(\omega_{\mu k}t)\sinh(2\alpha_{ k})}{\lambda_{k}} & 0&-\sin(\omega_{ k}t)\sinh(2\alpha_{ k}) \\
\frac{\cos(\omega_{ k}t)\sinh(2\alpha_{ k})}{\lambda_{k}} & \frac{\cosh(2\alpha_{k})}{\lambda_k} &-\sin(\omega_{ k}t)\sinh(2\alpha_{ k})&  0\\
0 & -\sin(\omega_{k}t)\sinh(2\alpha_{k}) & 
\lambda_{k}\cosh (2\alpha_{k})&-\lambda_{k}\cos(\omega_{ k}t)\sinh(2\alpha_{ k})\\
-\sin(\omega_{ k}t)\sinh(2\alpha_{k})& 0 & 
-\lambda_{k}\cos(\omega_{ k}t)\sinh(2\alpha_{ k})& \lambda_{k}\cosh (2\alpha_{ k})\\
\end{array} \right).
\end{align} 
By setting $x = y$, we can obtain the odd entanglement entropy of a subsystem consisting of a single degree of freedom on each left and right side of the TFD. By studying the asymptotic behavior of odd entanglement entropy, we can identify the contribution of the zero-mode in the massless limit, $m\rightarrow 0$; this exhibits itself as the logarithmic term. According to the above  explanations, we can act as follows: setting $x=y$ in the (\ref{EQ-Z1}) gives:
\begin{align}\label{EQ-Z2}
&G^{ab}_{x,x}(t) =\frac{1}{N}\sum_{k=0}^{N-1}
\nonumber\\\nonumber\\
&\hspace{-.2cm}\times\left(
\begin{array}{cccc}
\frac{\cosh(2\alpha_{ k})}{\lambda_k}    &\frac{\cos(\omega_{ k}t)\sinh(2\alpha_{k})}{\lambda_{k}} & 0&-\sin(\omega_{ k}t)\sinh(2\alpha_{ k}) \\
\frac{\cos(\omega_{ k}t)\sinh(2\alpha_{ k})}{\lambda_{k}} & \frac{\cosh(2\alpha_{k})}{\lambda_k} &-\sin(\omega_{ k}t)\sinh(2\alpha_{ k})&  0\\
0 & -\sin(\omega_{ k}t)\sinh(2\alpha_{k}) & 
\lambda_{k}\cosh (2\alpha_{k})&-\lambda_{k}\cos(\omega_{ k}t)\sinh(2\alpha_{ k})\\
-\sin(\omega_{ k}t)\sinh(2\alpha_{k})& 0 & 
-\lambda_{k}\cos(\omega_{ k}t)\sinh(2\alpha_{ k})& \lambda_{k}\cosh (2\alpha_{ k})\\
\end{array} \right).
\end{align} 
Considering (\ref{EE-Z}), in order to compute the odd entanglement entropy, one needs to evaluate the second Rényi entropy which from (\ref{RE-Z}) is related to the determinant in (\ref{EQ-Z2}), recall $\det(\tilde{G}^{ab}_{12})=\det(G_{12}^{ab})$. The details of this calculation is postponed to Appendix \ref{Ap-3} and we only report the final result here. 
In the range $t\gg \delta=\mathcal{L}/N$ and in the limit $N\rightarrow\infty$, $m\mathcal{L}\ll1$, $\beta/\mathcal{L}\ll1$ and $m\ll\delta^{-1}$,
the odd entanglement entropy becomes:
\begin{equation}\label{S-EE}
S_{\text{OEE}}(\rho_A(t))\sim 2(1-2 \log 2)+\log\Big(\frac{2e_{2}}{N\beta m}\Big)+\frac{1}{2}\log\Bigg[\frac{\sin^{2}(mt)+Q}{m ^{2}\mathcal{L}^{2}}\Bigg],
\end{equation}
where $Q$ is defined by
\begin{align}\label{EQ-}
Q=\frac{m^{2} \mathcal{L}^{2}}{ \pi^{2}}\sum_{k=1}^{\infty}\frac{1-\cos[(\frac{2\pi k}{\mathcal{L}} )t]}{k^{2}},
\end{align} 
and $e_2$ is given in equation (\ref{eq-2}). The logarithmic term in the (\ref{S-EE}) can be simplified in the three regimes as follows,
\begin{equation}
\frac{1}{2}\log\Bigg[\frac{\sin^{2}(mt)+Q}{m^{2}\mathcal{L}^{2}}\Bigg]\sim  \left\{\begin{array}{rl} \frac{1}{2}\log (t/\mathcal{L}) & \delta\ll t <\mathcal{L},\\\log (t/\mathcal{L}) & \mathcal{L} < t\ll m^{-1},\\\log (\frac{|\sin (mt)|}{m\mathcal{L}}) &\mathcal{L} \ll t. \end{array} \right.
\end{equation}
Hence, the asymptotic form of the odd entanglement entropy in these regimes is given by
\begin{equation}\label{analytic}
S_{\text{OEE}}(\rho_A(t))\sim 2(1-2 \log 2)+\log\Big(\frac{2e_{2}}{N\beta m}\Big)+\left\{\begin{array}{rl} \frac{1}{2}\log (t/\mathcal{L}) &\hspace{.5cm} \delta\ll t <\mathcal{L},\\\log (t/\mathcal{L}) &\hspace{.5cm} \mathcal{L} < t\ll m^{-1},\\\log (\frac{|\sin (mt)|}{m\mathcal{L}}) &\hspace{.5cm}\mathcal{L} \ll t, \end{array} \right.
\end{equation}
which matches with our numerical results (see below).
Of course, the second case is simply derived by the third case under the condition $t \ll m^{-1}$. In the large time asymptotics, the behavior is oscillatory with frequency $m$. This is expected due to the upper
bound for the growth of the entanglement which is provided by the thermal state.  Hence, only for times $t \ll m^{-1}$ we have the logarithmic behavior and for longer times we have the oscillatory behavior with frequency $m$.
It is worth noting that in order to determine the entanglement evolution in the regime $t < \delta$ we must account for the precise form of oscillations of various modes, hence, the above expression will fail in that regime.\\

\begin{figure}[H]
\centering	\includegraphics[scale=.39]{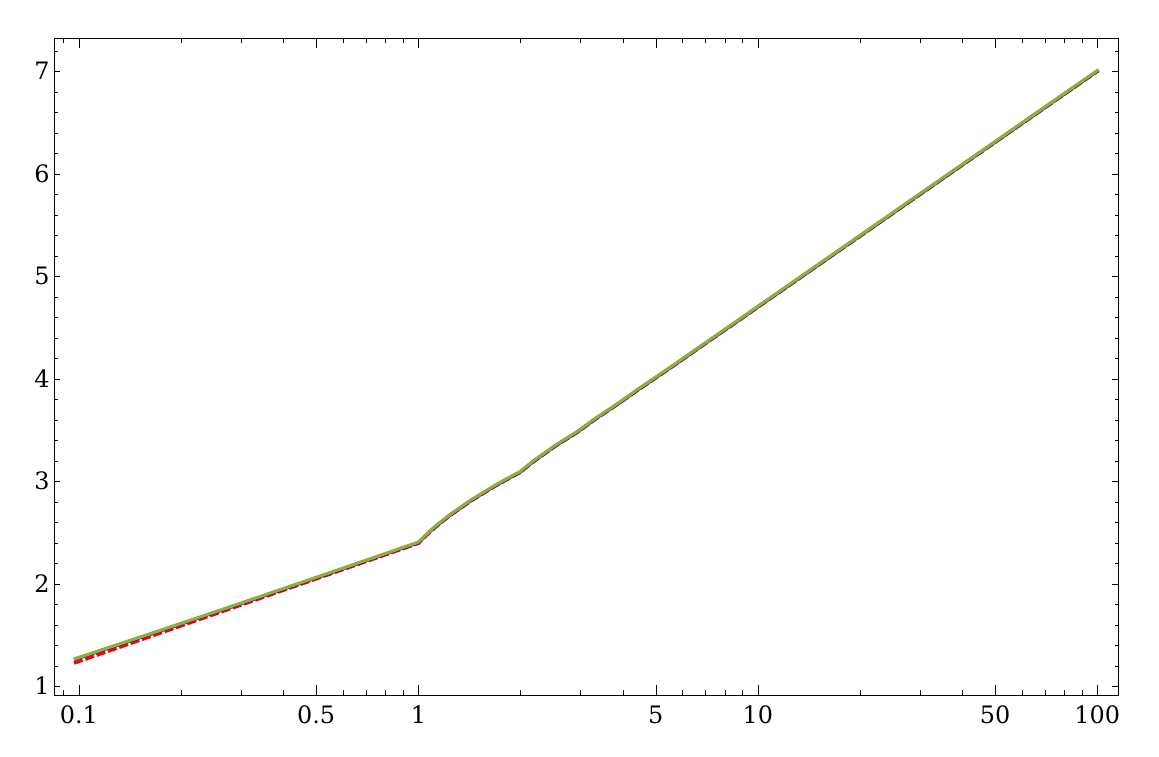}
\put(-225,0){\rotatebox{-270}{\fontsize{8}{8}\selectfont $S_{\text{EE}}(t)-S_{\text{EE}}(0) , S_{\text{OEE}}(t)-S_{\text{OEE}}(0)$}}		\put(-115,-5){{\fontsize{11}{11}\selectfont $t/\mathcal{L}$}}
\hspace{0.5cm}
\includegraphics[scale=.29]{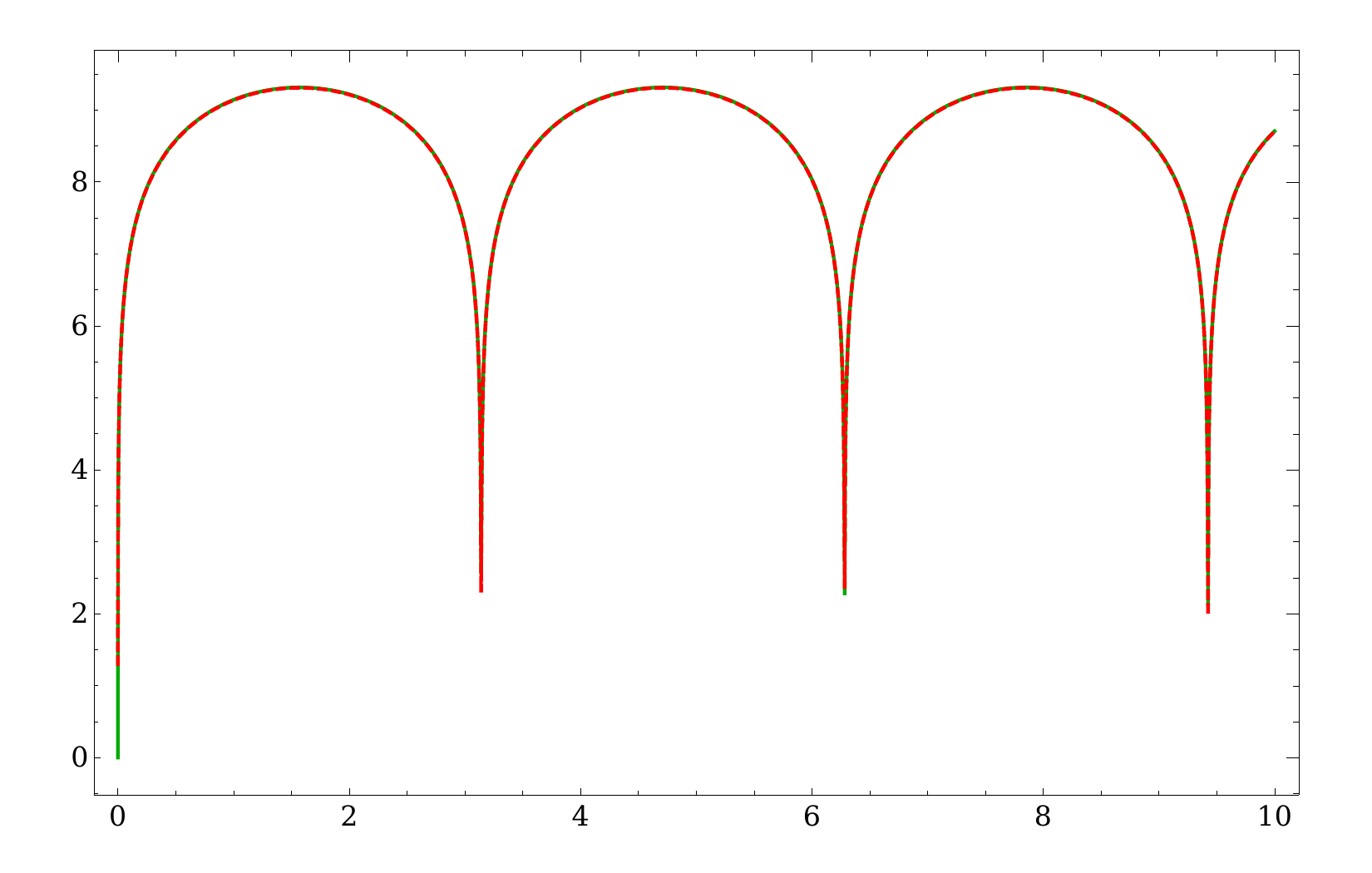}
\put(-120,-5){{\fontsize{11}{11}\selectfont $t\hspace{.5mm}m$}}
\vspace{.3cm}
\includegraphics[scale=.40]{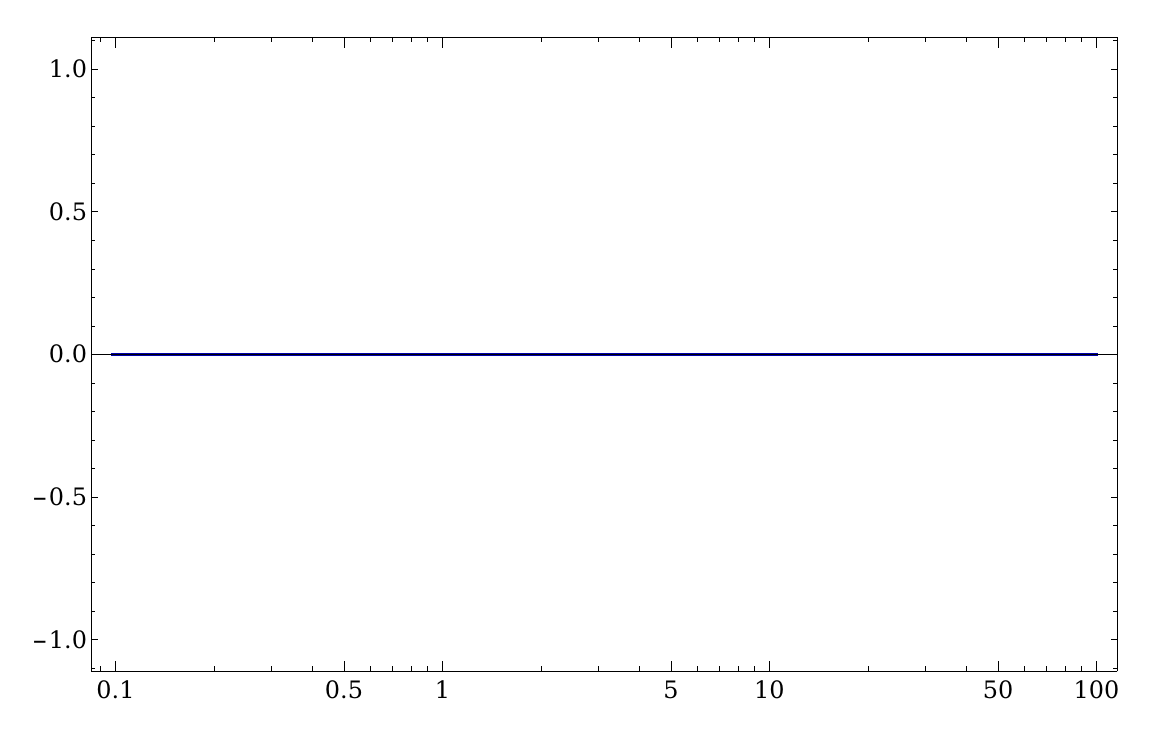}\put(-225,50){\rotatebox{-270}{\fontsize{10}{10}\selectfont $\mathcal{E}(t)-\mathcal{E}(0) $}}	\put(-120,-5){{\fontsize{11}{11}\selectfont $t/\mathcal{L}$}}
\hspace{.2cm}
\includegraphics[scale=.40]{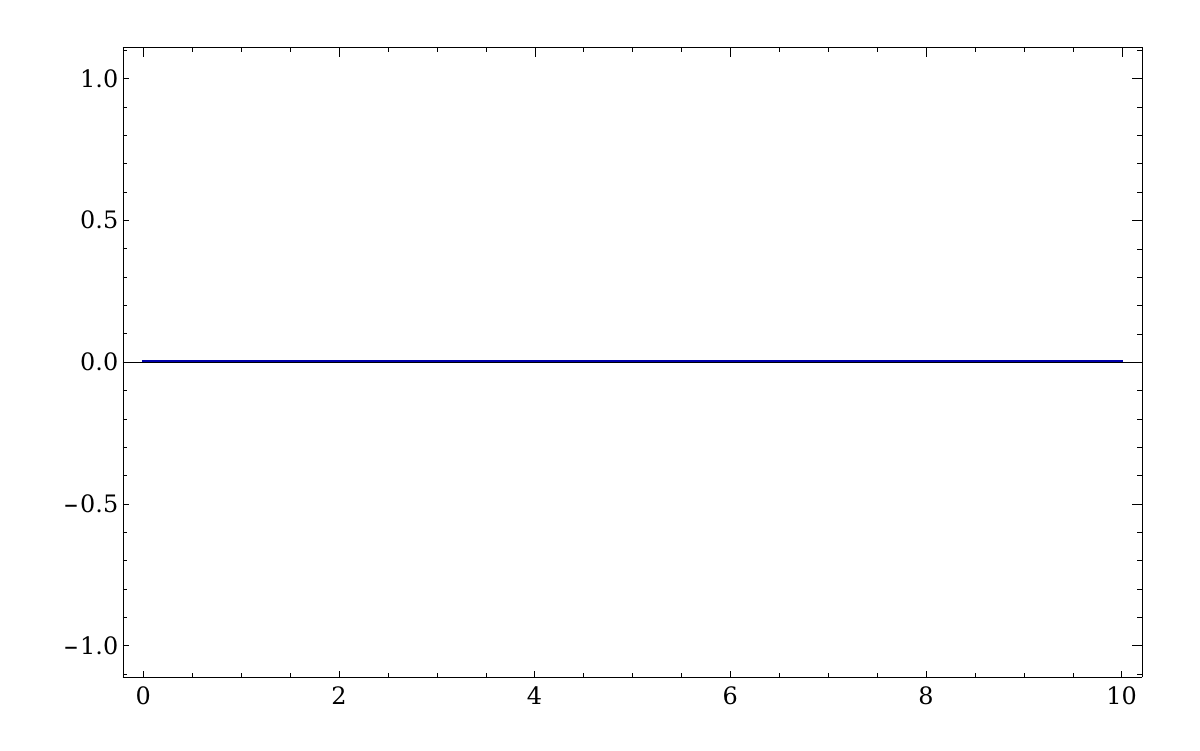}	\put(-120,-5){{\fontsize{11}{11}\selectfont $t\hspace{.5mm}m$}}
\caption{These plots correspond to the case of a subsystem consisting of a single site on each side for $m = 10^{-3}/\mathcal{L}$,  $\beta=10^{-2} \mathcal{L}$ and total sites $N=1501$. The analytical (dashed red) and numerical results are illustrated for the $S_{\text{EE}}$(solid blue) and $S_{\text{OEE}}$(solid green) in the top panels. The $\mathcal{E}$ is plotted in the bottom panels. In the top-right panel, the periodic behavior with respect to the parameter $t m$ is clear.}\label{Zeromode-1}	
\end{figure}

In the following, we present the numerically evaluated $S_{\text{EE}}$,  $S_{\text{OEE}}$ and $\mathcal{E}$ for a single site on each side of TFD in the limit $m\rightarrow 0$ and find a logarithmic contribution which is due to the presence of a zero-mode and compare it with  analytical result, (\ref{analytic}). In figure \ref{Zeromode-1}, the analytical (dashed red) and numerical results are illustrated for the $S_{\text{EE}}$(solid blue) and $S_{\text{OEE}}$ (solid green) in the top panels. The $\mathcal{E}$ is plotted in the bottom panel. The logarithmic growth is observed in variant time scales with different coefficients and it well-matches with the analytical result (\ref{analytic}). It is worth to mention that we also observe similar behavior for the case where $N_{A_{L(R)}}=N_{A_{L_{1}(R_{1})}}+N_{A_{L_{2}(R_{2})}}=1+1$, in the massless limit. The $S_{\text{EE}}$ matches exactly with $S_{\text{OEE}}$ whenever  $\mathcal{E}$  becomes zero. Since the entanglement of two sites does not have a well-defined continuum limit therefore in this figure we do not divide the result by the thermodynamic entropy $S_{\text{th}}$.
\section{Conclusion}\label{Sec-6}
In this manuscript, we have studied entanglement dynamics of TFD state for $1+1$-dimensional free scalar theory on a lattice by considering entanglement entropy, odd entanglement entropy and logarithmic negativity. The scalar field is discretized on two circles each with length $\mathcal{L}$ and the entangled subsystems are non-complementary regions each with two adjacent or two disjoint subregions. To compute desired entanglement measures we have used the covariance matrix formalism which is proper for Gaussian states. For evaluating entanglement entropy the eigenvalues of reduced covariance matrix are needed however for odd entanglement entropy and logarithmic negativity the eigenvalues of partial transposed reduced covariance matrix are required.
It is worth to mention that in the spirit of \cite{Hartman:2013qma} our setup is named an unusual quantum quench scenario\cite{Chapman:2018hou} in which two decoupled subsystems are entangled via their initial conditions.\\

The general perspective for time evaluation of entanglement entropy and odd entanglement entropy is 
an early linear growth then saturation for short times and then oscillatory behavior for longer times with periodicity of order the circle size  $\mathcal{L}$. Also, for  massless theory we have observed a logarithmic growth for the intermediate times due to presence of a zero-mode. However, this growth is limited by the upper bounds \ref{BO-1} and \ref{BO-2}. The linear growth and saturation can be understood using the quasi-particle picture of \cite{Calabrese:2006rx,Calabrese:2007rg}. According to the quasi-particle picture, the pre-quench initial state $|\psi_{0}\rangle$ acts as a source for independent entangled pairs each with an effective group velocity $v_{n}$ (\ref{vn}) that move ballistically in opposite directions through the system. The entanglement can spread when one quasi-particle is inside the subsystem and its partner is outside.
The initially linear growth arises due to the flux of quasi-particles going out of the interval while their partners are still inside. This linear growth lasts
until times $t\sim l$ ($l$ is the size of the entangling region) then it saturates at a value that is proportional to the thermal entropy of the system. Moreover, the time dependence of quantum correlations exhibits an oscillatory behavior periodically equal to the circle circumference, $\mathcal{L}$, due to finite size effects. This can also be understood by a quasi-particle picture where meeting quasi-particles on the opposite side of the circle leads to reducing
the correlations between the subsystem and its complement.  
We observed a regular oscillation in the massless limit since all the quasi-particles effectively move with the speed of light while for larger masses the different quasi-particles have different group velocities which lead to irregular oscillation\footnote{Actually, the quasi-particle picture provide a universal description for integrable models and it fails to capture dynamics of entanglement in more generic systems such as chaotic ones.}. Besides, we have found that correlations can be increased by increasing both the separation $d$ and the temperature. According to the quasi-particle picture, for the greater separation $d$, more quasi-particle pairs are produced in the region between the two subsystems which leads to  increasing the share of pairs that contribute to entanglement. Also enhancement by increasing the temperature indicates that odd entanglement entropy same as entanglement entropy is a measure of both classical and quantum correlations. To investigate the entanglement dynamics in the continuum limit, the lattice spacing $\delta$, mass $m$ and inverse temperature $\beta$ are fixed but the total number of lattice sites $N$ is increased. Effectively, in this limit the period of oscillation becomes larger and zero-mode contribution becomes less important at intermediate times. Furthermore, there are several unproved inequalities for odd entanglement entropy\cite{Mollabashi:2020ifv} where we have confirmed some of them numerically.
\\

It is worth emphasizing again that the logarithmic growth of odd entanglement entropy (and also entanglement entropy) can not be understood using the quasi-particle picture. To find a qualitative description and for tractability of computations, we 
focused on a subsystem consisting of a single site on each side of the TFD state. Accordingly, we intentionally suppress the linear regime by choosing a subsystem size that vanishes in the continuum limit. But we can restore the original system since zero-mode is non-local and can affect local subsystems, regardless of their size. In another word, zero-mode contribution is additive. The analytical results is derived by the approach of \cite{Chapman:2018hou}, which is based on a relation between the entanglement entropy and Rényi entropy of order 2 that is held for Gaussian states. By this approach, the Rényi entropy of order 2 for a configuration with a single site on each side can be evaluated by the determinant of a 4-by-4 covariance matrix and therefore its time dependency simply can be analyzed analytically. Interestingly, in our special decomposition of the subsystem, the determinants of the modified reduced density matrix $\tilde{G}_{A_{12}}(t)$ and reduced density matrix $G_{A_{12}}(t)$ became equal therefore evaluation of odd entanglement entropy become same as entanglement entropy\footnote{For entanglement entropy, see \cite{Chapman:2018hou}.}. The time  evolution pattern is composed of three time regimes: the first logarithmic regime: $S_{\text{OEE}}(\rho_A(t))\sim \frac{1}{2}\log (t/\mathcal{L})$ for $ t < \mathcal{L}$ , a second logarithmic region: $S_{\text{OEE}}(\rho_A(t))\sim\log (t/\mathcal{L})$ for $ \mathcal{L} < t\ll m^{-1}$ and finally an oscillating regime: $S_{\text{OEE}}(\rho_A(t))\sim \log (\sin mt)$ when $t$ is of the same order as $m^{-1}$.\\
  
Apart from the mentioned results, we have also studied logarithmic negativity for symmetric and non-symmetric subsystems $A_1$, $A_2$. In the symmetric case, at early times, it exists initial linear growth followed by an almost linear decreasing up to revival time. This is in agreement with previous studies \cite{Coser:2014gsa,Alba:2018hie} in scaling limit\footnote{The scaling limit means the long times and large subsystems with their ratio fixed.} and different setup. By comparison with mutual information in the same setup, we have discussed that this implies that the decreasing behavior is a peculiarity of the entanglement and is not reflected by the correlation measures. Moreover, in the decompactification limit the rate of decreasing of logarithmic negativity becomes more sharply which it might be the sign of \emph{sudden death of entanglement} before the trivial (finite size effect) revival. Overall, the non-symmetric case has the same characteristics but with a  difference which is the appearance of a (narrow) plateau after an early linear growth. We have also observed that whenever the size of subsystems is greater than separation distance $d$, logarithmic negativity is non-zero for early times. It is consistent with a dual holographic picture, in which  logarithmic negativity is dual to the geometric object named as entanglement wedge cross-section. In the holographic prescription, for two disjoint intervals, whenever the disjoint separation between two intervals $d$ is less than the size of subsystems, one has a non-zero entanglement wedge cross-section and when $d$ increases one can see a phase transition via vanishing the  entanglement wedge cross-section. On the other hand, according to the result of \cite{Kusuki:2019rbk}, the logarithmic negativity is proportional to $\Delta S= S_{\text{OEE}}-S_{\text{EE}}$. Therefore, whenever logarithmic negativity vanishes, the entanglement entropy and odd entanglement entropy became equal which also matches with our numerical results. Moreover, for the adjacent case, we observe that by decreasing the temperature the delay time for starting the growth of logarithmic negativity becomes larger.
For the future direction, it would be very interesting to obtain analytical forms for evaluation of odd entanglement entropy and logarithmic negativity. It is worth to mention that no analytic result is found for entanglement entropy of \emph{bosonic} QFTs on a lattice even in the ground state. Last but not least, it is exciting to explore the logarithmic growth at intermediate times for these measures in the holographic context .\\

\vspace{0cm}
{\large{\bf Acknowledgment}}\\
Special thanks to Mohsen Alishahiha, Pascuale Calabrese, Lucas Fabian Hackl, Yuya Kusuki, Ali Mollabashi, Behrad Taghavi, Kotaro Tamaoka and Erik Tonni for useful comments and fruitful discussions. Authors also thank Behrad Taghavi and Erik Tonni for carefully reading the draft and IPM-Grid computing group for providing computing and storage facilities. Part of MG work is supported by Iran Science Elites Federation (ISEF). 
\appendix 
\section{Temperature effects: various entanglement measures }\label{Ap-2}
In this appendix, we consider the effect of changing temperature on $S_{\text{EE}}$, $S_{\text{OEE}}$, $\mathcal{E}$, and $\Delta S = S_{\text{OEE}}-S_{\text{EE}}$. As we will see, the behavior of OEE is same as EE and both of them  approach thermal entropy by increasing temperature. This confirms that OEE is a measure of both classical and quantum correlations. However, the effect of decreasing temperature on LN behavior is only a delay in its initial growth which indicates that the LN is a measure of quantum correlation. The adjacent and disjoint subsystem configurations will be considered separately.
\subsection{Two adjacent intervals on each side}
\begin{figure}[H]	\centering\includegraphics[scale=.38]{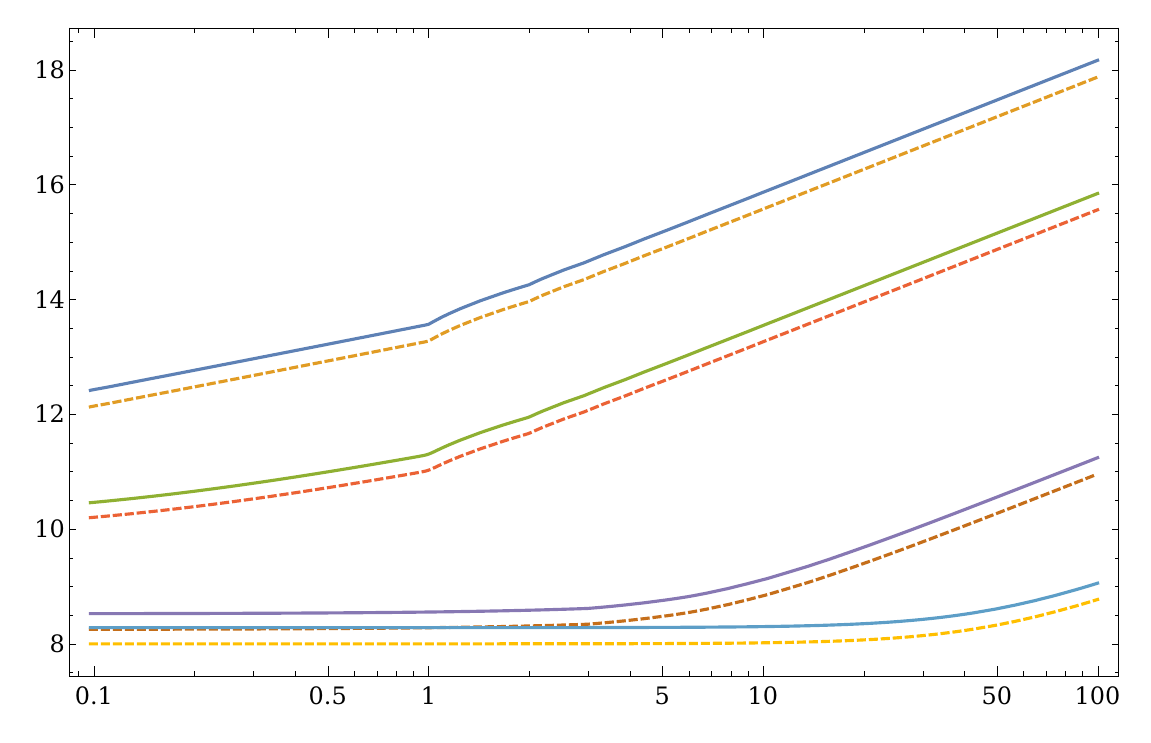}\put(-222,28){\rotatebox{-270}{\fontsize{12}{12}\selectfont $S_{\text{EE}}(t) , S_{\text{OEE}}(t)$}}		\put(-110,-10){{\fontsize{11}{11}\selectfont $t/\mathcal{L}$}}\hspace{.8cm} \includegraphics[scale=.38]{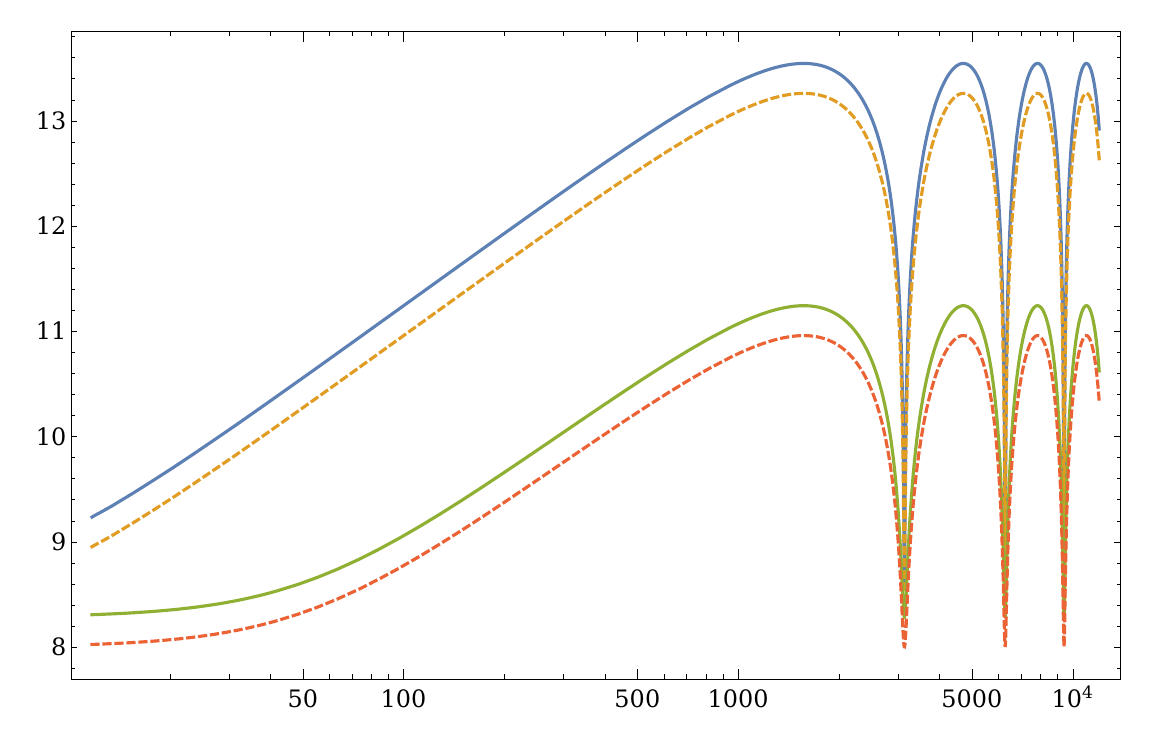}\put(-225,28){\rotatebox{-270}{\fontsize{12}{12}\selectfont $S_{\text{EE}}(t) , S_{\text{OEE}}(t)$}}		\put(-110,-10){{\fontsize{11}{11}\selectfont $t/\mathcal{L}$}}
\caption{The time dependence of $S_{\text{OEE}}$ and $S_{\text{EE}}$ are presented for $N=1501$, $N_{A_{L(R)}}=N_{A_{L_{1}(R_{1})}}\hspace{-.1cm}+\hspace{-.05cm}N_{A_{L_{2}(R_{2})}}\hspace{-.1cm}=\hspace{-.1cm}1+1$, and $m \mathcal{L}=10^{-3}$.
The solid and dashed lines show the $S_{\text{EE}}$ and $S_{\text{OEE}}$, respectively for temperatures ranging from $\beta=10^{-2} \mathcal{L}$ to $10^{2}\mathcal{L}$. The right plot presents a longer period of time to see the finite size effects.}\label{Temp-adjac1}
\end{figure}

\begin{figure}[H]
\hspace{.8cm}\includegraphics[scale=.38]{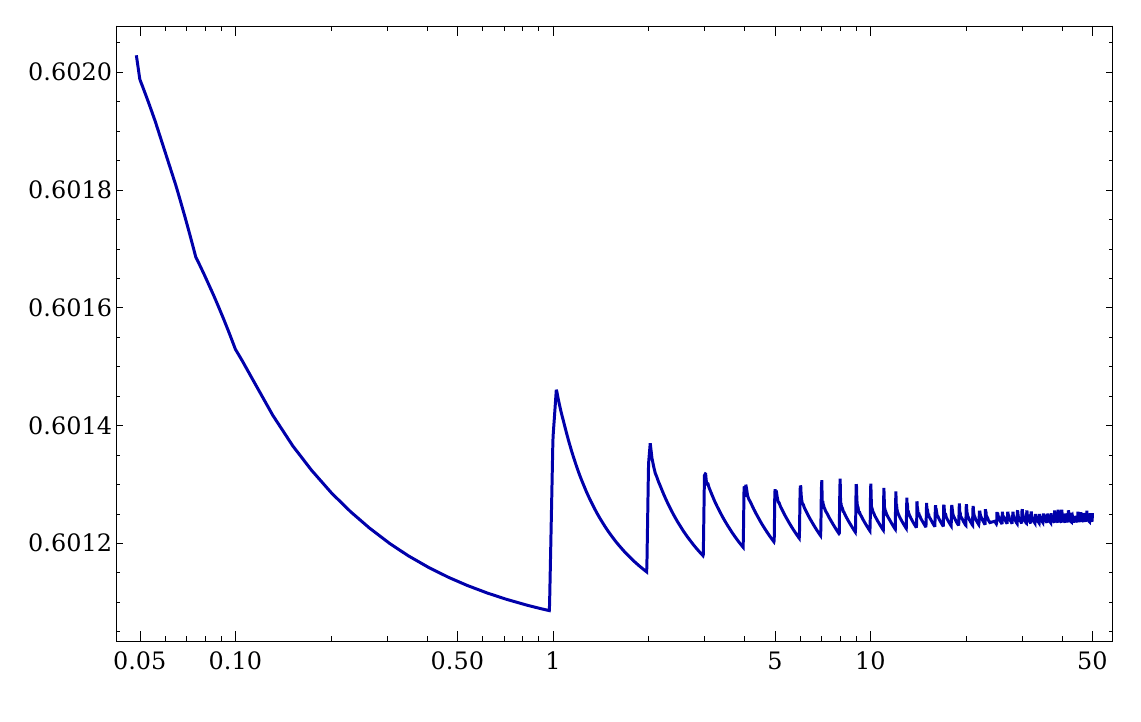}
\put(-220,60){\rotatebox{-270}{\fontsize{13}{13}\selectfont $\mathcal{E}(t)$}}		\put(-110,-10){{\fontsize{11}{11}\selectfont $t/\mathcal{L}$}}	
\hspace{.6cm}
\includegraphics[scale=.38]{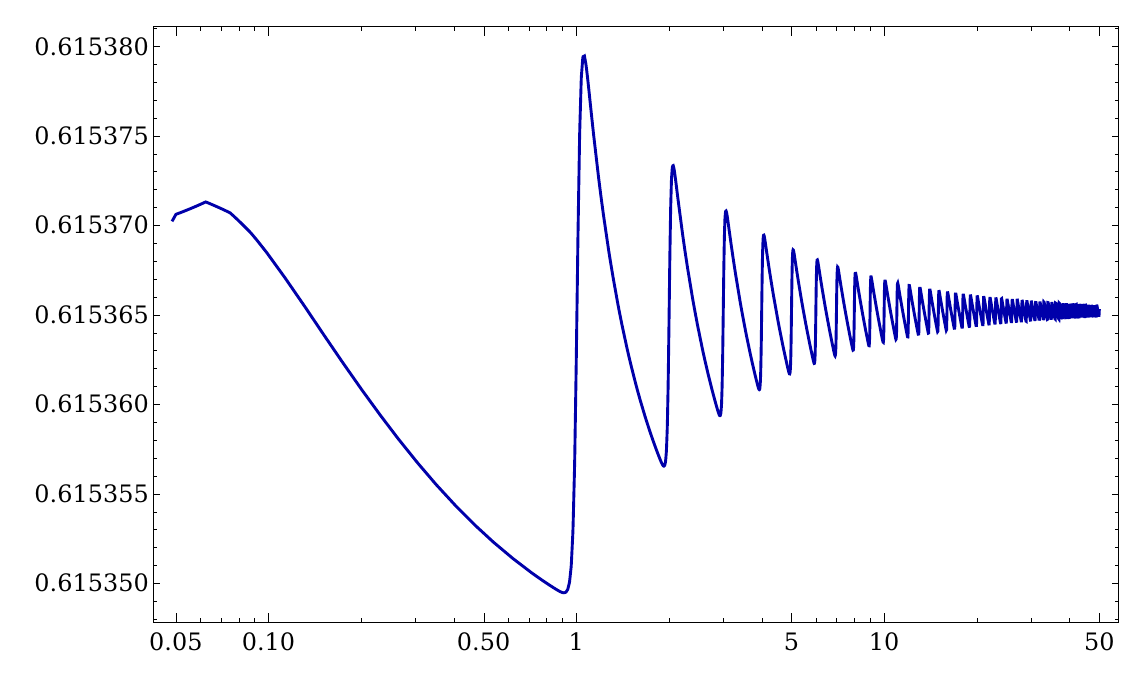}
\put(-220,60){\rotatebox{-270}{\fontsize{13}{13}\selectfont $\mathcal{E}(t)$}}		\put(-110,-10){{\fontsize{11}{11}\selectfont $t/\mathcal{L}$}}	
\vspace{.5cm}

\hspace{.75cm}\includegraphics[scale=.38]{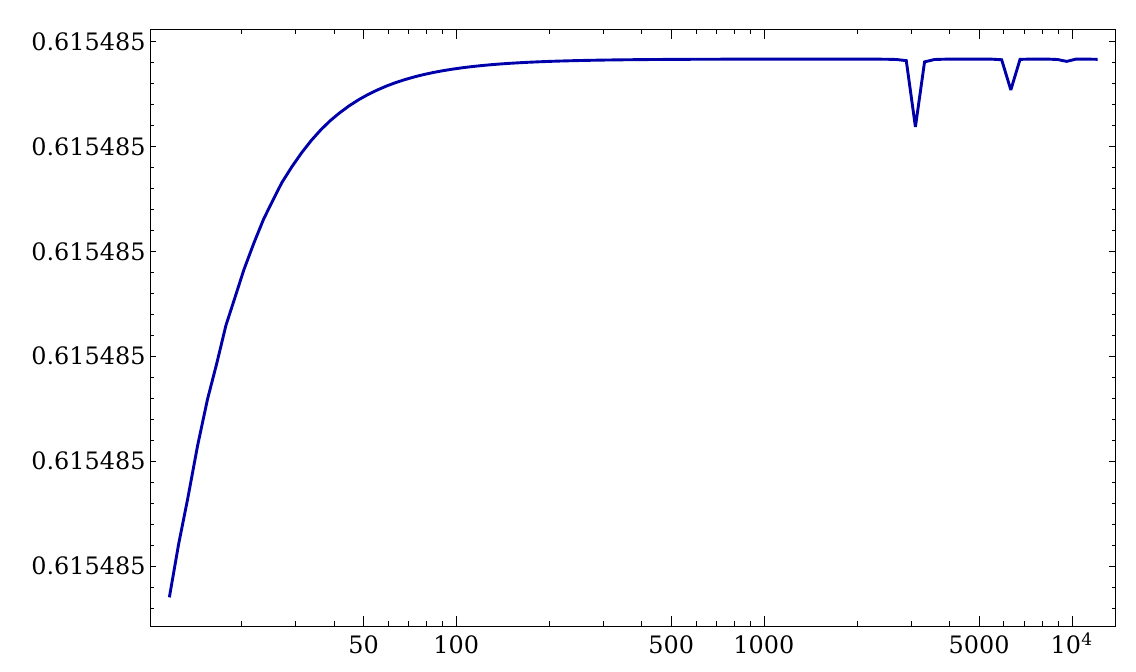}
\put(-220,60){\rotatebox{-270}{\fontsize{13}{13}\selectfont $\mathcal{E}(t)$}}		\put(-110,-10){{\fontsize{11}{11}\selectfont $t/\mathcal{L}$}}	
\hspace{.7cm}
\includegraphics[scale=.38]{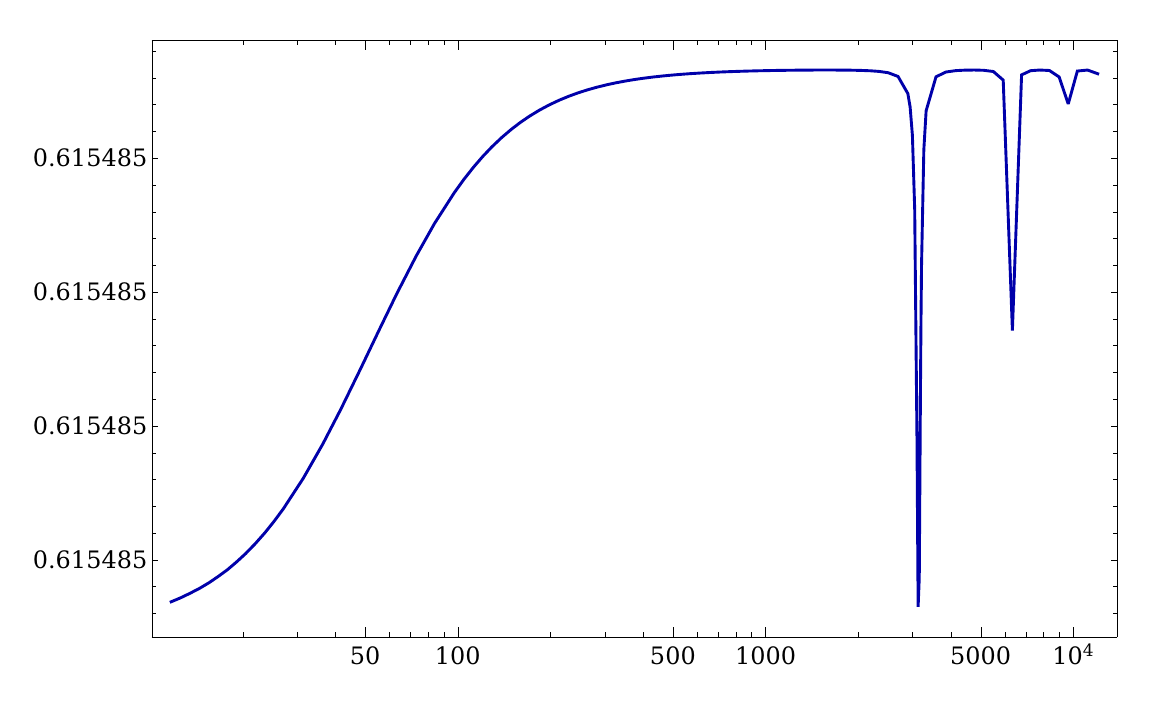}
\put(-223,60){\rotatebox{-270}{\fontsize{13}{13}\selectfont $\mathcal{E}(t)$}}		\put(-110,-10){{\fontsize{11}{11}\selectfont $t/\mathcal{L}$}}	
\caption{The long time behavior of logarithmic negativity $\mathcal{E}$ is presented for $N=1501$, $N_{A_{L(R)}}=N_{A_{L_{1}(R_{1})}}\hspace{-.1cm}+\hspace{-.05cm}N_{A_{L_{2}(R_{2})}}\hspace{-.1cm}=\hspace{-.1cm}1+1$, and $m \mathcal{L}=10^{-3}$, for temperatures $\beta=10^{-2}\mathcal{L}$ (upper-left), $\beta=10^{-1}\mathcal{L}$ (upper-right), $\beta=10\mathcal{L}$ (bottom-left), and $10^{2}\mathcal{L}$(bottom-right).}\label{TempLN-adjac4}	
\end{figure}
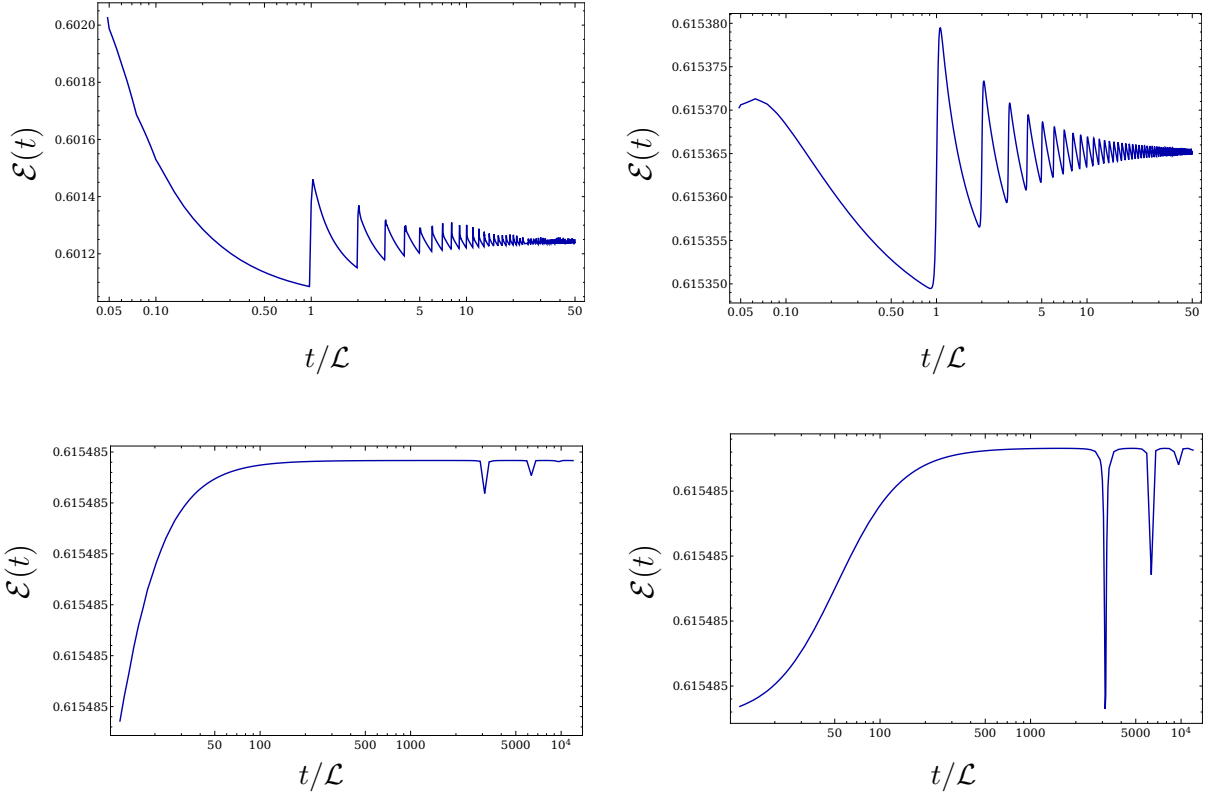

In figure \ref{Temp-adjac1}, left panel, time evolution of the $S_{\text{EE}}$ and $S_{\text{OEE}}$ are presented  for temperatures ranging from $\beta=10^{-2} \mathcal{L}$ to $10^{2}\mathcal{L}$.
The left plot presents $S_{\text{EE}}$ for $\beta=10^{-2} \mathcal{L}$(solid blue), $\beta=10^{-1} \mathcal{L}$(solid green), $\beta=10 \mathcal{L}$(solid purple) and $\beta=10^{2} \mathcal{L}$(solid  light blue). Also, in this plot $S_{\text{OEE}}$ is depicted from $\beta=10^{-2} \mathcal{L}$ to $10^{2}$ with dashed orange, dashed red, dashed brown and dashed yellow, respectively. The upper two curves present high-temperature limit where the   thermal correlation dominates. For low temperatures, the initial growth happens after a delay. The right panel denotes the finite size effects for temperatures $\beta=10 \mathcal{L}$ (solid blue, dashed orange), and  $\beta=10^{2} \mathcal{L}$ (solid green and dashed red). 
The time dependence of $\Delta S=S_{\text{OEE}}-S_{\text{EE}}$ is presented in figure \ref{TempSS-adjac2}, for various ranges of temperatures, $\beta=10^{-2} \mathcal{L}$ ( dark blue), $\beta=10^{-1} \mathcal{L}$ (orange), $\beta=10 \mathcal{L}$  (green), and  $\beta=10^{2} \mathcal{L}$ (red) curves. The left panel shows the initial decreasing and the left one shows saturation. In initial times, the difference is bigger for high temperatures in comparison with low temperatures. 
\begin{figure}[H]
\centering
\centering\includegraphics[scale=.38]{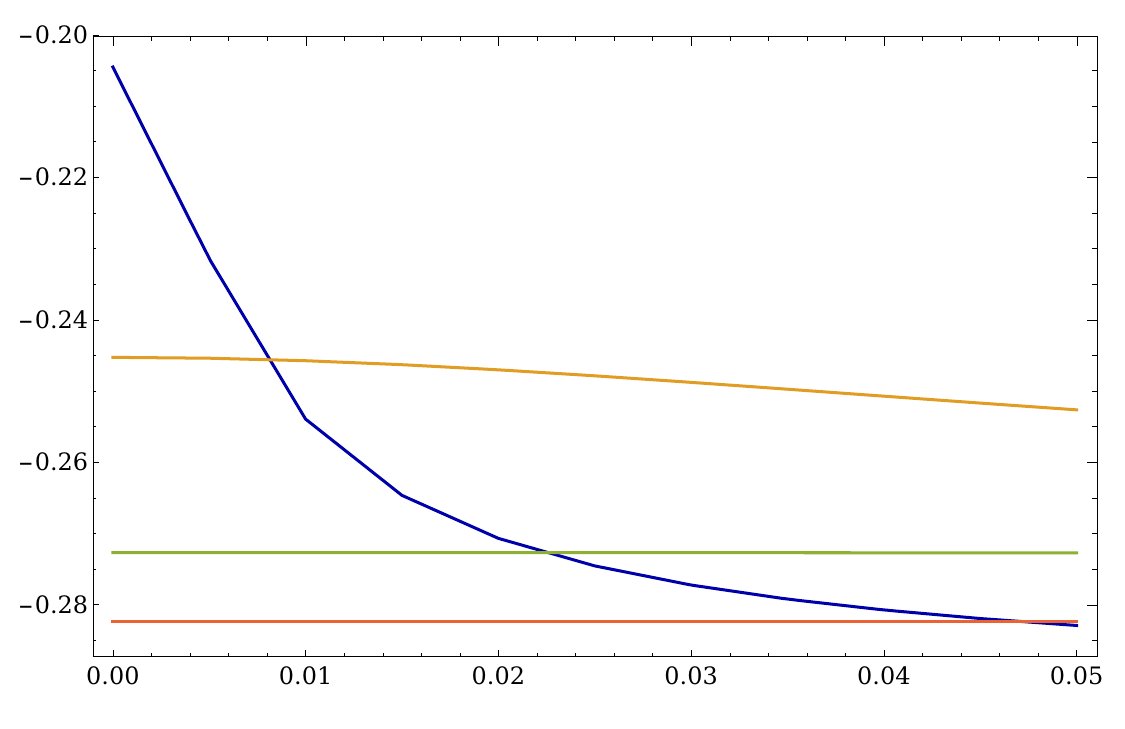}
\put(-220,60){\rotatebox{-270}{\fontsize{13}{13}\selectfont $\Delta S(t)$}}		\put(-110,-10){{\fontsize{11}{11}\selectfont $t/\mathcal{L}$}}	
\hspace{.7cm}
\includegraphics[scale=.38]{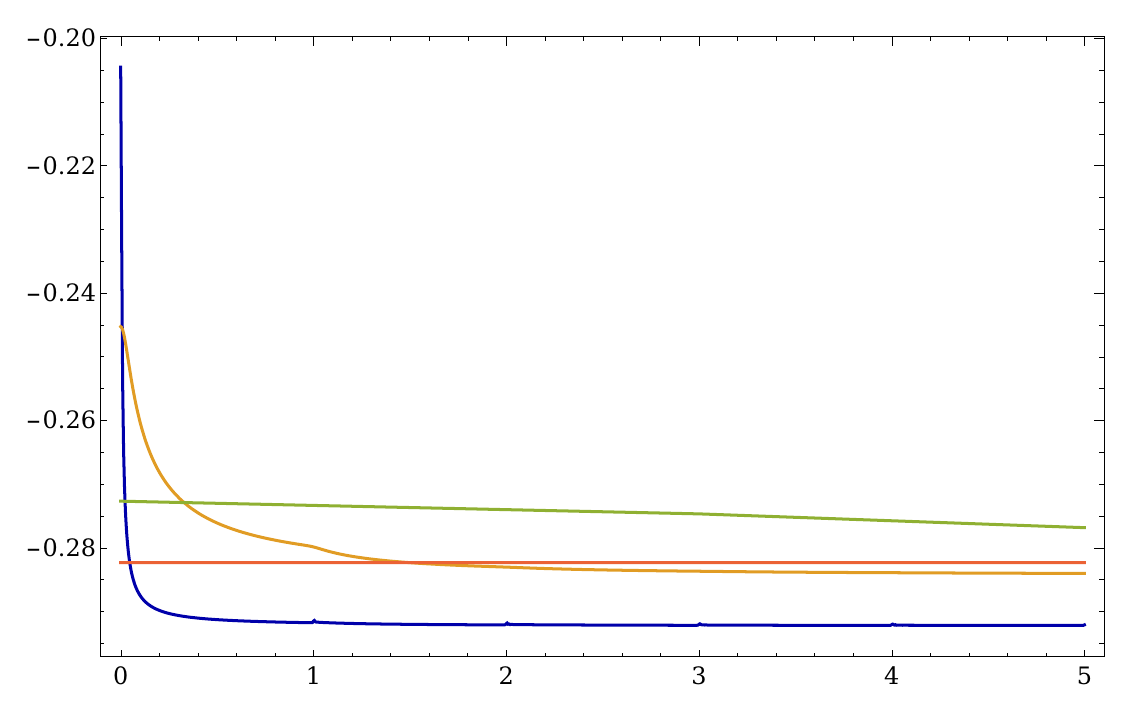}	\put(-220,60){\rotatebox{-270}{\fontsize{13}{13}\selectfont $\Delta S(t)$}}		\put(-110,-10){{\fontsize{11}{11}\selectfont $t/\mathcal{L}$}}	
\caption{The time dependence of $\Delta S= S_{\text{OEE}}-S_{\text{EE}}$ is presented for $N=1501$, $N_{A_{L(R)}}=N_{A_{L_{1}(R_{1})}}\hspace{-.1cm}+\hspace{-.05cm}N_{A_{L_{2}(R_{2})}}\hspace{-.1cm}=\hspace{-.1cm}1+1$ and $m \mathcal{L}=10^{-3}$.}\label{TempSS-adjac2}
\end{figure}
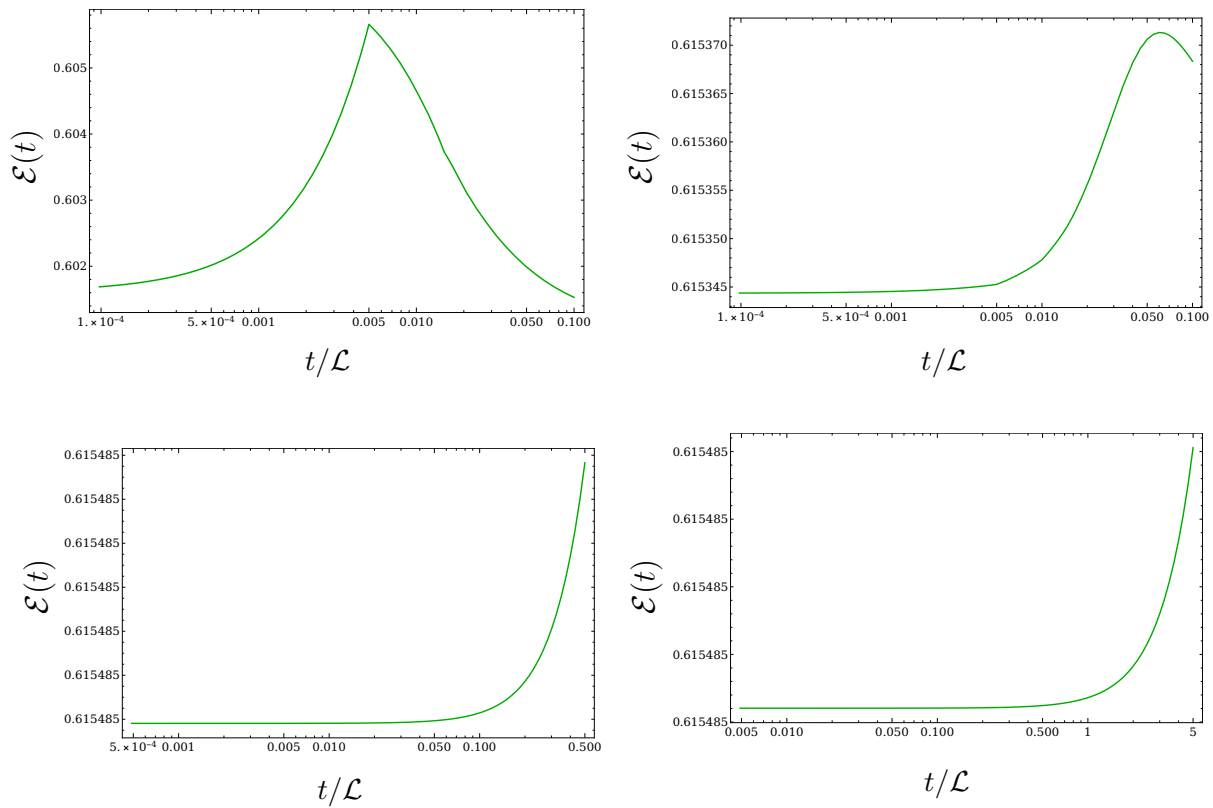
\begin{figure}[H]	
\hspace{.75cm}\includegraphics[scale=.38]{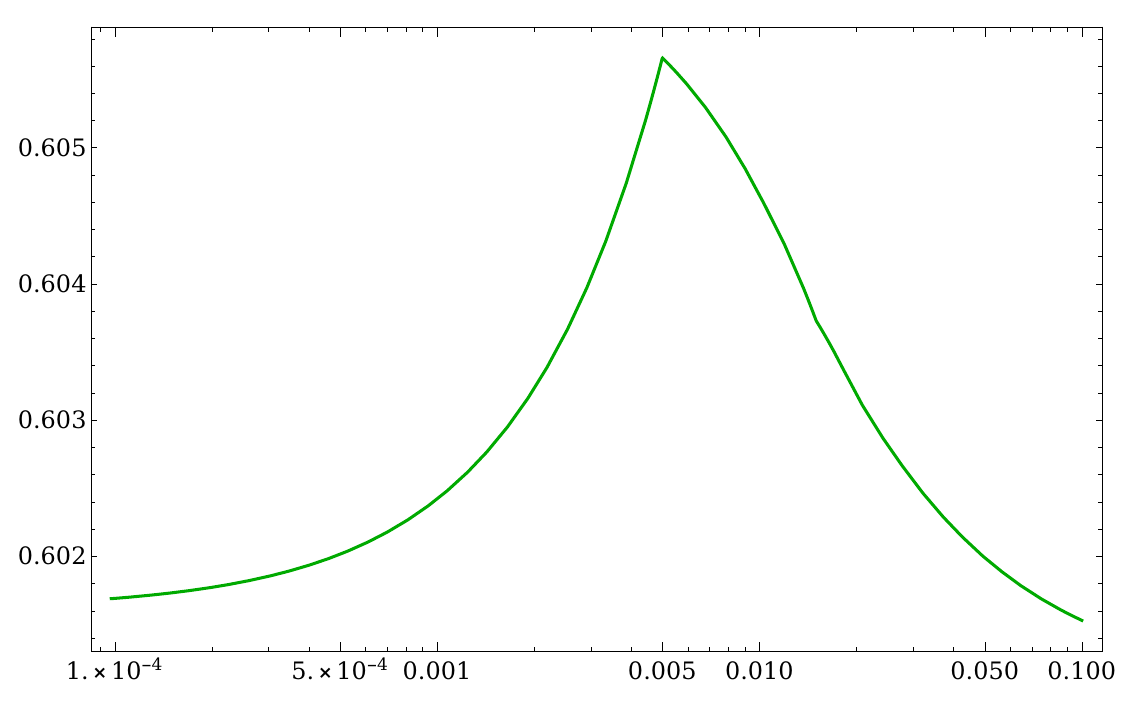}
\put(-220,60){\rotatebox{-270}{\fontsize{13}{13}\selectfont $\mathcal{E}(t)$}}		\put(-110,-10){{\fontsize{11}{11}\selectfont $t/\mathcal{L}$}}	
\hspace{.6cm}
\includegraphics[scale=.38]{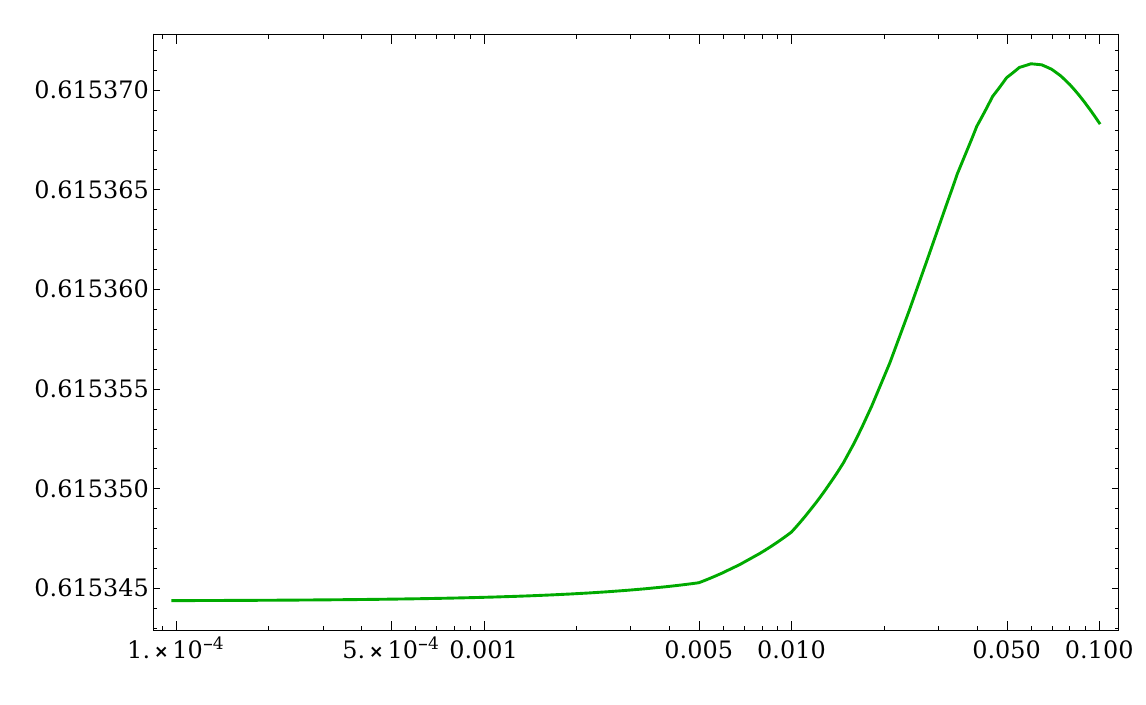}
\put(-220,60){\rotatebox{-270}{\fontsize{13}{13}\selectfont $\mathcal{E}(t)$}}		\put(-110,-10){{\fontsize{11}{11}\selectfont $t/\mathcal{L}$}}
\vspace{.4cm}
	
\hspace{.75cm}\includegraphics[scale=.38]{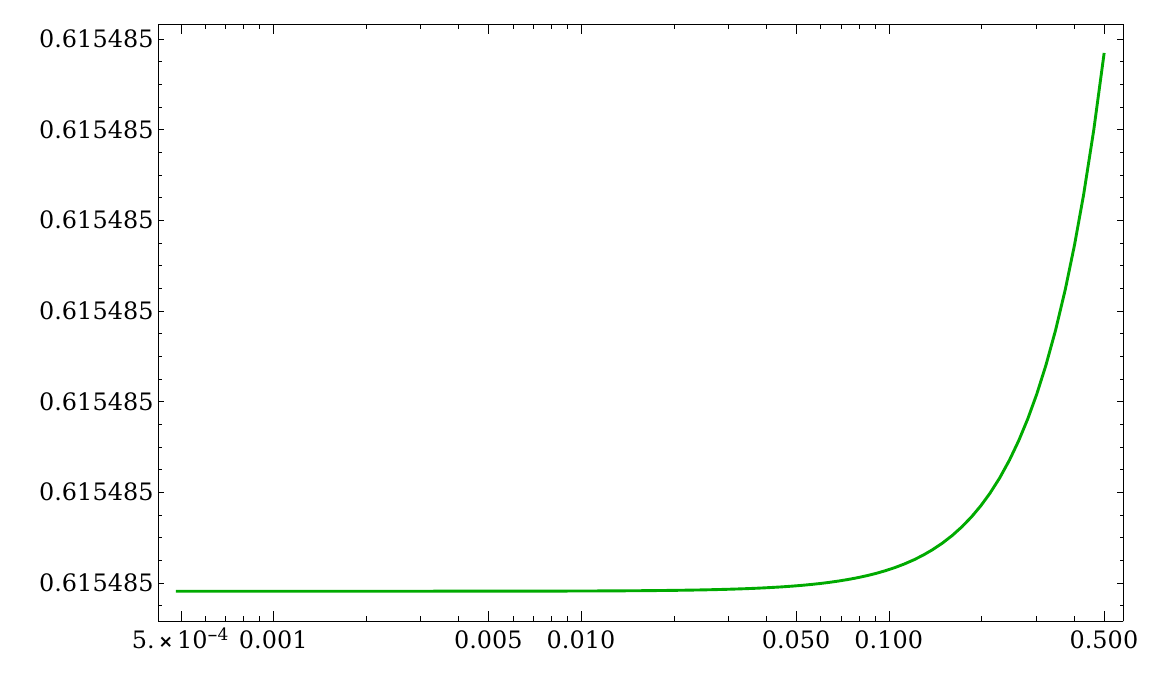}
\put(-220,60){\rotatebox{-270}{\fontsize{13}{13}\selectfont $\mathcal{E}(t)$}}		\put(-110,-10){{\fontsize{11}{11}\selectfont $t/\mathcal{L}$}}
\hspace{.3cm}
\includegraphics[scale=.38]{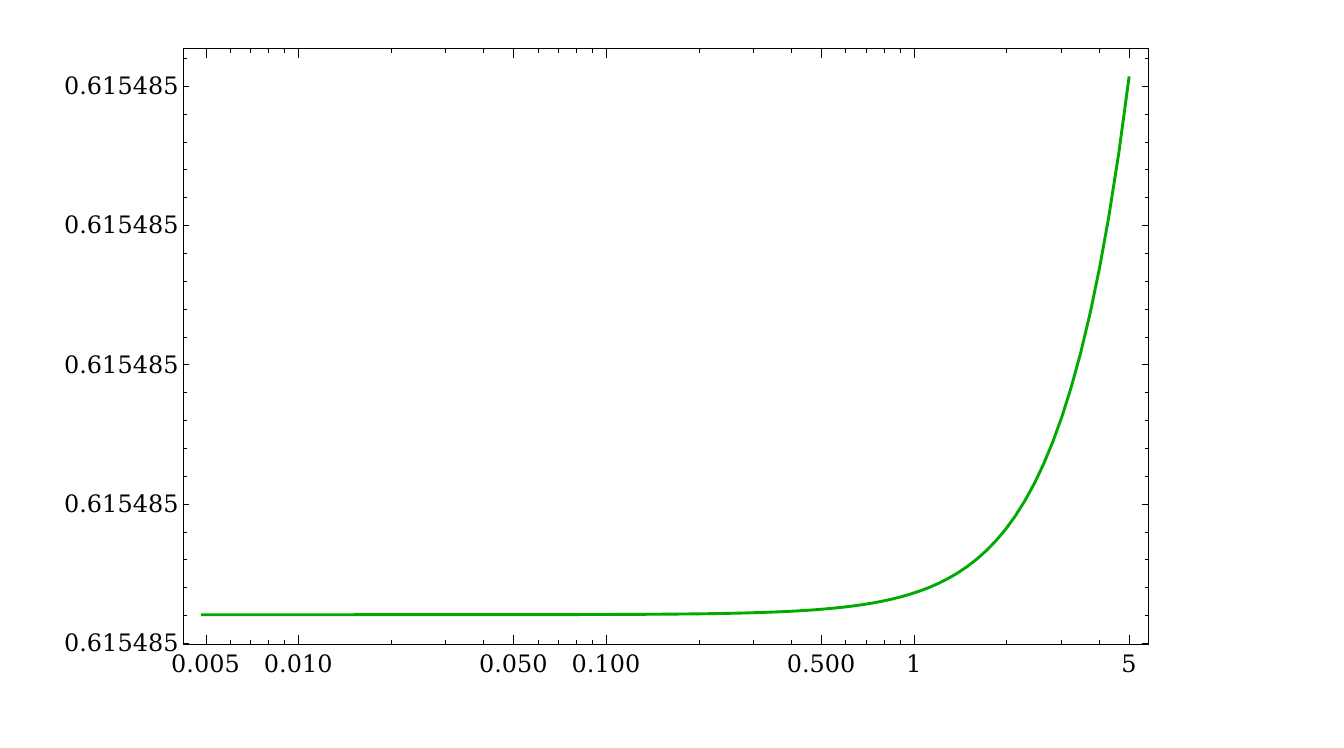}
\put(-250,60){\rotatebox{-270}{\fontsize{13}{13}\selectfont $\mathcal{E}(t)$}}		\put(-140,-7){{\fontsize{11}{11}\selectfont $t/\mathcal{L}$}}
\caption{The initial growth of  logarithmic negativity is presented for $N=1501$, $N_{A_{L(R)}}=N_{A_{L_{1}(R_{1})}}\hspace{-.1cm}+\hspace{-.05cm}N_{A_{L_{2}(R_{2})}}\hspace{-.1cm}=\hspace{-.1cm}1+1$, and $m \mathcal{L}=10^{-3}$ for temperatures $\beta=10^{-2}\mathcal{L}$ (upper-left), $\beta=10^{-1}\mathcal{L}$ (upper-right), $\beta=10\mathcal{L}$ (bottom-left), and $10^{2}\mathcal{L}$(bottom-right).}\label{TempLN-adjac3}	
\end{figure}
In figures \ref{TempLN-adjac3} and \ref{TempLN-adjac4}, the time evolution of $\mathcal{E}$ is studied for temperatures ranging from $\beta=10^{-2} \mathcal{L}$ to $10^{2} \mathcal{L}$. The temperature decreases by a factor of $10$ from the upper-left panel to the lower-right panel. In figure \ref{TempLN-adjac3}, the upper panels denote the time evolution for high temperatures and the bottom panels are related to low temperatures. Interestingly, the LN is zero for times $t < \beta/2$ and after that, it begins to grow linearly\footnote{This delay is also reported in \cite{Coser:2014gsa}.}. Followed by figure \ref{TempLN-adjac3}, the long time behavior of logarithmic negativity for various temperatures are presented in figure \ref{TempLN-adjac4}. From upper  panels, which are correspond to high temperatures $\beta=10^{-2} \mathcal{L}$(upper-left) and  $\beta=10^{-1} \mathcal{L}$(upper-right), we see the reduction of logarithmic negativity for time scales of order $t\sim \mathcal{L}$ and sudden quantum revival after this time. The finite size effects induce a periodic behavior.
For bottom panels,  which correspond to the low temperatures, $\beta=10 \mathcal{L}$(bottom-left) and $\beta=10^{2} \mathcal{L}$(bottom-right), we observe the growth of logarithmic negativity followed by a plateau and then an oscillatory behavior due to the finite size effect. 
\subsection{Two disjoint intervals on each side}
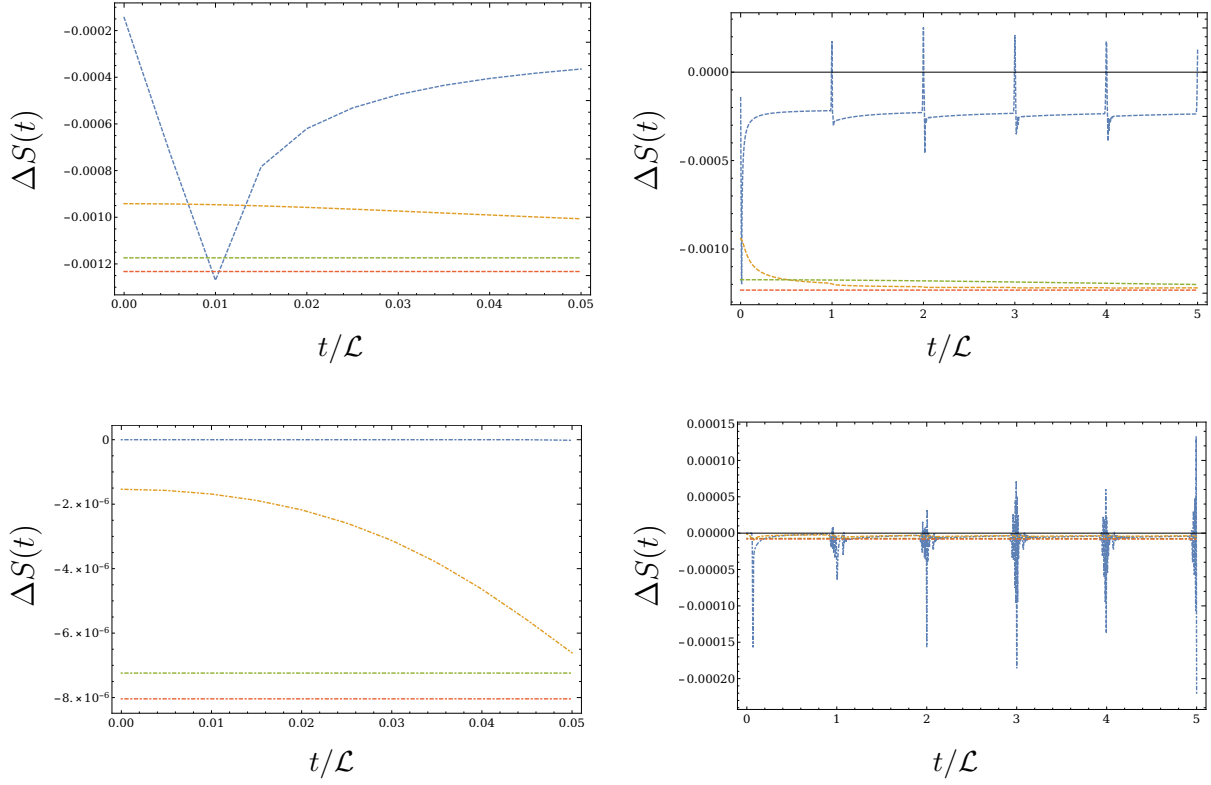
\begin{figure}[H]
\centering
\includegraphics[scale=.38]{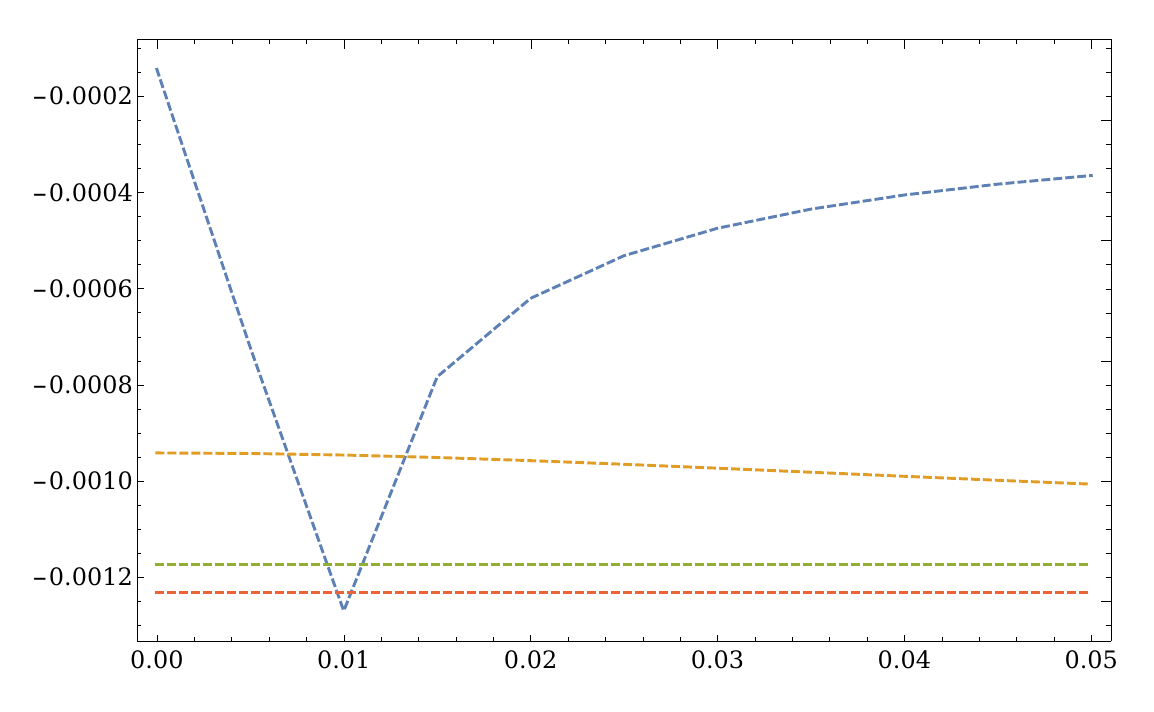}
\put(-225,52){\rotatebox{-270}{\fontsize{13}{13}\selectfont $\Delta S(t) $}}		\put(-110,-10){{\fontsize{11}{11}\selectfont $t/\mathcal{L}$}}
\hspace{.6cm}
\includegraphics[scale=.38]{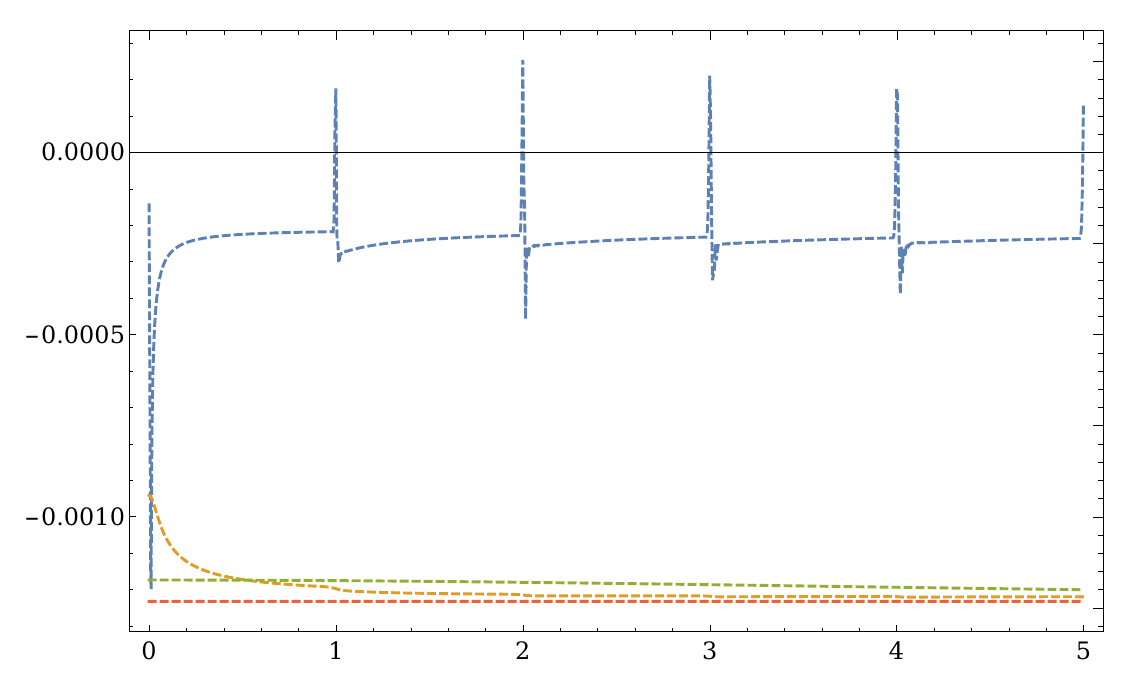}\put(-220,52){\rotatebox{-270}{\fontsize{13}{13}\selectfont $\Delta S(t) $}}		\put(-110,-10){{\fontsize{11}{11}\selectfont $t/\mathcal{L}$}}
\vspace{.5cm}

\includegraphics[scale=.38]{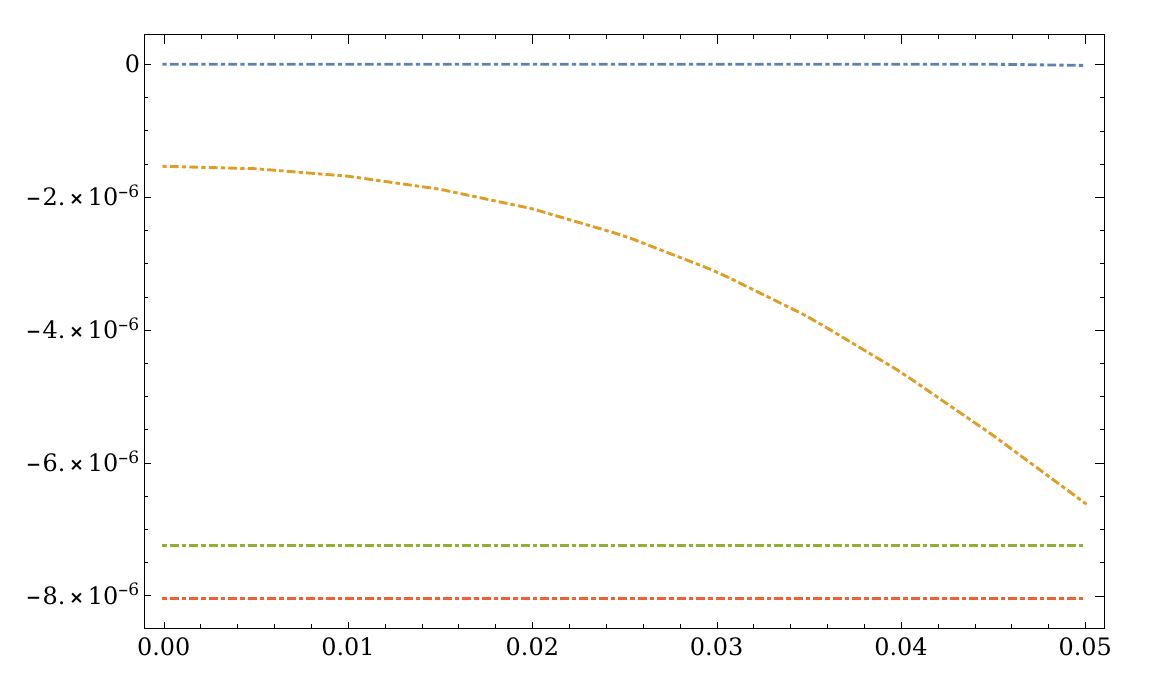}\put(-222,52){\rotatebox{-270}{\fontsize{13}{13}\selectfont $\Delta S(t) $}}		\put(-110,-10){{\fontsize{11}{11}\selectfont $t/\mathcal{L}$}}
\hspace{.6cm}
\includegraphics[scale=.38]{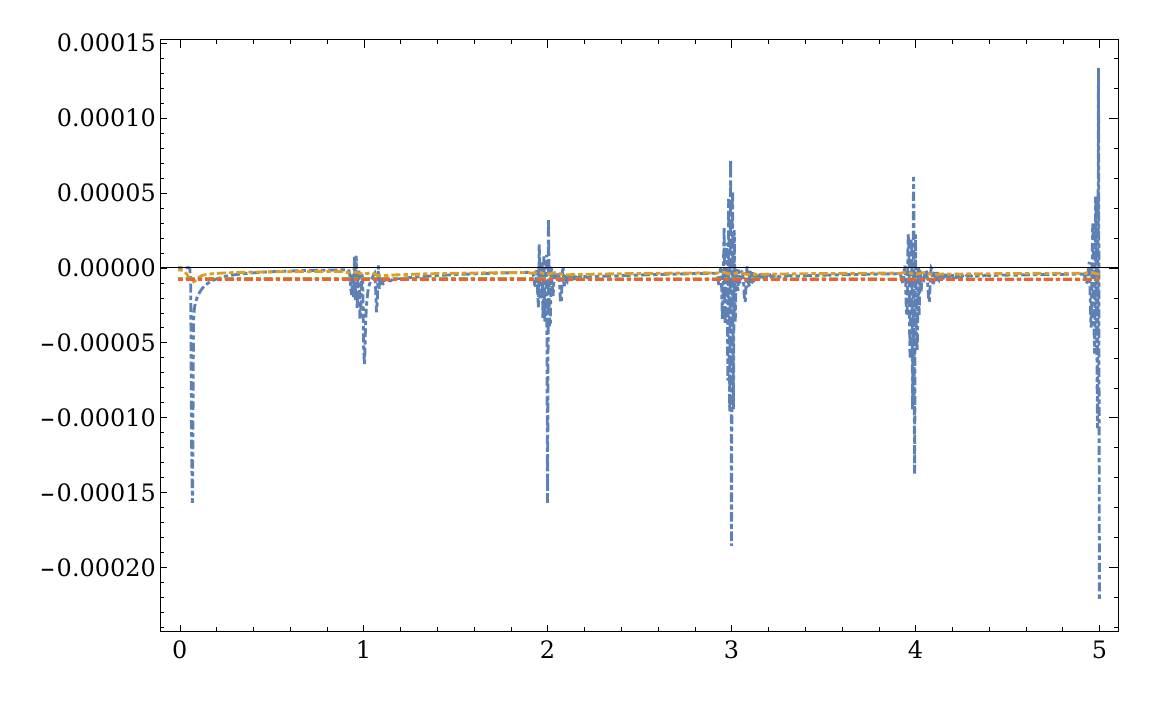}\put(-222,52){\rotatebox{-270}{\fontsize{13}{13}\selectfont $\Delta S(t) $}}		\put(-110,-10){{\fontsize{11}{11}\selectfont $t/\mathcal{L}$}}
\caption{Time dependence of $\Delta S = S_{\text{OEE}}-S_{\text{EE}}$ is presented for $N=1501$, $N_{A_{L(R)}}=N_{A_{L_{1}(R_{1})}}+N_{A_{L_{2}(R_{2})}}\hspace{-.1cm}=\hspace{-.1cm}1+1$, $m \mathcal{L}=10^{-3}$, $d=10$ (upper panels) and $d=100$ (bottom panels). The blue, orange, green, and red curves correspond to temperatures $\beta=10^{-2}\mathcal{L}, 10^{-1}\mathcal{L}, 10\mathcal{L}, 10^{2}\mathcal{L}$, respectively.} 
\label{TempSS-distant3}
\end{figure}

In this subsection, we will study the effect of changing separation $d$ and temperature together on the time evolution of $S_{\text{EE}}$, $S_{\text{OEE}}$ and $\Delta S= S_{\text{OEE}}-S_{\text{EE}}$. 
\begin{figure}[H]
\centering
\includegraphics[scale=.38]{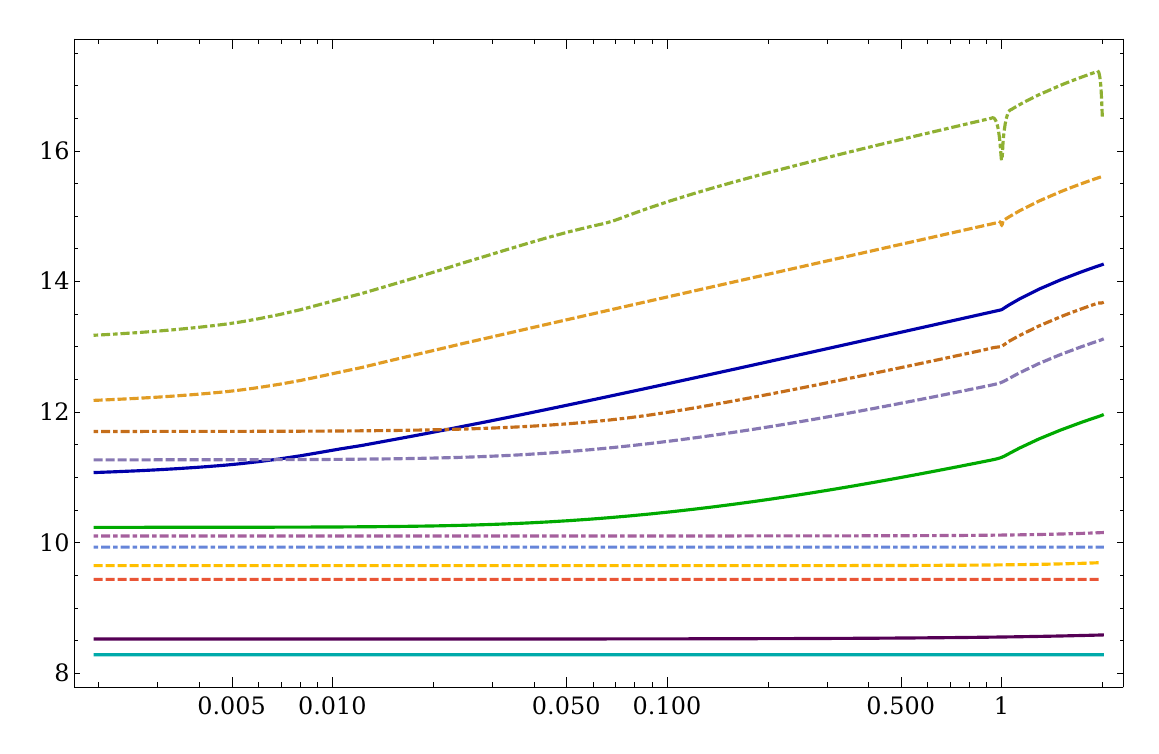}	
\put(-222,52){\rotatebox{-270}{\fontsize{13}{13}\selectfont $S_{\text{EE}}(t) $}}		\put(-110,-10){{\fontsize{11}{11}\selectfont $t/\mathcal{L}$}}
\hspace{.6cm}
\includegraphics[scale=.38]{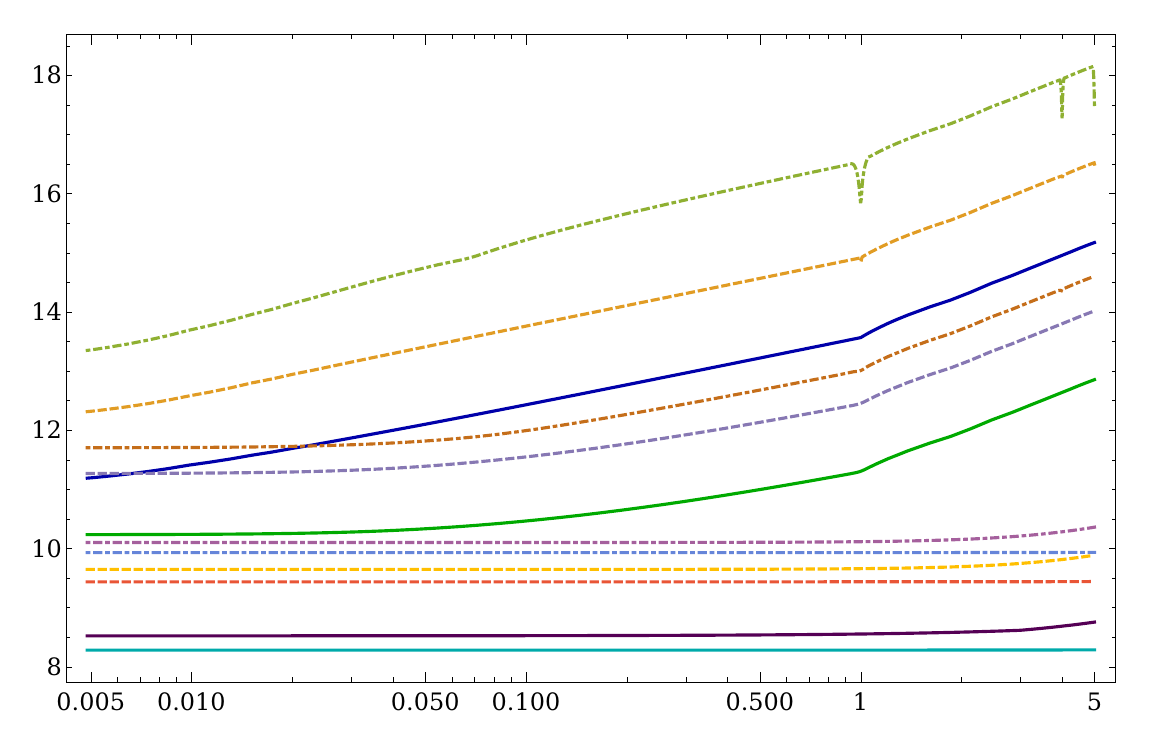}\put(-222,52){\rotatebox{-270}{\fontsize{13}{13}\selectfont $S_{\text{EE}}(t) $}}		\put(-110,-10){{\fontsize{11}{11}\selectfont $t/\mathcal{L}$}}
\vspace{.5cm}
\includegraphics[scale=.38]{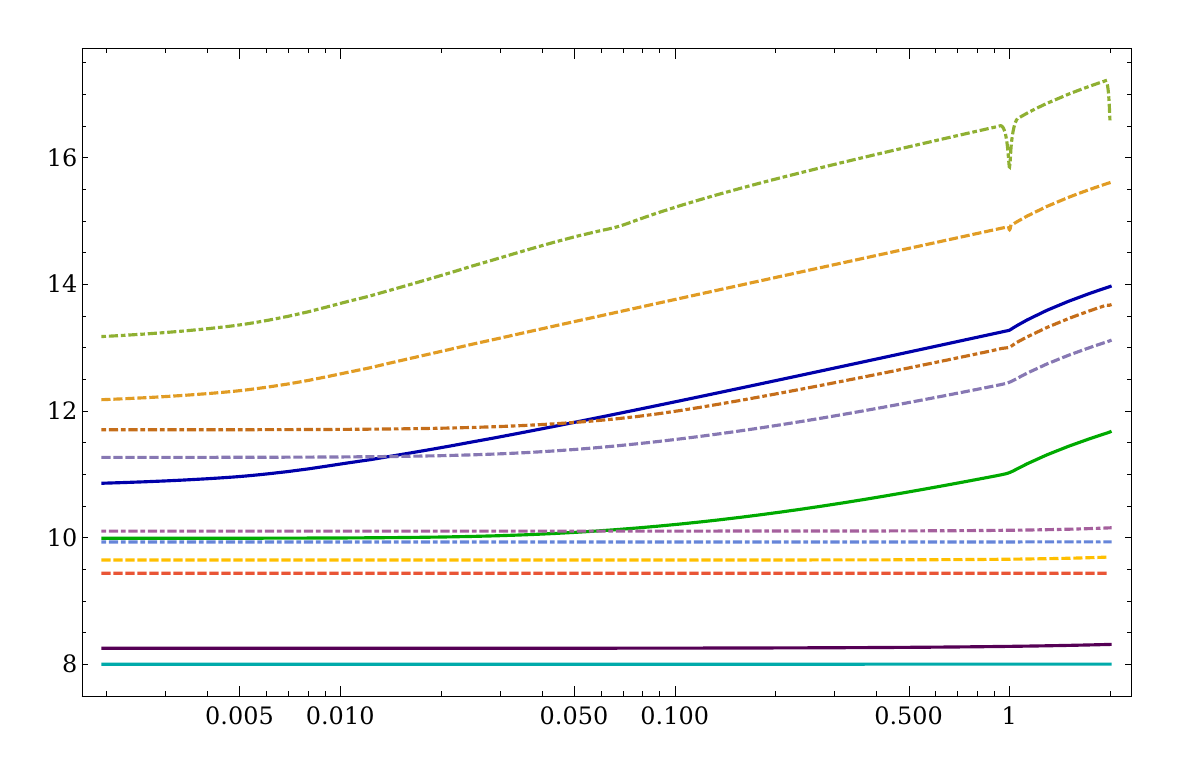}\put(-225,52){\rotatebox{-270}{\fontsize{13}{13}\selectfont $S_{\text{OEE}}(t) $}}		\put(-110,-10){{\fontsize{11}{11}\selectfont $t/\mathcal{L}$}}
\hspace{.6cm}
\includegraphics[scale=.38]{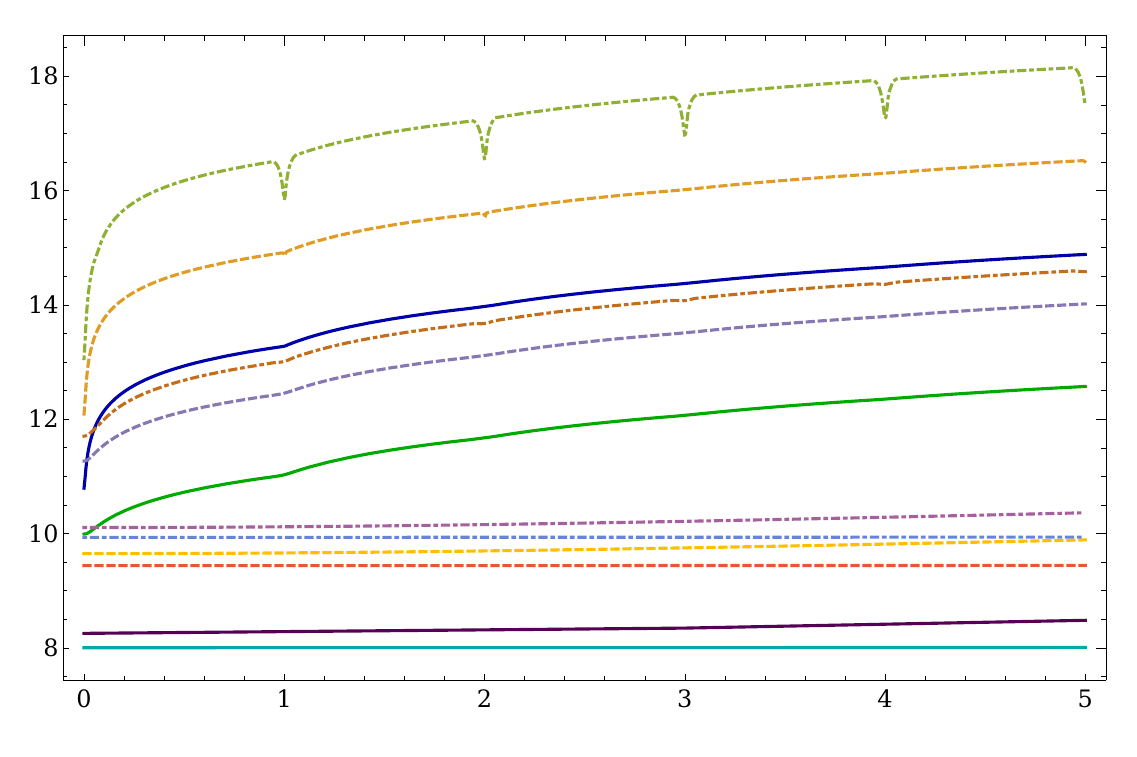}\put(-223,52){\rotatebox{-270}{\fontsize{13}{13}\selectfont $S_{\text{OEE}}(t) $}}		\put(-110,-10){{\fontsize{11}{11}\selectfont $t/\mathcal{L}$}}
\caption{The time evolution of $S_{\text{EE}}$ and $S_{\text{OEE}}$ for $N=1501$, $N_{A_{L(R)}}\hspace{-.1cm}=\hspace{-.1cm}N_{A_{L_{1}(R_{1})}}\hspace{-.1cm}+\hspace{-.05cm}N_{A_{L_{2}(R_{2})}}\hspace{-.1cm}=\hspace{-.1cm}1+1$, and $m \mathcal{L}=10^{-3}$.
\{The solid dark blue($\beta=10^{-2}\mathcal{L}$), solid green($\beta=10^{-1}\mathcal{L}$), solid dark purple($\beta=10\mathcal{L}$), solid cyan($\beta=10^{2}\mathcal{L}$)\}, \{dashed orange($\beta=10^{-2}\mathcal{L}$),  dashed light purple($\beta=10^{-1}\mathcal{L}$), dashed yellow($\beta=10\mathcal{L}$), dashed red ($10^{2}\mathcal{L}$)\}, and \{dot-dashed light green($\beta=10^{-2}\mathcal{L}$), dot-dashed brown ($\beta=10^{-1}\mathcal{L}$),  dot-dashed purple ($\beta=10\mathcal{L}$), and dot-dashed light blue $(\beta=10^{2}\mathcal{L})$\} correspond to \{$d=0,10, 100$\}, respectively.}\label{TempS-distant2}
\end{figure}
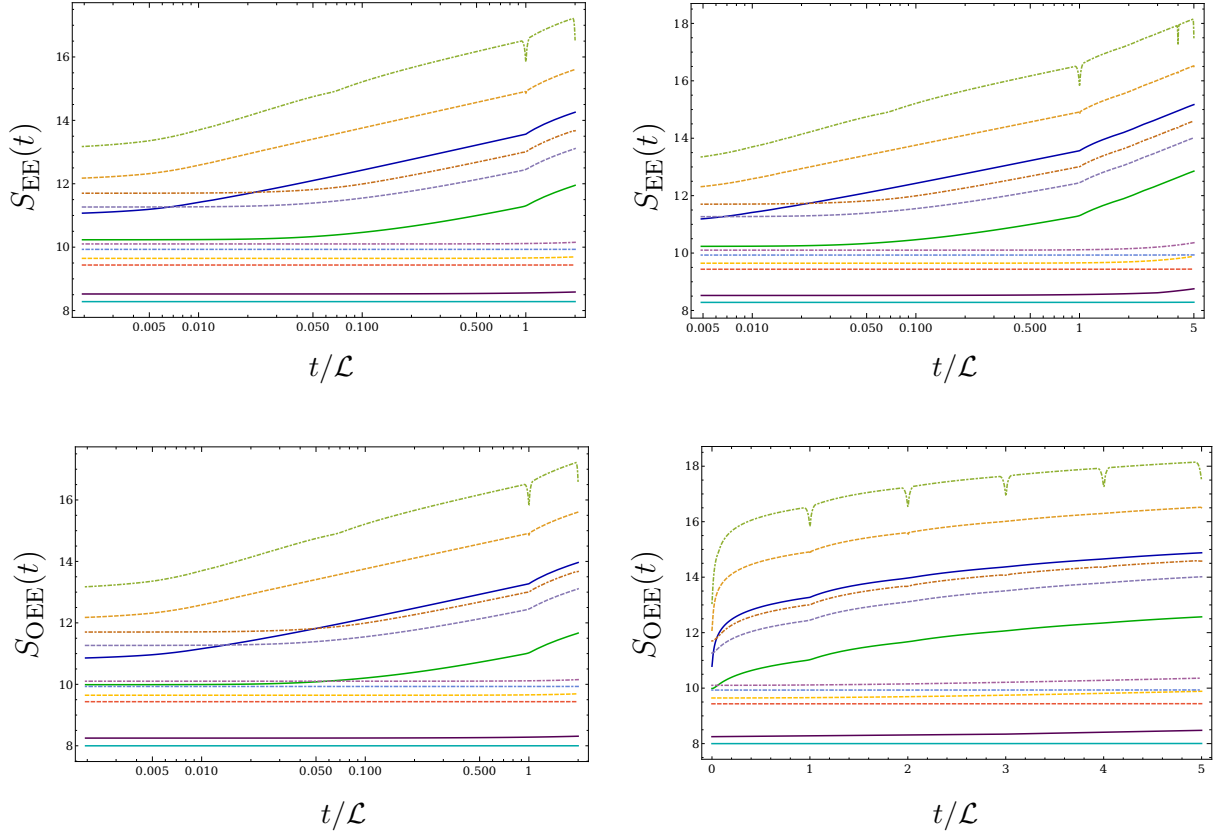
The figure \ref{TempS-distant2} denotes the time-dependent behavior of $S_{\text{EE}}$(upper panels) and $S_{\text{OEE}}$(bottom panels). The solid lines correspond to $d=0$.  The dashed and dot-dashed lines correspond to  $d=10, 100$, respectively.
Effectively, $S_{\text{OEE}}$ and $S_{\text{EE}}$ behave the same and there exists a competition between increasing the distance $d$ and decreasing the temperature. By increasing the distance,  $(S_{\text{OEE}}, S_{\text{EE}})$ increase and by decreasing the temperature both of them decrease. Therefore, in general, it exists a critical ratio $d_{\text{cr}}/\beta{_{\text{cr}}}$ which around it increasing distance or decreasing the temperature wins the competition.   Figure \ref{TempSS-distant3} represents the time dependence of $\Delta S= S_{\text{OEE}}-S_{\text{EE}}$ with initial fluctuation around zero value followed by saturation. The effects of finite size can be seen as oscillatory behavior.
\section{Zero-mode computations}\label{Ap-3}
In this Appendix, we provide details of calculations in section \ref{Sec-5}. Using (\ref{EQ-Z1}),  the related covariance matrix in the position basis  $\xi_{x}^{a}=(q_{x}^{L},q_{x}^{R},p_{x}^{L},p_{x}^{R})$ takes the following form:
\begin{align}\label{}
&G^{ab}_{x,x}(t) =\frac{1}{N}\sum_{k}^{N-1}
\nonumber\\\nonumber\\
&\hspace{-.2cm}\times\left(
\begin{array}{cccc}
\frac{\cosh(2\alpha_{ k})}{\lambda_k}    &\frac{\cos(\omega_{ k}t)\sinh(2\alpha_{ k})}{\lambda_{k}} & 0&-\sin(\omega_{ k}t)\sinh(2\alpha_{ k}) \\
\frac{\cos(\omega_{ k}t)\sinh(2\alpha_{ k})}{\lambda_{k}} & \frac{\cosh(2\alpha_{k})}{\lambda_k} &-\sin(\omega_{ k}t)\sinh(2\alpha_{ k})&  0\\
0 & -\sin(\omega_{ k}t)\sinh(2\alpha_{k}) & 
\lambda_{k}\cosh (2\alpha_{k})&-\lambda_{k}\cos(\omega_{ k}t)\sinh(2\alpha_{ k})\\
-\sin(\omega_{ k}t)\sinh(2\alpha_{k})& 0 & 
-\lambda_{k}\cos(\omega_{ k}t)\sinh(2\alpha_{ k})& \lambda_{k}\cosh (2\alpha_{ k})\\
\end{array} \right).
\end{align}
The schematic form of this matrix is
\begin{align}\label{EQ-}
G^{ab}_{x,x}(t) =
\left(
\begin{array}{cccc}
e_{1}  &\mathcal{M}_{1}(t) & 0&\mathcal{M}_{2}(t)  \\
\mathcal{M}_{1}(t) & e_{1} &\mathcal{M}_{2}(t) &  0\\
0 & \mathcal{M}_{2}(t)  & 
e_{2} &\mathcal{M}_{3}(t) \\
\mathcal{M}_{2}(t) & 0 & 
\mathcal{M}_{3}(t) & e_{2}\\
\end{array} \right),
\end{align} 
whose determinant becomes
\begin{align}\label{EQ2-2}
\det \Big(G^{ab}_{x,x}(t)\Big) = \underset{Leading \hspace{.6mm} order}{\underbrace{e_{1}^{2}e_{2}^{2}-e_{2}^{2}F_{1}^{2}(t)}}+\underset{Subleading \hspace{.6mm} order}{\underbrace{(\mathcal{M}_{2}^{2}(t)-\mathcal{M}_{1}(t)\mathcal{M}_{2}(t))^{2}-2e_{1}e_{2}\mathcal{M}_{2}^{2}(t)-e_{1}^{2}\mathcal{M}_{3}^{2}(t)}}.
\end{align} 
When $t\gg \delta=\frac{\mathcal{L}}{N}$ and $N\rightarrow\infty$, $m\mathcal{L}\ll1$,  $\beta/\mathcal{L}\ll1$ and  $m\ll\delta^{-1}$, one can see that
\begin{align}\label{eq-1}
&\sinh(2\alpha_{k})=1/\sinh(\frac{\beta\omega_{ k}}{2})\underset{k\ll N, \beta\ll \mathcal{L}}{\longrightarrow}\frac{2}{\beta\omega_{ k}}
\nonumber\\
&\omega_{k}=\omega_{N-k}\sim\frac{2\pi k }{\mathcal{L}}
\end{align}
which altogether imply that  
\begin{align}\label{eq-1}
&e_{1}=\frac{1}{N}\sum_{k=0}^{N-1}\frac{\cosh(2\alpha_{k})}{\lambda_k}= \frac{1}{N}\Big(\frac{\cosh(2\alpha_{ 0})}{\lambda_0} +\sum_{k=1}^{N-1}\frac{\cosh(2\alpha_{ k})}{\lambda_k} \Big)=\frac{1}{N}\Big(\frac{\cosh(2\alpha_{ 0})}{\lambda_0} +2\sum_{k=1}^{\frac{N}{2}}\frac{\cosh(2\alpha_{ k})}{\lambda_k} \Big)
\nonumber\\
&\quad=\frac{2}{N\beta \mathcal{L} m^{2}}\Big(1+\frac{m^{2} \mathcal{L}^{2}}{2 \pi^{2}}\sum_{k=1}^{\infty}\frac{1}{k^{2}} \Big)\sim \frac{2}{N\beta \mathcal{L} m^{2}}\Big(1+\frac{m^{2} \mathcal{L}^{2}}{12} \Big),
\end{align}
\begin{align}\label{eq-2}
&e_{2}=\frac{1}{N}\sum_{k=0}^{N-1}\lambda_{k}\cosh (2\alpha_{ k})=\frac{1}{N}\Big(\lambda_{0}\cosh (2\alpha_{ 0})+\sum_{k=1}^{N-1}\lambda_{k}\cosh (2\alpha_{ k})\Big)=\frac{1}{N}\Big(\frac{2 \mathcal{L}}{\beta}+2\sum_{k=1}^{\frac{N}{2}}\frac{2 \mathcal{L}}{\beta}\Big)\sim\frac{2 \mathcal{L}}{\beta},
 \end{align}
 
 \begin{align}\label{eq-1}
 	&\mathcal{M}_{1}(t)=\frac{1}{N}\sum_{k=0}^{N-1}\frac{\cos(\omega_{k}t)\sinh(2\alpha_{ k})}{\lambda_{k}}= \frac{1}{N}\Big(\frac{\cos(\omega_{ 0}t)\sinh(2\alpha_{ 0})}{\lambda_0} +\sum_{k=1}^{N-1}\frac{\cos(\omega_{k}t)\sinh(2\alpha_{ k})}{\lambda_{k}} \Big)
 	\nonumber\\
 	&\qquad=\frac{1}{N}\Big(\frac{\cos(\omega_{ 0}t)\sinh(2\alpha_{ 0})}{\lambda_0} +2\sum_{k=1}^{\frac{N}{2}}\frac{\cos(\omega_{ k}t)\sinh(2\alpha_{k})}{\lambda_{k}} \Big)
 	\nonumber\\
 	&\qquad\sim\frac{2}{N\beta \mathcal{L} m}\Big(\frac{\cos(\omega_{ 0}t)}{m} +\frac{m \mathcal{L}^{2}}{2 \pi^{2}}\sum_{k=1}^{\infty}\frac{\cos(\omega_{ k}t)}{k^{2}} \Big)\sim\frac{2}{N\beta \mathcal{L} m}\Big(\frac{\cos(mt)}{m} +\frac{m \mathcal{L}^{2}}{2 \pi^{2}}\sum_{k=1}^{\infty}\frac{\cos(\omega_{ k}t)}{k^{2}} \Big),
 \end{align}

 \begin{align}\label{eq-1}
 	&\mathcal{M}_{2}(t)=\frac{-1}{N}\sum_{k=0}^{N-1}\sin(\omega_{ k}t)\sinh(2\alpha_{ k})= \frac{-1}{N}\Big(\sin(\omega_{ 0}t)\sinh(2\alpha_{ 0}) +\sum_{k=1}^{N-1}\sin(\omega_{ k}t)\sinh(2\alpha_{ k}) \Big)
 	\nonumber\\
 	&\qquad=\frac{-1}{N}\Big(\sin(\omega_{ 0}t)\sinh(2\alpha_{ 0}) +2\sum_{k=1}^{\frac{N}{2}}\sin(\omega_{ k}t)\sinh(2\alpha_{ k}) \Big)
 	\nonumber\\
 	&\qquad\sim\frac{-2}{N\beta}\Big(\frac{\sin(\omega_{ 0}t)}{m} +\frac{\mathcal{L}}{ \pi}\sum_{k=1}^{\infty}\frac{\sin(\omega_{ k}t)}{k} \Big)\sim\frac{-2}{N\beta}\Big(\frac{\sin(mt)}{m} +\frac{\mathcal{L}}{ \pi}\sum_{k=1}^{\infty}\frac{\sin(\omega_{ k}t)}{k} \Big),
 \end{align}

 \begin{align}\label{eq-1}
 	&\mathcal{M}_{3}(t)=\frac{-1}{N}\sum_{k}^{N-1}\lambda_{k}\cos(\omega_{ k}t)\sinh(2\alpha_{ k})= \frac{-1}{N}\Big(\lambda_0\cos(\omega_{ 0}t)\sinh(2\alpha_{ 0}) +\sum_{k=1}^{N-1}\lambda_{k}\cos(\omega_{ k}t)\sinh(2\alpha_{ k}) \Big)
 	\nonumber\\
 	&\qquad=\frac{-1}{N}\Big(\lambda_0\cos(\omega_{ 0}t)\sinh(2\alpha_{ 0}) +2\sum_{k=1}^{\frac{N}{2}}\lambda_{k}\cos(\omega_{ k}t)\sinh(2\alpha_{ k}) \Big)
 	\nonumber\\
 	&\qquad\sim\frac{-2 \mathcal{L} m}{N\beta  }\Big(\frac{\cos(\omega_{ 0}t)}{m} +\frac{2}{m}\sum_{k=1}^{\infty}\frac{k\cos(\omega_{ k}t)}{k} \Big)\sim\frac{-2 \mathcal{L} m}{N\beta  }\Big(\frac{\cos(\omega_{ 0}t)}{m} +\frac{2}{m}\sum_{k=1}^{\infty}\cos(\omega_{ k}t) \Big),
\end{align}
According to the above asymptotic results, the leading order contribution for $\det \Big(G^{ab}_{x,x}(t)\Big)$ in (\ref{EQ2-2}) is given by
\begin{align}\label{eq-1}
&\hspace{-1.5cm}e_{1}^{2}e_{2}^{2}-e_{2}^{2}\mathcal{M}_{1}^{2}(t)=\frac{4e_{2}^{2}}{N^{2}\beta^{2}m^{4}\mathcal{L}^{2}}\Bigg[(1 +\frac{m^{2} \mathcal{L}^{2}}{2 \pi^{2}}\sum_{k=1}^{\infty}\frac{1}{k^{2}} \Big)^{2}
+\Big(\cos(mt)+ \frac{m^{2} \mathcal{L}^{2}}{2 \pi^{2}}\sum_{k=1}^{\infty}\frac{\cos(\omega_{ k}t)}{k^{2}} \Big)^{2}\Bigg]
\nonumber\\
&\hspace{1cm}=\frac{4e_{2}^{2}}{N^{2}\beta^{2}m^{2}(m)^{2}\mathcal{L}^{2}}\Bigg[\sin^{2}(mt)+\frac{m^{2} \mathcal{L}^{2}}{ \pi^{2}}\sum_{k=1}^{\infty}\frac{1-\cos[(\frac{2\pi k}{\mathcal{L}} )t]}{k^{2}}\Bigg]
\nonumber\\
&\hspace{1cm}=\frac{4e_{2}^{2}}{N^{2}\beta^{2}m^{2}}\Bigg[\frac{\sin^{2}(mt)+Q}{m^{2}\mathcal{L}^{2}}\Bigg],
\end{align}
where we have defined $Q$ as
\begin{align}\label{EQ3-}
Q=\frac{m^{2} \mathcal{L}^{2}}{ \pi^{2}}\sum_{k=1}^{\infty}\frac{1-\cos[(\frac{2\pi k}{\mathcal{L}} )t]}{k^{2}}.
\end{align} 
Therefore, (\ref{EE-Z}) implies that
\begin{equation}
S_{\text{OEE}}(\rho_A(t))\sim 2(1-2 \log 2)+\log\Big(\frac{2e_{2}}{N\beta m}\Big)+\frac{1}{2}\log\Bigg[\frac{\sin^{2}[mt]+Q}{m^{2}\mathcal{L}^{2}}\Bigg].
\end{equation}

\end{document}